  \documentclass[
  aps,
  prd,
  twocolumn,
  showpacs,
  superscriptaddress
  ]{revtex4-2}

  \usepackage[english]{babel}
  \usepackage[dvips]{color}
  \usepackage{natbib}
  \usepackage{latexsym}
  \usepackage{epsfig}
  \usepackage{amsmath}
  \usepackage{amssymb}
  \usepackage{multirow}
  \usepackage{tabularx}
  \usepackage{multirow}
  \usepackage{longtable}
  \usepackage{wasysym}
  \usepackage{hyperref}
  \usepackage{eufrak}

  \usepackage[section]{placeins}

  \usepackage{xcolor}

\begin{document}

  \newcolumntype{C}{>{\centering}X}
  \newcolumntype{Y}{>{\centering\arraybackslash}X}
  \newcolumntype{K}{>{\centering\arraybackslash\arraybackslash}X}

  \newcommand{\MC}[3]{\multicolumn{#1}{#2}{#3}}
  \newcommand{\GS}[1]{\phantom{#1}}
  \newcommand{\MR}[2]{\multirow{#1}*{#2}}

  \setcounter{MaxMatrixCols}{13}

  \title{Electromagnetic nucleon form factors \\
         in the extended vector meson dominance model}

  \author{K.~S.~Kuzmin}
  \email{kkuzmin@theor.jinr.ru}
  \affiliation{Bogoliubov Laboratory of Theoretical Physics,
               Joint Institute for Nuclear Research, \\
               RU-141980 Dubna, Russia}
  \affiliation{National Research Centre ``Kurchatov Institute'', \\
               RU-123182 Moscow, Russia}

  \author{N.~M.~Levashko}
  \email{levashko$\_$nm@mail.ru}
  \affiliation{National Research Centre ``Kurchatov Institute'', \\
               RU-123182 Moscow, Russia}

  \author{M.~I.~Krivoruchenko}
  \email{mikhail.krivoruchenko@itep.ru}
  \affiliation{National Research Centre ``Kurchatov Institute'', \\
               RU-123182 Moscow, Russia}

  \date{\today}

  \begin{abstract}
  An extended vector meson dominance model is developed to describe electromagnetic nucleon form factors.
  The model includes families of the $\rho$- and $\omega$-mesons
  with the associated radial excitations.
  The free parameters of the model are determined using a global statistical analysis of experimental data
  on the electromagnetic nucleon form factors in space- and timelike regions of transferred momenta.
  The vector meson masses and widths are equal to their empirical values, while the residues of form factors at
  the poles corresponding to the ground states of the $\rho$- and $\omega$-mesons are consistent
  with the findings of both the Frazer-Fulco unitarity relations and the Bonn potential for
  coupling constants of the $\rho$- and $\omega$-mesons with nucleons. 
  Theoretical constraints imposed on the model include the quark counting rules,
  the Okubo-Zweig-Iizuka rule,
  the scaling law of Sachs form factors at moderate momentum transfers,
  and the suppression of Sachs form factors near the nucleon--antinucleon threshold.
  A reasonable description of the nucleon form factors in the experimentally accessible range
  of transferred momenta, as well as the electric and magnetic nucleon radii and Zemach radii, is obtained.

  \end{abstract}


  \maketitle

  \section{Introduction}
  \renewcommand{\theequation}{1.\arabic{equation}}
  \setcounter{equation}{0}

  The investigation of the electromagnetic characteristics of nucleons
  remains an active topic in elementary particle physics.
  The electromagnetic form factors of nucleons provide information on the structure
  and interactions of nucleons with other fundamental particles.
  The form factors are analytical functions of the four-momentum transfer squared $t= q^2$.
  Their analytical properties are simple enough to allow the effective use of dispersion theories.
  Unitarity relations for electromagnetic pion and nucleon
  form factors~\cite{Frazer:1959gy,Frazer:1960zzb,Frazer:1960zza,Gounaris:1968mw},
  as well as quark counting rules
  (QCR's)~\cite{Matveev:1973ra,Brodsky:1973kr,Brodsky:1974vy,Vainshtein:1977db}
  \begin{eqnarray}
  F_{1N}(t)&=&O\left(1/t^2\right),                                \label{Eq:F1N} \\
  F_{2N}(t)&=&O\left(1/t^3\right) \ \ \ \text{as}~t \to - \infty, \label{Eq:F2N}
  \end{eqnarray}
  are important frameworks for gaining a deeper understanding of the form factor properties.
  The unitarity relations lead to the vector meson dominance (VMD) model.
  The final-state interaction theorem implies that the model admits
  an extension to include higher radial excitations of vector mesons.
  The extended vector meson dominance (eVMD) model
  can be merged with QCR's.
  The eVMD model maintains gauge invariance of electromagnetic interactions~\cite{Santini:2008pk}
  and thoroughly describes various kinds of processes~\cite{Feynman:1972,Korner:1976hv,%
  Faessler:1999de,Faessler:2000md,Faessler:2009tn,Krivoruchenko:2001jk,Lin:2021xrc}.

  In the 1960s, measurements of the nucleon form factors for the spacelike region
  with moderate momentum transfers $Q^2=-q^2~\lesssim~1~\text{GeV}^2$ revealed
  that Sachs form factors follow the scaling relations (SR's)
  \begin{equation}\label{Eq:scaling_ff}
  G_{Ep}(t)= G_{Mp}(t)/\mu_p= G_{Mn}(t)/\mu_n,
  \end{equation}
  where $\mu_N$ are the magnetic moments of nucleons in nuclear magnetons.
  The measured form factors can be parameterized using the dipole function
  \begin{equation}\label{Eq:dipole_f}
  G_D(t)= \left(1-\dfrac{t}{M^2_V}\right)^{-2},
  \end{equation}
  which corresponds to exponential radial charge and megnetization density distributions,
  where $M_V= 0.84$ GeV is the mass parameter of the vector channel.
  As momentum transfers expand, both SR's and dipole dependence decrease in accuracy.
  The newest solutions involve advanced parameterizations~\cite{Kelly:2004hm,Galster:1971kv,Bradford:2006yz,Bodek:2007ym},
  with the most accurate ones integrated into the Monte Carlo event generators 
  to numerically simulate various types of processes for experiments.

  The measurements of nucleon form factors for the timelike region of $q^2$
  were a significant milestone in the study of nucleon form factors~%
  \cite{Castellano:1973wh,Voci:1997ku,Bassompierre:1977ks,Conversi:1965nn,%
  Hartill:1969ahu,Bisello:1983at,Delcourt:1979ed,Antonelli:1993vz,Antonelli:1994kq}.

  There are numerous theoretical schemes for describing the form factors of nucleons
  (see~\cite{Punjabi:2015bba,Denig:2012by,Pacetti:2014jai} for the review).
  Theoretical models are primarily phenomenological and have a variety of free parameters.
  The analyticity of form factors, QCR's, and unitarity relations are essential for the development of models.

  Up to isotopic factors, the vector component of the weak nucleon current corresponds to the photon vertex
  allowing the weak process amplitudes to be constructed
  using data on the electromagnetic nucleon form factors.
  A thorough description of lepton interactions with nucleons is also required
  for processing and interpreting the results of experiments involving lepton scattering on nucleons,
  particularly for accelerator experiments
  (FNAL MiniBooNE, NO$\nu$A, DUNE, projects at Kamiokande laboratory)
  to study the neutrino properties.

\begin{figure*}[t]
  \includegraphics[width=0.97\linewidth]{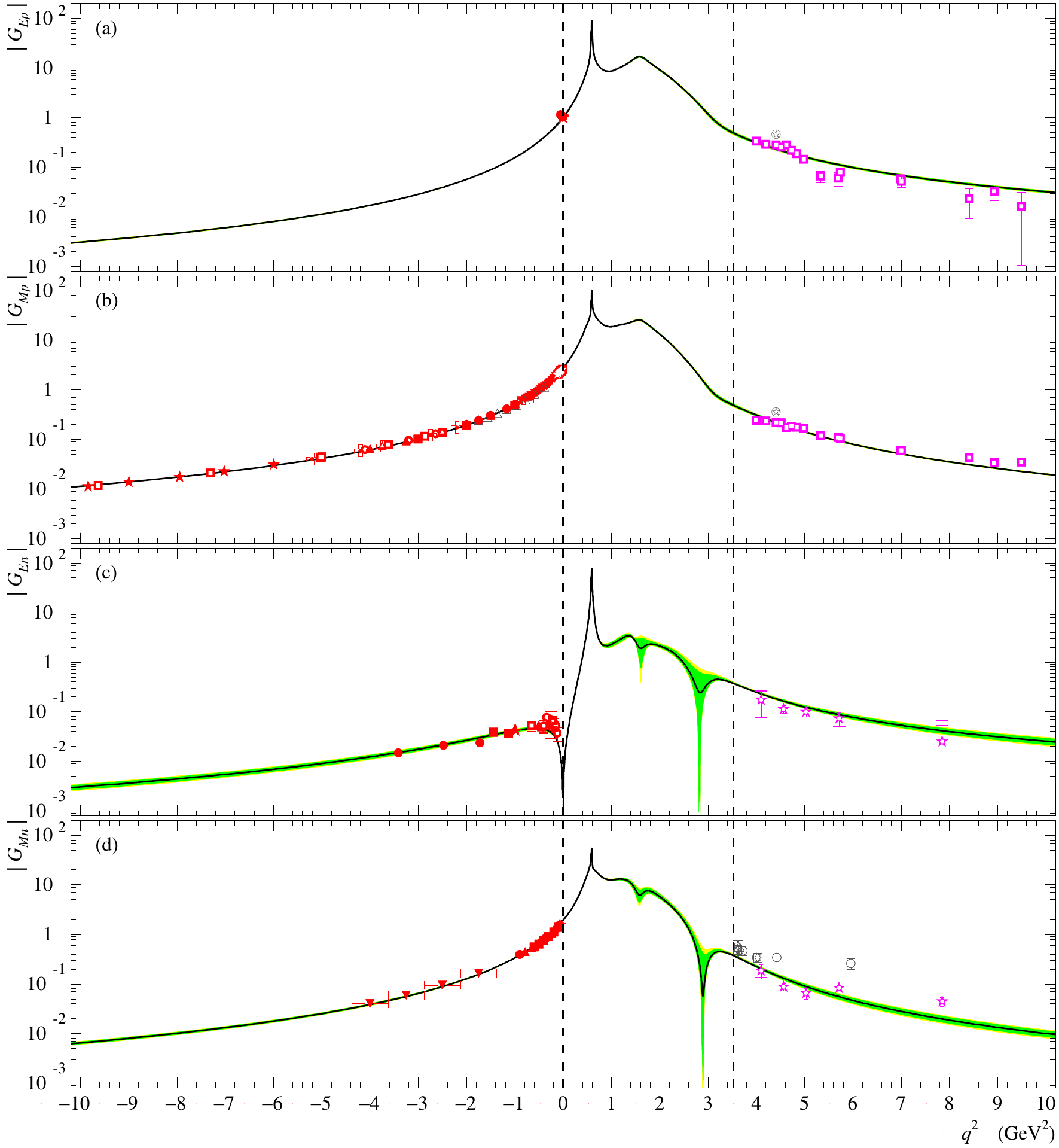}
  \caption{Modules of the electric and magnetic form factors of the proton (a, b) and
           neutron (c, d) for the space- and timelike regions
           of the four-momentum transfer squared
           predicted within the eVMD-VI model
           in comparison with the experimental data.
           Vertical dashed lines limit the experimentally unexplored
           region $0 < q^2 < 4m^2_N$.}
  \label{Fig:GpE_GnE_GpM_GnM_011_2_3_PRD}
  \end{figure*}

  New measurements of nucleon form factors have been reported in recent years,
  with particular emphasis on the timelike region. 
  The eVMD model~\cite{Faessler:2009tn} successfully reproduced earlier datasets with two free parameters.
  We will find that modifying only two parameters is insufficient to reproduce the data.
  The model parameters need to be updated to account for the larger dataset.
  In Sec.~\ref{Sec:eVMDmodel}, we describe the eVMD model in general and develop its advanced version.
  Section~\ref{Sec:Numerical_results} details the experimental data fitting methods and
  reports the new model parameter values, predictions for electric and magnetic nucleon radii,
  and vector meson-nucleon coupling constants.
  The model predictions for the form factors in comparison with the experimental data
  are shown in Figs.~\ref{Fig:GpE_GnE_GpM_GnM_011_2_3_PRD}~--~\ref{Fig:GEp_GMp_GEn_GMn_011_2_1_PRD}.
  The results are summarized in Conclusion.

  \section{Extended \\ vector meson dominance model}
  \label{Sec:eVMDmodel}
  \renewcommand{\theequation}{2.\arabic{equation}}
  \setcounter{equation}{0}

  Decomposition of the electromagnetic nucleon current by Lorentz covariant structures
  determines the electromagnetic form factors of nucleons, $F_{1N}(t)$ and $F_{2N}(t)$:
  \begin{widetext}
  \begin{equation}\label{Eq:current}
  \langle{p_f,s_f\left|J_{\textrm{em}}^\mu(q)\right|p_i,s_i\rangle}=
  \overline{u}\left(p_f,s_f\right)
  \left(F_{1N}(t)\gamma^\mu +\dfrac{1}{2m_N}F_{2N}(t)i\sigma^{\mu\nu}q_\nu\right)u\left(p_i,s_i\right),
  \end{equation}
  \end{widetext}
  where $N= p$ for the proton and $n$ for the neutron, $m_N$ is the nucleon mass.
  The nucleon momenta in the final and initial states are represented by $p_f$ and $p_i$,
  while the polarization vectors of the nucleons are represented by $s_f$ and $s_i$.
  The momentum transfer is $q= p_f-p_i$, and $t= q^2$.
  The equation
  \begin{equation*}
  F_{iI}(t)= \dfrac{1}{2}\left(F_{ip}(t) \pm F_{in}(t)\right)
  \end{equation*}
  determines the isotopic components of the form factors;
  the upper and lower signs correspond to the isoscalar and isovector channels with $I= 0$ and $1$, respectively.
  The matrices $\gamma^{\mu}$ and $\sigma^{\mu\nu}$ are defined as in~\cite{Bjorken:1964}.

  Sachs form factors
  \begin{eqnarray}
  G_{EN}(t) &=& F_{1N}(t) +\dfrac{t}{4m^2_N}F_{2N}(t), \label{Eq:GEN} \\
  G_{MN}(t) &=& F_{1N}(t) +F_{2N}(t)                  \label{Eq:GMN}
  \end{eqnarray}
  are commonly used in analysing experimental data and their theoretical interpretation.
  The value $G_{MN}(0)$ is the nucleon magnetic moment in nuclear magnetons,
  while $F_{2N}(0)$ is its anomalous part.
  The value $F_{1N}(0)$ specifies the nucleon charge $e_N$ in the proton charge units.

  \subsection{Constraints imposed on the eVMD models}

  \subsubsection{Threshold identities}

  At the nucleon--antinucleon threshold ($N\overline{N}$),
  the electric $\left(E\right)$ and magnetic $\left(M\right)$ form factors satisfy the threshold identities (TI's):
  \begin{equation} \label{Eq:th}
  G_{EN}\left(4m^2_N\right)= G_{MN}\left(4m^2_N\right).
  \end{equation}
  The identities are a direct consequence of definitions~\eqref{Eq:GEN} and \eqref{Eq:GMN},
  and they are obtained automatically when the parameterization is carried out using $F_{TN}(t)$.
  If form factors are parameterized in terms of $G_{TN}(t)$, then 
  limitations~\eqref{Eq:th} are applied as an extra requirement.

  \subsubsection{Quark counting rules}

  Hadron form factors obey QCR's~\cite{Matveev:1973ra,Brodsky:1973kr,Brodsky:1974vy}.
  Taking into account additional suppression of amplitudes,
  generated by spin flip of quarks~\cite{Vainshtein:1977db},
  QCR's take the form of Eqs.~\eqref{Eq:F1N}~and~\eqref{Eq:F2N} and give rise to
  \begin{equation}
  G_{EN}(t) \sim G_{MN}(t)= O\left(1/t^2\right) \ \ \ \text{as}\ \ \ t\rightarrow-\infty.
  \end{equation}
  The spin flip has a particularly substantial impact on the electromagnetic transitions
  of high-spin hadrons~\cite{Vainshtein:1977db,Krivoruchenko:2001jk,Faessler:2000md}.

  Electron--proton scattering experiments confirm QCR's up to $Q^2\sim 10~\textrm{GeV}^2$.
  The same holds true for the electromagnetic pion form factor. 
  Quantum chromodynamics (QCD) predicts that the asymptotic regime
  for hadron form factors starts at $Q^2\sim 100~\text{GeV}^2$
  (cf., e.g., the leading term and the next power term
  of the asymptotic decomposition of the pion form factor~\cite{Geshkenbein:1988ef}).
  At intermediate momentum transfers, QCR's are still considered a phenomenological limitation.

  \subsubsection{Scaling relations}

  In the spacelike region for low momentum transfers Sachs form factors fulfill SR's~\eqref{Eq:scaling_ff}.
  With acceptable accuracy $G_{Ep}(t)$, $G_{Mp}(t)/\mu_p$ and $G_{Mn}(t)/\mu_n$
  are described by the dipole function~\eqref{Eq:dipole_f}.
  The asymptotic behavior of $G_D(t)$ is in agreement with QCR's.

  Identity~\eqref{Eq:th} does not contradict SR's, provided the Sachs form factors vanish at $t= 4m^2_N$.
  This assumption seemed to R.~Feynman highly improbable~\cite{Feynman:1972},
  and it contradicts the present-day experiments:
  $\left|G_{Ep}\left(4m^2_N\right)\right|= 0.401 \pm 0.021$~\cite{BESIII:2021rqk}~and~%
  $\left|G_{En}\left(4m^2_N\right)\right|= 0.322 \pm 0.016$~\cite{SND:2023fos} 
  and $0.454 \pm 0.062$~\cite{SND:2022wdb};
  the errors describe the statistical and systematic uncertainties.

  It is noteworthy to mention, however, that the threshold values of nucleon form factors
  are small compared to their values at zero momentum transfer:
  \begin{eqnarray}
  \left|G_{Ep}(4m^2_N)\right| &\sim& \dfrac{1}{2} \, G_{Ep}(0) \sim \dfrac{1}{7} \, G_{Mp}(0), \\
  \left|G_{En}(4m^2_N)\right| &\sim& \dfrac{1}{5} \, \left|G_{Mn}(0)\right|,
  \end{eqnarray}
  implying that SR's may still have an impact on the $N\overline{N}$ threshold. 

  In the spacelike region, when the momentum transfer increases up to $Q^2 \sim 3$ GeV$^2$, 
  the SR's violations become particularly obvious with regard to the ratio $G_{Ep}(t)/G_{Mp}(t)$
  (see Sec.~\ref{Sec:Numerical_results}).

  Although SR's are performed approximately, they allow an accurate
  nonparametric description of an extensive array of experimental data at low momentum transfers.

  \subsubsection{Okubo-Zweig-Iizuka rule}

  According to the Okubo-Zweig-Iizuka (OZI) rule,
  mesons with valence strange quark--antiquark pairs 
  are only weakly coupled to non-strange hadrons~\cite{Griffiths:2008}.
  The theoretical explanation of this rule is based on the fact that
  annihilation of a strange quark--antiquark pair causes the appearance
  of a large momentum in the QCD coupling constant,
  resulting in amplitude suppression.
  The OZI rule suppresses contributions of strange vector mesons $\phi(1020)$ and
  their radial excitations to electromagnetic form factors of non-strange hadrons, including nucleons.

  \subsubsection{Analyticity}

  Analyticity is a universal attribute of quantum field theory,
  arising from causality and Lorentz covariance.
  Form factors are functions with the established analytical properties.
  The functions are analytical in the complex $t$-plane
  with a cut $\left(4m^2_\pi,+\infty\right)$ in the isovector channel and
  a cut $\left(9m^2_\pi,+\infty\right)$ in the isoscalar channel,
  where $m_\pi$ is the pion mass. 
  The class of analytical functions that meet these conditions is relatively vast,
  making it hard to severely constrain form factors based just on analyticity. 

  \subsubsection{Frazer-Fulco unitarity}

  The formulation of two-body unitarity relations for the electromagnetic
  pion form factor~\cite{Frazer:1959gy,Gounaris:1968mw}
  and nucleon form factors~\cite{Frazer:1959gy,Frazer:1960zza,Frazer:1960zzb} became a noteworthy achievement. 
  The formalism used provides form factors with the correct analytical properties.
  The two-body unitarity is justified for spectral functions with the pion energy $t<1$ GeV$^2$.
  The pion form factor is expressed in terms of the mass and decay width of the $\rho$-meson.
  In the experimentally accessible energy range, it drops as $\sim 1/t$ in line with QCR's
  but asymptotically behaves like $1/\ln(-t)$ as $t\to-\infty$.
  In~\cite{Geshkenbein:1988ef}, a model was proposed to correct the behavior of the pion form factor
  by taking into consideration the QCD asymptotics in the leading order together with preasymptotic power corrections.
  To evaluate spectral functions of nucleon form factors, experimental data on the 
  pion--nucleon scattering amplitudes are required,
  and these amplitudes should be extrapolated into the non-physical region of positive
  $t$ values~\cite{Shirkov:1969,Hoehler:1983} (for a recent review, see~\cite{Lin:2021umz}).
  The spectral functions exhibit a $\rho$-meson resonant contribution, validating the application
  of the VMD model in the isovector channel.
  According to quark models, the isoscalar $\omega$-meson is the isotopic companion of the $\rho$-meson,
  so the isoscalar channel can  be included in the VMD models.

  \subsection{eVMD$_1$ model}

  The VMD models violate the QCR's for nucleon form factors.
  Consequently, the model is modified by including radial excitations of $\omega$- and
  $\rho$-mesons, the existence of which was first predicted in the Veneziano model~\cite{Collins:1977jy}.
  The eVMD models that include isotopic families of vector mesons
  are widely used~\cite{Sakurai:1972wk,Hohler:1976ax,Korner:1976hv,%
  Belushkin:2006qa,Lin:2021xrc,Faessler:2009tn,Krivoruchenko:2001jk,Yan:2023yff}.
  This subsection describes a class of eVMD models named the eVMD$_1$
  that precisely match the requirements of TI's, QCR's, and SR's and are utilized
  as a first-order approximation in constructing realistic models of nucleon form factors.

  In the no-width limit of vector mesons, the eVMD representation for Sachs form factors
  can be expressed as follows:
  \begin{equation}\label{Eq:mult}
  G_{TN}(t)= P_{n-2}^{TN}(t){\prod\limits_V}\dfrac{m^2_V}{m^2_V -t},
  \end{equation}
  where
  $T= E$, $M$ indicates the type of form factors,
  $P_{n-2}^{TN}(t)$ is a polynomial of order $n-2$, 
  $n$ is the number of vector mesons, and
  $V= 1, \ldots, n$.
  QCR's are satisfied identically owing to the multiplicative representation,
  whereas TI's impose the constraint $P_{n-2}^{EN}\left(4m^2_N\right)= P_{n-2}^{MN}\left(4m^2_N\right)$.

  The multiplicative representation of the form factors is completely equivalent to the additive one:
  \begin{equation*} \label{EQ}
  G_{TN}(t)= P_{n-2}^{TN}(t) \sum_V \dfrac{1}{t -m^2_V}\underset{~u= m^2_V}{\text{res}} 
             \prod \limits_{V^{\prime}}\dfrac{m_{V^{\prime}}^2}{m_{V^{\prime}}^2 - u}.
  \end{equation*}
  The function \eqref{Eq:mult} is rational, 
  which ensures the decomposition into partial fractions.
  The residues of $F_{iI}(t)$ at $t= m^2_V$ determine the contributions of vector mesons to the form factors.
  The coupling constants $f_{iI}^{VNN}$ of nucleons and vector mesons with isotopic spin $I$
  are identified as follows:
  \begin{equation}\label{Eq:couplings}
  \dfrac{f_{i I}^{VNN}}{g_{\gamma V}}=-\dfrac{1}{m^2_V}\underset{~t= m^2_V}{\text{res}}F_{iI}(t),
  \end{equation}
  where $g_{\gamma V}$ stands for the vector meson coupling constants with photons.

  The finite values of nucleon form factors at $t=~4m^2_N$ within the context of the considered model 
  are attributed to corrections related to vector meson widths which lead to deviations from SR's. 
  Incomplete degeneracy in masses of isoscalar and isovector mesons and a possible 
  contribution of $\phi$-mesons to nucleon form factors
  due to the approximate character of the OZI rule also violate SR's.

  In the eVMD$_1$ model, the threshold values of the form factors are assumed to be zero. 
  The polynomials $P_{n-2}^{TN}(t)$ are factorized to give
  \begin{align*}
  P_{n-2}^{TN}(t) = \left(1 -\dfrac{t}{4m^2_N}\right) P_{n-3}^{TN}(t).
  \end{align*}
  The normalization conditions read $P_{n-3}^{EN}(0)= e_N$ and $P_{n-3}^{MN}(0)= \mu_N$. 
  The polynomials of the magnetic form factors follow SR's and are expressed as
  \[
  P_{n-3}^{Mp}(t)/\mu_p= P_{n-3}^{Mn}(t)/\mu_n= P_{n-3}^{Ep}(t).
  \]
  The representations satisfy TI's, QCR's and SR's.
  There are $n-3$ parameters for each of the polynomials $P_{n-3}^{Ep}(t)$ and $P_{n-3}^{En}(t)$ 
  to fit the experimental data.

  The isotopic components of the form factors equal
  \begin{widetext}
  \begin{eqnarray}
  F_{1I}(t) &=& \left(\left(e_I -\mu_I\dfrac{t}{4m^2_N}\right)
                 P_{n-3}^{Ep}(t) \pm \dfrac{1}{2} P_{n-3}^{En}(t)\right)
                 {\displaystyle\prod\limits_V}\dfrac{m^2_V}{m^2_V -t}, \label{Eq:F1I} \\
  F_{2I}(t) &=& \left(\left(-e_I +\mu_I\right)
                 P_{n-3}^{Ep}(t) \mp \dfrac{1}{2} P_{n-3}^{En}(t)\right)
                 {\displaystyle\prod\limits_V}\dfrac{m^2_V}{m^2_V -t}, \label{Eq:F2I}
  \end{eqnarray}
  \end{widetext}
  where $e_I= \left(e_p \pm e_n\right)/2$ and $\mu_I= \left(\mu_p \pm \mu_n\right)/2$;
  the upper and lower signs indicate the isotopic scalars and vectors, respectively.
  The form factors $G_{TN}(t)$ include contributions of the isoscalar and isovector mesons.
  To eliminate contributions of isoscalar mesons from the isovector channel and vice versa,
  we could demand $P_{n-3}^{Ep}\left(m^2_V\right)=P_{n-3}^{En}\left(m^2_V\right)= 0$,
  which implies, however, that the meson $V$ makes no contribution to the form factors.
  A non-trivial solution is conceivable for
  \[
  \det\left\Vert
  \begin{array}[c]{ll}
   e_I - \mu_I\dfrac{m_V^2}{4m_N^2} & \pm \dfrac{1}{2} \\
  -e_I + \mu_I                      & \mp \dfrac{1}{2}
  \end{array}
  \right\Vert =0.
  \]
  This equation results in $m_V^2= 4m^2_N$ and $G_{TN}\left(4m^2_N\right)\neq 0$, which contradicts SR's.
  To reconcile the representation with TI's, QCR's and SR's, the full $\omega$/$\rho$ degeneracy in masses
  of the ground and excited states of the $\omega$- and $\rho$-mesons is required.
  In this case, the number of $\omega$-mesons is equal to the number of $\rho$-mesons.
  To keep poles simple in the multiplicative representation, the parameter $n$
  should be regarded as the number of mesons in the family of either isoscalar or isovector mesons.

  Equations~\eqref{Eq:F1I} and \eqref{Eq:F2I} with two polynomials $P_{n-3}^{Ep}(t)$ and $P_{n-3}^{En}(t)$
  define the eVMD$_1$ model that strictly meets
  the requirements of TI's, QCR's and SR's.
  The fundamental aspects are as follows: the presence of $n$ vector mesons in isoscalar
  and isovector channels, the mass degeneracy of isoscalar and isovector mesons, 
  and the meson zero widths.
  Strange vector mesons do not find a place in the eVMD$_1$ model since each channel
  has the same number of vector mesons, which is in line with the OZI rule.

  The approximate nature of the OZI rule limits the precision of SR's to
  $O\left(f^{{\phi}NN}_{i 0}/f^{{\omega}NN}_{i 0}\right)$.
  Of course, realistic models account for the widths of vector mesons.
  In the model discussed 
  in the next section, vector meson widths vanish below the two- or three-pion threshold. 
  The meson widths, as a result, produce no effect on spatial behavior of the form factors. 
  When the form factors are analytic functions of $t$, the widths of vector mesons typically 
  modify spacelike asymptotes via the dispersion integrals.
  Restoring QCR's requires imposing conditions of convergence and superconvergence
  of the dispersion integrals on the model parameters.

  In the most basic scenario of $n= 3$, the key eVMD$_1$ model 
  parameters are determined without directly addressing the experimental data.
  The normalization condition $P_{n-3}^{EN}(0)= e_N$ becomes the identity $P_{n-3}^{EN}(t)= e_N$,
  which results in
  \begin{eqnarray}
  F_{1N}(t) &=& \left( e_N -\mu_N\dfrac{t}{4m^2_N}\right)
                {\displaystyle\prod\limits_V}\dfrac{m^2_V}{m^2_V -t}, \label{Eq:3n1} \\
  F_{2N}(t) &=& \left(-e_N +\mu_N\right)
                {\displaystyle\prod\limits_V}\dfrac{m^2_V}{m^2_V -t}. \label{Eq:3n2}
  \end{eqnarray}

  \subsection{eVMD models}

  The MFK model~\cite{Faessler:2009tn} with three meson states in each isotopic channel 
  incorporates two free parameters $c_N$ to account for deviations from SR's.
  It gives the numerator of $F_{1N}(t)$ a less specific structure: $e_N-c_N t$.
  Equation~\eqref{Eq:3n1} suggests $c_N=-\mu_N/\left(4m^2_N\right)$.
  The parameter values of the $n= 3$ eVMD$_1$ model for
  $c_p=-0.79$ GeV$^{-2}$ and
  $c_n= 0.54$ GeV$^{-2}$ are qualitatively comparable to the values
  $c_p=-0.46$ GeV$^{-2}$ and
  $c_n= 0.30$ GeV$^{-2}$ extracted in~\cite{Faessler:2009tn} from the data.
  Table~\ref{Tab:Table} compares the predicted coupling constants of the $\omega$- and $\rho$-mesons
  with nucleons to those of other models. 
  It also displays the electric and magnetic nucleon radii given by Eqs.~\eqref{Eq:rEN} and
  \eqref{Eq:rMN} for $h_N= 0$.

  In recent years, an extensive array of new experimental data on nucleon form factors was accumulated,
  while obvious deviations from the predictions of the MFK model emerged in the timelike region.
  The model has two free parameters only because of the implementation of the basic 
  assumptions of the eVMD model for $n= 3$,
  while parameterizations and phenomenological models of nucleon form actors typically contain
  more than a dozen free parameters~\cite{Ye:2017gyb,Belushkin:2006qa,Lin:2021xrc,Kelly:2004hm,%
  Galster:1971kv,Bradford:2006yz,Bodek:2007ym}.
  Our calculations indicate that adding the latest experimental data and
  simply redefining the parameters $c_p$ and $c_n$ is not enough to reach satisfactory agreement. 
  A more thorough generalization of the MFK model should be provided.
  The current generalization addresses the following aspects:

  i) To cope with the problem of timelike momenta, we introduce $\omega$- and $\rho$-mesons with 
     masses of 1.70 GeV, which are closest to the $N\overline{N}$ threshold.
     The timelike behavior of the form factors is expected to be sensitive to these mesons,
     although their influence on the spacelike region is less obvious.
     Adding one pair of vector mesons increases the number of free parameters in the model from two to six.

     To describe small variations of nucleon form factors near the threshold, 
     two more hypothetical mesons in each of the isotopic channels were introduced 
     in~\cite{Lin:2021xrc}. These mesons are almost twice as heavy as nucleons and 
     have widths of about 1 GeV.

  ii) The second adjustment concerns the parameterization of vector meson widths.
      To guarantee that the form factors are continuous functions, the widths become dependent
      on the four-momentum transfer squared.
      The parameterization~\cite{Fuchs:2002vs} of isoscalar vector meson widths, $\Gamma_V(t)$,
      at $t < t_0 < m_V^2$ employs the Gell-Mann, Sharp and Wagner formula~\cite{Gell-Mann:1962hpq}.
      In the isovector channel, the width is proportional to the two-pion phase space volume
      multiplied by the pion four-momentum squared in the center-of-mass frame.
      At $t>m^2_V$, the widths are regarded constant.
      The value of $t_0$ in both the cases is defined by the condition of matching continuously
      the function and its first derivative at $t= t_0$ and $t=m_V^2$
      with the interpolating second-degree polynomial at $t_0 < t < m_V^2$.
      Below the three-pion threshold, the isoscalar vector mesons have zero widths.
      The isovector mesons have zero widths below the two-pion threshold.
      At $t= m_V^2$, the meson widths are adjusted to the experimental values.

  In the modified model, the most general expression for the form factors takes the form
  \begin{widetext}
  \begin{eqnarray*}
  F_{1I}(t) &=& \sum_V
                P^{1I}_{n - 2}(m_V^2)\dfrac{1}{t -m^2_V +i\sqrt{t}\Gamma_V(t)}\underset{~u= m^2_V}
                {\text{res}}{\prod\limits_{V^{\prime}}}\dfrac{m^2_{V^{\prime}}}{m^2_{V^{\prime}} -u}, \\
  F_{2I}(t) &=& \sum_V
                P^{2I}_{n-3}(m_V^2) \dfrac{1}{t -m^2_V +i\sqrt{t}\Gamma_V(t)}\underset{~u= m^2_V}
                {\text{res}}{\prod\limits_{{V^{\prime}}}}\dfrac{m^2_{V^{\prime}}}{m^2_{V^{\prime}} -u},
  \end{eqnarray*}
  \end{widetext}
  where $P^{iI}_k(t)$ are polynomials of order $k$ that can be expressed in terms of $P^{TN}_{n-2}(t)$
  with the use of Eqs.~\eqref{Eq:GEN} and \eqref{Eq:GMN}. 
  The pole at $t= 4m_N^2$, which emerges when inverting Eqs.~\eqref{Eq:GEN} and \eqref{Eq:GMN}, 
  is fictitious due to the TI's.
  The summation runs over the ground and excited states of vector mesons with isospin $I$. 

  The polynomials for the four vector meson states have the parameterization 
  \begin{eqnarray}
  P^{1I}_2(t) &=& \phantom{-}e_I +c_I t +d_I t^2,  \label{Eq:p1} \\
  P^{2I}_1(t) &=&          - e_I +\mu_I +h_I t,    \label{Eq:p2}
  \end{eqnarray}
  where $c_I= \left(c_p \pm c_n\right)/2$ for isotopic scalars ($+$) and vectors ($-$), respectively,
  and the similarly for $d_I$ and $h_I$.
  This model is referred to as the eVMD-VI model,
  where the number VI denotes the number of fitted model parameters.

  \subsection{Moments of nucleon radii}

  Electric and magnetic moments of nucleon radii are determined by charge and
  magnetization density distributions in the Breit frame of nucleons:
  \begin{equation}\label{Eq:IIIC1}
  \langle{r^s\rangle}_{TN}= \int d\mathbf{r}\,r^s\rho_{TN}(\mathbf{r}).
  \end{equation}
  The identification of $\rho_{TN}(\mathbf{r})$ with the Fourier transforms
  of the electric form factors $\mathcal{G}_{EN}(t) \equiv G_{EN}(t)$ and
  the normalized magnetic form factors $\mathcal{G}_{MN}(t) \equiv G_{MN}(t)/G_{MN}(0)$,
  \begin{equation}\label{Eq:IIIC2}
  \rho_{TN}(\mathbf{r})= \int\dfrac{d\mathbf{Q}}{\left(2\pi\right)^3}
  e^{i\mathbf{Qr}}\mathcal{G}_{TN}\left(-\mathbf{Q}^2\right),
  \end{equation}
  leads to the equations
  \begin{eqnarray}
  \langle{r^{2s+1}\rangle}_{TN}&=&
  \dfrac{\left(2s+2\right)!}{\pi}
  \int_{-\infty}^0\dfrac{\mathcal{G}_{TN}^{\text{reg}}(t)}{t^{s+1}\sqrt{-t}}dt, \label{Eq:odd_powers}\\
  \langle{r^{2s}\rangle}_{TN}  &=&
  \dfrac{(2s+1)!}{s!}\left.
  \dfrac{d^s\mathcal{G}_{TN}(t)}{dt^s}\right\vert_{t= 0},                       \label{Eq:even_powers}
  \end{eqnarray}
  where $s= 1,2,\ldots$ and
  \begin{equation*}
  \mathcal{G}_{TN}^{\text{reg}}(t)=
  \mathcal{G}_{TN}(t)-\sum_{k= 0}^s\dfrac{t^k}{k!}\left. 
  \dfrac{d^k\mathcal{G}_{TN}(t)}{dt^k}\right\vert_{t= 0}.
  \end{equation*}

  The expansion at $t= 0$ goes over the integer powers of $t$ because the form factors
  are analytical functions in a neighborhood of $t= 0$.
  The derivation of Eqs.~\eqref{Eq:odd_powers} and \eqref{Eq:even_powers} is given in Appendix A.

  In the region $\Re(t)<0$ of the complex $t$-plane, the form factors are analytical functions, too.
  The analytical continuation of the form factors in the region $\Re(t) \geq 0$ leads to the eVMD 
  model with zero widths of vector mesons.
  The only singularities of the form factors are
  the thresholds and simple poles corresponding to the vector meson masses.
  The integration in Eq.~\eqref{Eq:odd_powers} is assumed to run along the upper edge of the cut $(-\infty,0)$.
  Bypassing the square root singularity at $t= 0$,
  we complement it with an integral in the opposite direction along the lower edge of the cut.
  We close the contour at infinity and, using Cauchy's theorem, obtain in the model under consideration
  \begin{eqnarray}
  \langle{r^{2s+1}\rangle}_{TN} &=&
  \left(2s+2\right)!\sum_V
  \dfrac{\tilde{f}_{TN}^{VNN}}{g_{\gamma V}}\dfrac{1}{m_V^{2s+1}}, \\ 
  \langle{r^{2s}\rangle}_{TN}   &=&
  \left(2s+1\right)!\sum_V
  \dfrac{\tilde{f}_{TN}^{VNN}}{g_{\gamma V}}\dfrac{1}{m_V^{2s}},   \label{Eq:r2sTNmod}
  \end{eqnarray}
  where (cf.~\eqref{Eq:couplings})
  \begin{equation}
  \dfrac{\tilde{f}_{TN}^{VNN}}{g_{\gamma V}}=
  -\dfrac{1}{m_V^2}\underset{~t= m_V^2}{\text{res}}\mathcal{G}_{TN}(t).
  \end{equation}

  The important case of the second moments allows the result to be presented
  in terms of the model parameters in a simple way:
  \begin{eqnarray}
  \dfrac{1}{6}\langle{r^2\rangle}_{EN} &=&
  c_N+\dfrac{\mu_N -e_N}{4m_N^2} +e_N\sum_V\dfrac{1}{m_V^2}, \label{Eq:rEN} \\
  \dfrac{1}{6}\langle{r^2\rangle}_{MN} &=&
  \dfrac{c_N +h_N}{\mu_N} +\sum_V\dfrac{1}{m_V^2}.           \label{Eq:rMN}
  \end{eqnarray}

  Zemach moments of the charge and magnetization distributions of the proton
  arise in the theory of Lamb shift and hyperfine interactions.
  They are defined by convolution of the charge distribution density
  with the charge or magnetization distribution density of the proton:
  \begin{equation*}
  \langle{r^s\rangle}_{ET}= \int d\mathbf{r}\,r^s\varrho_{ET}(\mathbf{r}),
  \end{equation*}
  where 
  \begin{equation*}
  \varrho_{ET}(\mathbf{r})=
  \int d\mathbf{r}^\prime\rho_{Ep}
  \left(\mathbf{r}-\mathbf{r}^\prime\right)
  \rho_{Tp}\left(\mathbf{r}^\prime\right).
  \end{equation*}

  In terms of proton form factors, using the Fourier transform, the moments take the form:
  \begin{eqnarray}
  \langle{r^{2s+1}\rangle}_{ET} &=&
  \dfrac{\left(2s+2\right)!}{\pi}
  \int_{-\infty}^0
  \dfrac{\left(G_{Ep}(t)\mathcal{G}_{Tp}(t)\right)^{\text{reg}}}{t^{s+1}\sqrt{-t}} dt,~~~~~~~~              \label{Eq:firApp} \\
  \langle{r^{2s}\rangle}_{ET}   &=&
  \sum_{k =0}^s\dfrac{\left(2s+1\right)!\langle{r^{2k}\rangle}_{Ep}\langle{r^{2s-2k}\rangle}_{Tp}}
  {\left(2s-2k+1\right)!\left(2k+1\right)!},                                         \label{Eq:r2s}
  \end{eqnarray}
  where
  \begin{eqnarray*}
  \left(G_{EN}(t)\mathcal{G}_{TN}(t)\right)^{\text{reg}}
  &=& G_{EN}(t)\mathcal{G}_{TN}(t) \\
  &-& \sum_{k= 0}^s\dfrac{t^k}{k!}\left.\dfrac{d^kG_{EN}(t)\mathcal{G}_{TN}(t)}{dt^k}\right\vert_{t= 0}.
  \end{eqnarray*}

  In the eVMD models under consideration,
  \begin{eqnarray*}
  \langle{r^{2s+1}\rangle}_{ET} &=&
  {\left(2s+2\right)!}\sum_{V\neq {V^{\prime}}}
  \dfrac{\tilde{f}_{Ep}^{VNN}\tilde{f}_{Tp}^{{V^{\prime}}NN}}{g_{\gamma V}g_{\gamma {V^{\prime}}}}\times \\
  & & \dfrac{m_V^{2s+3}-m_{V^{\prime}}^{2s+3}}{m_V^{2s+1}m_{V^{\prime}}^{2s+1}\left(m_V^2-m_{V^{\prime}}^2\right)} \\
  &+& \dfrac{\left(2s+3\right)!}{2}\sum_V
      \dfrac{\tilde{f}_{Ep}^{VNN}\tilde{f}_{Tp}^{VNN}}{g_{\gamma V}g_{\gamma V} m_V^{2s+1}},
  \end{eqnarray*}
  while the even moments $\langle{r^{2s}\rangle}_{ET}$ are determined by
  Eqs.~\eqref{Eq:r2sTNmod} and \eqref{Eq:r2s}.

  \section{Global fit procedure and results}
  \label{Sec:Numerical_results}
  \renewcommand{\theequation}{3.\arabic{equation}}
  \setcounter{equation}{0}

  The model parameters $c_I$, $d_I$, and $h_I$ 
  are determined by fitting the experimental data.
  Masses and widths and other physical inputs are taken according to the most recent data,
  as suggested by PDG~\cite{ParticleDataGroup:2024}.
  The evidence for a $1.25$ GeV resonances
  is reported in~\cite{Frenkiel:1972ngp,Bartalucci:1978gy,LAMP2Group:1979ibr,Hammoud:2020aqi}.
  The following set of vector meson masses (given in GeV) is used:
  $m_{\rho}= m_\omega= 0.77$,
  $m_{\rho^{\prime}}= m_{\omega^{\prime}}= 1.25$,
  $m_{\rho^{\prime\prime}}= m_{\omega^{\prime\prime}}= 1.45$, and
  $m_{\rho^{\prime\prime\prime}}= m_{\omega^{\prime\prime\prime}}= 1.70$.
  The widths of vector meson states (given in GeV) are 
  $\Gamma_\rho= 0.15$,
  $\Gamma_\omega= 0.0085$,
  $\Gamma_{\rho^\prime}= 0.3$,
  $\Gamma_{\omega^\prime}= 0.13$,
  $\Gamma_{\rho^{\prime\prime}}= 0.4$,
  $\Gamma_{\omega^{\prime\prime}}= 0.29$,
  $\Gamma_{\rho^{\prime\prime\prime}}= 0.25$, and
  $\Gamma_{\omega^{\prime\prime\prime}}= 0.315$.

  \subsection{Data collection}

  The proton and neutron electromagnetic form factors
  were measured for space- and timelike regions of $q^2$ with the
  Thomas Jefferson National Accelerator Facility (JLab)~%
  \cite{E94110:2004lsx,Qattan:2004ht,JeffersonLabE93-026:2003tty,%
        Punjabi:2005wq,Hu:2006fy,ResonanceSpinStructure:2006oim,%
        E93-038:2003ixb,JeffersonLaboratoryE93-038:2005ryd,%
        JeffersonLabE95-001:2006dax,MacLachlan:2006vw,Riordan:2010id,%
        GEp2gamma:2010gvp,Puckett:2010ac,Paolone:2010qc,Zhan:2011ji,%
        JeffersonLabHallA:2011yyi,Puckett:2017flj,Sulkosky:2017prr,%
        SANE:2018cub,Christy:2021snt,JeffersonLabHallA:1999epl},
  National Accelerator Laboratory
  Stanford Linear Accelerator Center (SLAC)~%
  \cite{Janssens:1965kd,Litt:1969my,Walker:1989af,Sill:1992qw,Lung:1992bu,%
        Andivahis:1994rq,Andreotti:2003bt,%
        E760:1992rvj,E835:1999ml,BaBar:2013ves},
  Massachusetts Institute of Technology Bates Linear Accelerator Center (MIT BLAC)~%
  \cite{Eden:1994ji,BatesFPP:1997rpw,Crawford:2006rz,BLAST:2008bub},
  Cornell Electron Storage Ring (CESR)~%
  \cite{CLEO:2005tiu},
  Brookhaven National Laboratory (BNL)~%
  \cite{Hartill:1969ahu},
  Nationaal Instituut voor Kernfysica en Hoge Energie-Fysica (NIKHEF)~%
  \cite{Anklin:1994ae,Passchier:1999cj,Passchier:1999ju},
  Deutsches Elektronen-SYnchrotron (DESY)~%
  \cite{Bartel:1973rf},
  Bonn University Electron Synchrotron (BUES)~%
  \cite{Berger:1971kr},
  Institute for Nuclear Physics of the University of Mainz,
  Mainz MIcrotron (MAMI)~%
  \cite{Borkowski:1974mb,Anklin:1998ae,Herberg:1999ud,Rohe:1999sh,A1:2001xxy,%
        Kubon:2001rj,Schlimme:2013eoz,Mihovilovic:2016rkr},
  Mainz Electron Linear Accelerator (MELA)~%
  \cite{Borkowski:1974tm,Simon:1980hu},
  Laboratoire de l'Acc\'el\'erateur Lin\'eaire
  Universite de Paris-Sud
  Orsay Colliding Beam Facility (ODCI)~%
  \cite{Delcourt:1979ed,Bisello:1983at,DM2:1988rej,DM2:1990tut,Biagini:1990nb},
  Adone $e^+e^-$ Storage Ring in Frascati (FSR)~%
  \cite{Castellano:1973wh,Antonelli:1992ha,Antonelli:1993vz,Antonelli:1994kq,%
        Voci:1997ku,Antonelli:1998fv},
  European Organization for Nuclear Research (CERN)~%
  \cite{Conversi:1965nn,Bassompierre:1977ks,Bassompierre:1983kt,Bardin:1991rz},
  Budker INP VEPP-2000 $e^+e^-$ Collider (BINP)~%
  \cite{Achasov:2014ncd,CMD-3:2015fvi,Druzhinin:2019gpo,SND:2023fos,%
        SND:2022wdb,Achasov:2023hic,Korol:2023bop}, and
  Beijing Spectrometer at the Electron Positron double-ring collider (BES)~%
  \cite{BES:2005lpy,BESIII:2015axk,BESIII:2019tgo,BESIII:2019hdp,%
        BESIII:2021tbq,BESIII:2021rqk,BESIII:2022rrg}.

  The data of
  $G_{Ep}/G_D$~\cite{Mihovilovic:2016rkr,%
                     Simon:1980hu} (40 data points),
  $G_{Mp}/\left(\mu_p G_D\right)$~\cite{Qattan:2004ht,%
                                        Christy:2021snt,%
                                        E94110:2004lsx,%
                                        Andivahis:1994rq,%
                                        Sill:1992qw,%
                                        Walker:1989af,%
                                        Litt:1969my,%
                                        Janssens:1965kd,%
                                        Borkowski:1974tm,%
                                        Borkowski:1974mb,%
                                        Bartel:1973rf,Brash:2001qq} (84),
  $\mu_p G_{Ep}/G_{Mp}$~\cite{Zhan:2011ji,%
                              JeffersonLabHallA:2011yyi,%
                              Punjabi:2005wq,%
                              Puckett:2017flj,%
                              Puckett:2017flj,%
                              Paolone:2010qc,%
                              MacLachlan:2006vw,%
                              SANE:2018cub,%
                              ResonanceSpinStructure:2006oim,%
                              Hu:2006fy,%
                              BatesFPP:1997rpw,%
                              Crawford:2006rz} (57),
  $G_{En}$~\cite{Sulkosky:2017prr,%
                 Riordan:2010id,%
                 JeffersonLaboratoryE93-038:2005ryd,%
                 JeffersonLabE93-026:2003tty,%
                 BLAST:2008bub,%
                 Eden:1994ji,%
                 Herberg:1999ud,%
                 Rohe:1999sh,%
                 Passchier:1999ju,Passchier:1999cj} (22)
  $G_{Mn}/\left(\mu_n G_D\right)$~\cite{JeffersonLabE95-001:2006dax,%
                                        Kubon:2001rj,%
                                        Anklin:1998ae,%
                                        Anklin:1994ae,%
                                        Lung:1992bu} (19),
  $\mu_n G_{En}/G_{Mn}$~\cite{Schlimme:2013eoz,%
                             Riordan:2010id,%
                             JeffersonLaboratoryE93-038:2005ryd,%
                             BLAST:2008bub} (11),
  $\left|G_p\right|$~\cite{Bardin:1991rz,%
                           Bassompierre:1983kt,%
                           Bassompierre:1977ks,%
                           BESIII:2019hdp,%
                           CMD-3:2015fvi,%
                           BaBar:2013ves} (73)
  $\left|G_{Ep}\right|$~\cite{BESIII:2019hdp} (16),
  $\left|G_{Mp}\right|$~\cite{BESIII:2019hdp} (16),
  $\left|G_{Ep}/G_{Mp}\right|$~\cite{BESIII:2019hdp,%
                                   CMD-3:2015fvi,%
                                   BaBar:2013ves} (23),
  $\left|G_n\right|$~\cite{Achasov:2014ncd,%
                           SND:2023fos,Achasov:2023hic} (19),
  $\left|G_{En}\right|$~\cite{BESIII:2022rrg} (5),
  $\left|G_{Mn}\right|$~\cite{BESIII:2022rrg} (5), and
  $\left|G_{En}/G_{Mn}\right|$~\cite{BESIII:2022rrg} (5)
  were selected for statistical analysis.
  The total set of 395 points consists of 233 and 162 points for the space- and time-like region, respectively. 
  The proton form factors are studied significantly better than the neutron form factors.
  The global fit data set consists of 78\% and 22\% of the proton and neutron data points, respectively.

  The data measured with 
  SLAC 1966~\cite{Janssens:1965kd},
       1970~\cite{Litt:1969my},
       1989~\cite{Walker:1989af},
       1993~\cite{Sill:1992qw}, 
  DESY 1973~\cite{Bartel:1973rf} and
  BUES 1971~\cite{Berger:1971kr},
  SLAC 1994~\cite{Andivahis:1994rq}
  were revisited by Brash {\emph{et al}}.~\cite{Brash:2001qq} and
  Arrington~\cite{Arrington:2003df}, respectively.

  \begin{table*}[t]
  \caption{The parameters $c_p$, $d_p$, $h_p$, $c_n$, $d_n$, and $h_n$
           of the eVMD$_1$, MFK, and the eVMD-VI models.
           The parameter values supplied with $1\sigma$ ($2\sigma$) errors correspond to the eVMD-VI model.
           The values of the coupling constants of vector mesons with nucleons are given
           in comparison with the estimates based on the unitarity relations for isovector
           nucleon form factors~\cite{Lendel:1966,Hohler:1974ht,Krivoruchenko:1997nb}
           and predictions of the boson exchange $NN$ interaction Bonn model~\cite{Machleidt:1987hj}.
           The values of electric and magnetic nucleon radii squared are also provided
           (see explanations in the text).}
  \label{Tab:Table}
  \center{
  \begin{tabularx}{\linewidth}{c|cCcccccc} \hline\hline\noalign{\smallskip}
  Parameters,                   &
  eVMD$_1$ model                &
  MFK~model                     &
  eVMD-VI model                 &
  PDG 2024                      &
  Ref.                          &
  Ref.                          &
  Ref.                          &
  Bonn model                     \\
  couplings, radii              &
                                &
  \cite{Faessler:2009tn}        &
                                &
  \cite{ParticleDataGroup:2024} &
  \cite{Lendel:1966}            &
  \cite{Hohler:1974ht}          &
  \cite{Krivoruchenko:1997nb}   &
  \cite{Machleidt:1987hj}        \\
  \noalign{\smallskip}\hline\noalign{\smallskip}
  \MC{1}{r|}{$c_p\left[\text{GeV}^{-2}\right]$~~}                     &$-0.792$          &$-0.463$     &    $- 0.812{\pm}0.008(0.011)$&                               &           &               &                     &           \\ \noalign{\smallskip}
  \MC{1}{r|}{$d_p\left[\text{GeV}^{-4}\right]$~~}                     &$-$               &$-$          &$\GS{-}0.221{\pm}0.003(0.004)$&                               &           &               &                     &           \\ \noalign{\smallskip}
  \MC{1}{r|}{$h_p\left[\text{GeV}^{-2}\right]$~~}                     &$-$               &$-$          &    $- 0.552{\pm}0.009(0.011)$&                               &           &               &                     &           \\ \noalign{\smallskip}
  \MC{1}{r|}{$c_n\left[\text{GeV}^{-2}\right]$~~}                     &$\GS{-}0.542$     &$\GS{-}0.297$&$\GS{-}0.256{\pm}0.016(0.022)$&                               &           &               &                     &           \\ \noalign{\smallskip}
  \MC{1}{r|}{$d_n\left[\text{GeV}^{-4}\right]$~~}                     &$-$               &$-$          &    $ -0.110{\pm}0.010(0.013)$&                               &           &               &                     &           \\ \noalign{\smallskip}
  \MC{1}{r|}{$h_n\left[\text{GeV}^{-2}\right]$~~}                     &$-$               &$-$          &$\GS{-}0.650{\pm}0.039(0.052)$&                               &           &               &                     &           \\ \noalign{\smallskip}
  \hline\noalign{\smallskip}
  $f_{10}^{\omega NN}$                                                &$~~16.4$          &$~~17.2$     &    $~~17.1$                  &                               &           &               &                     &$15.9$     \\ \noalign{\smallskip}
  $f_{20}^{\omega NN}$                                                &$-2.31$           &$-2.47$      &    $-1.50$                   &                               &           &               &                     &$0$        \\ \noalign{\smallskip}
  $f_{11}^{\rho NN}$                                                  &$\GS{-}1.18$      &$\GS{-}3.02$ &$\GS{-}3.43$                  &                               &$\GS{1}5.8$&$2.61{\pm}0.40$&$-0.15{\pm}1.81$     &$\GS{1}3.2$\\ \noalign{\smallskip}
  $f_{21}^{\rho NN}$                                                  &$~~20.9$          &$~~20.8$     &    $~~21.3$                  &                               &$15.5$     &$15.8{\pm}2.87$&$\GS{-}14.0{\pm}0.60$&$19.8$     \\ \noalign{\smallskip}
  \hline\noalign{\smallskip}
  \MC{1}{r|}{$\langle{r^2\rangle}^{1/2}_{Ep}\left[\text{fm}\right]$~~}&$\GS{-0}0.77$     &$\GS{-}0.82$ &$\GS{-}0.815$                 &$\GS{-}0.8409{\pm}0.0004$      &           &               &                     &           \\ \noalign{\smallskip}
  \MC{1}{r|}{$\langle{r^2\rangle}_{En}\left[\text{fm}^2\right]$~~}    &$\GS{-0}0\GS{.77}$&$-0.06$      &    $- 0.067$                 &     $-0.1155{\pm}0.0017$      &           &               &                     &           \\ \noalign{\smallskip}
  \MC{1}{r|}{$\langle{r^2\rangle}^{1/2}_{Mp}\left[\text{fm}\right]$~~}&$\GS{-0}0.77$     &$\GS{-}0.78$ &$\GS{-}0.788$                 &$\GS{-}0.851{\pm}0.026$        &           &               &                     &           \\ \noalign{\smallskip}
  \MC{1}{r|}{$\langle{r^2\rangle}^{1/2}_{Mn}\left[\text{fm}\right]$~~}&$\GS{-0}0.77$     &$\GS{-}0.79$ &$\GS{-}0.791$                 &$\GS{-}0.864_{-0.008}^{+0.009}$&           &               &                     &           \\ \noalign{\smallskip}
  \noalign{\smallskip}\hline\hline
  \end{tabularx}}
  \end{table*}

  To distinguish the values of $\left|G_{EN}(t)\right|$ and $\left|G_{MN}(t)\right|$
  above the $N\overline{N}$ threshold, the angular distribution of the final particles is measured.
  Most of the existing data are given for the effective form factor 
  \begin{equation}
  \label{Eq:Geff}
  G^{\text{eff}}_N(t)= \sqrt{\dfrac{\left|G_{EN}(t)\right|^2 +
                       \eta\left|G_{MN}(t)\right|^2}{1+\eta}},
  \end{equation}
  where $\eta= t/\left(2m_N^2\right)$,
  extracted from the total cross sections.
  The current analysis uses the data that are not based on the assumption
  $\left|G_{EN}(t)\right|= \left|G_{MN} (t)\right|$.

  Data analyses mainly use the single-photon exchange approximation,
  which is justified for low momentum transfers.
  The corrections associated with the two-photon exchange approach $(0.2 \div 0.6)\%$ at
  $Q^2~\sim~4$~GeV$^2$ for the proton and increase with $Q^2$,
  causing some uncertainty in the measurement data. 
  Data analysis including a theoretical description of two-photon exchange and radiative corrections
  provides an improved evaluation of the form factor values and measurement errors~\cite{Ye:2017gyb}.

  \subsection{Data analysis}

  \begin{figure}[htb!]
  \includegraphics[width=0.97\linewidth]{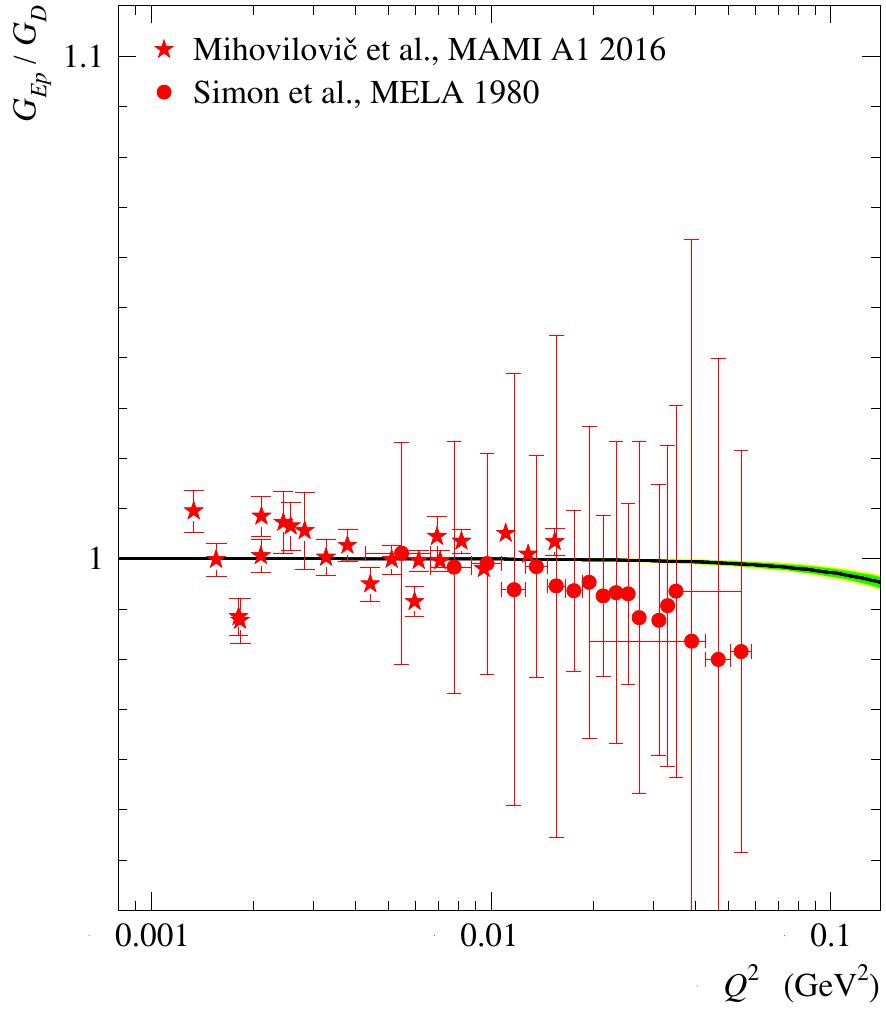}
  \caption{Normalized electric form factor of the proton
           measured for the spacelike region with
           MAMI A1 2016~\cite{Mihovilovic:2016rkr} and
           MELA 1980~\cite{Simon:1980hu}
           in comparison with the eVMD-VI model.}
  \label{Fig:GpE_GD_011_2_1_PRD}
  \end{figure}
  \begin{figure*}[htb!]
  \includegraphics[width=0.97\linewidth]{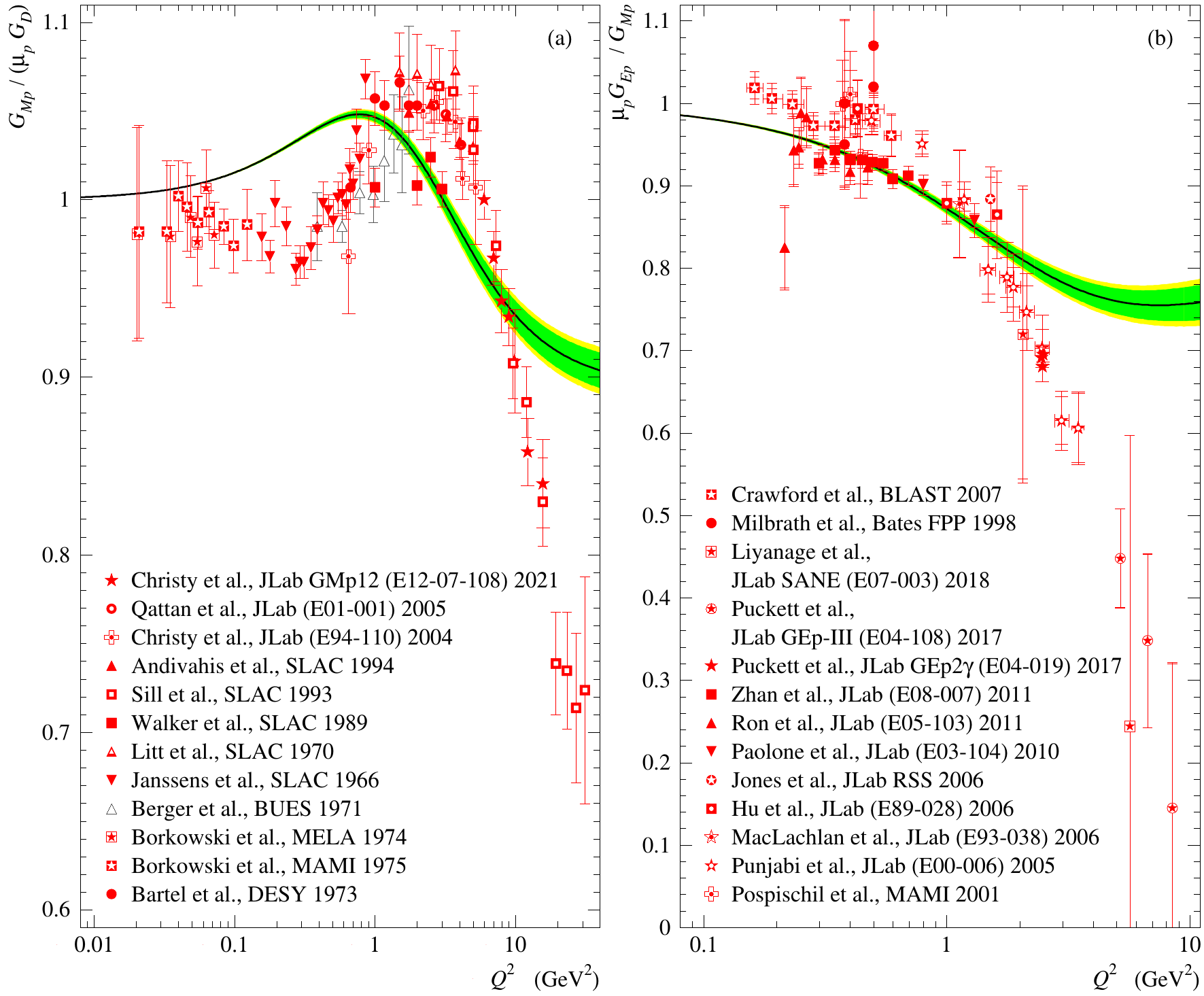}
  \caption{Normalized magnetic form factor of the proton (a) and
           the ratio of normalized electric to magnetic form factors of the proton (b)
           measured for the spacelike region with
           BLAST 2007~\cite{Crawford:2006rz},
           MIT Bates FPP 1998~\cite{BatesFPP:1997rpw},
           JLab Hall A GMp12 (E12-07-108) 2021~\cite{Christy:2021snt},
                Hall C SANE (E07-003) 2018~\cite{SANE:2018cub},
                       GEp-III (E04-108) 2017~\cite{Puckett:2017flj},
                       GEp2$\gamma$ (E04-019) 2017~\cite{Puckett:2010ac,Puckett:2017flj},
                                    (E03-104) 2010~\cite{Paolone:2010qc},
                Hall A (E05-103) 2011~\cite{JeffersonLabHallA:2011yyi},
                       (E08-007) 2011~\cite{Zhan:2011ji},
                Hall C RSS 2006~\cite{ResonanceSpinStructure:2006oim},
                Hall A (E89-028) 2006~\cite{Hu:2006fy},
                       (E93-038) 2006~\cite{MacLachlan:2006vw},
                       (E01-001) 2005~\cite{Qattan:2004ht},
                       (E00-006) 2005~\cite{Punjabi:2005wq},
                       (E94-110) 2004~\cite{E94110:2004lsx},
           SLAC 1994~\cite{Andivahis:1994rq},
                1993~\cite{Sill:1992qw},
                1989~\cite{Walker:1989af},
                1970~\cite{Litt:1969my},
                1966~\cite{Janssens:1965kd},
           BUES 1971~\cite{Berger:1971kr},
           MELA 1974~\cite{Borkowski:1974tm},
           MAMI 2001~\cite{A1:2001xxy},
                1975~\cite{Borkowski:1974mb}, and
           DESY 1973~\cite{Bartel:1973rf}
           in comparison with the eVMD-VI model.}
  \label{Fig:GpM_mGD_mGpE_GpM_011_2_1_PRD}
  \end{figure*}
  \begin{figure*}[htb!]
  \begin{minipage}[htb!]{0.97\linewidth}
  \center{\includegraphics[width=\linewidth]{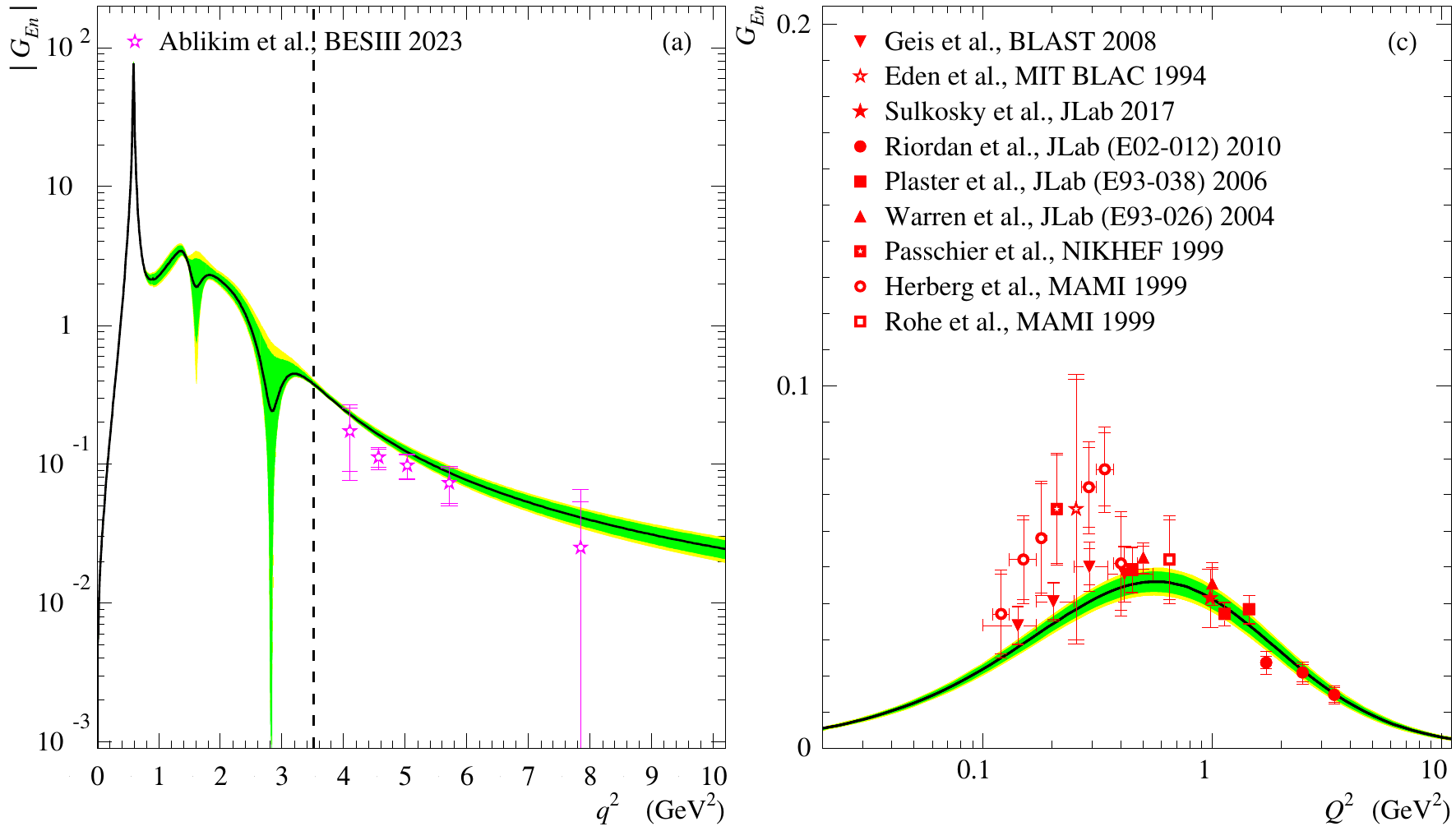}} \\
  \end{minipage}
  \begin{minipage}[htb!]{0.97\linewidth}
  \center{\includegraphics[width=\linewidth]{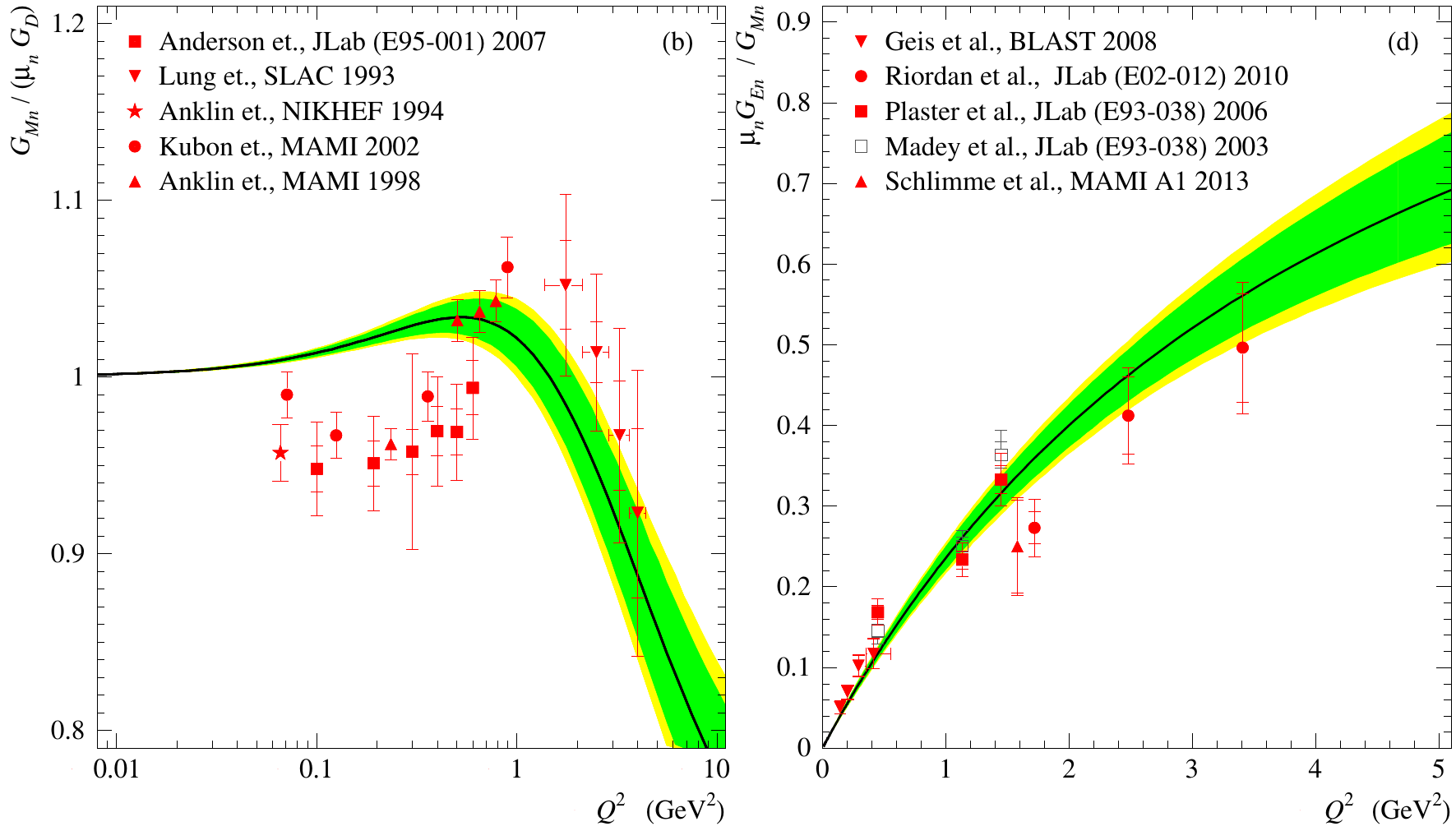}} \\
  \end{minipage}
  \caption{Module of electric form factor of the neutron (a)
           measured for the timelike region with BESIII 2023~\cite{BESIII:2022rrg};
           normalized magnetic form factor of the neutron (b),
           electric form factor of the neutron (c), and
           the ratio of normalized electric to magnetic form factors of the neutron (d)
           measured for the spacelike region with
           BLAST 2008~\cite{BLAST:2008bub},
           MIT BLAC 1994~\cite{Eden:1994ji},
           JLab Hall A 2017~\cite{Sulkosky:2017prr},
                       (E02-012) 2010~\cite{Riordan:2010id},
                       (E95-001) 2007~\cite{JeffersonLabE95-001:2006dax},
                Hall C (E93-038) 2006~\cite{JeffersonLaboratoryE93-038:2005ryd}
                       (previous rapidly published result~\cite{E93-038:2003ixb} is also shown),
                      (E93-026) 2004~\cite{JeffersonLabE93-026:2003tty},
           NIKHEF 1999~\cite{Passchier:1999cj}
           MAMI 1999~\cite{Herberg:1999ud,%
                           Rohe:1999sh},
           SLAC 1993\cite{Lung:1992bu},
           NIKHEF 1994~\cite{Anklin:1994ae},
           MAMI A1 2013~\cite{Schlimme:2013eoz},
           MAMI 2002~\cite{Kubon:2001rj}, and
                1998~\cite{Anklin:1998ae}
           in comparison with the eVMD-VI model.
           The data from~\cite{E93-038:2003ixb,Schlimme:2013eoz}
           are normalized on $\mu_n$.}
  \label{Fig:GnE_GnM_mGD_mGnE_GnM_011_2_3_PRD}
  \end{figure*}
  \begin{figure*}[htb!]
  \includegraphics[width=0.97\linewidth]{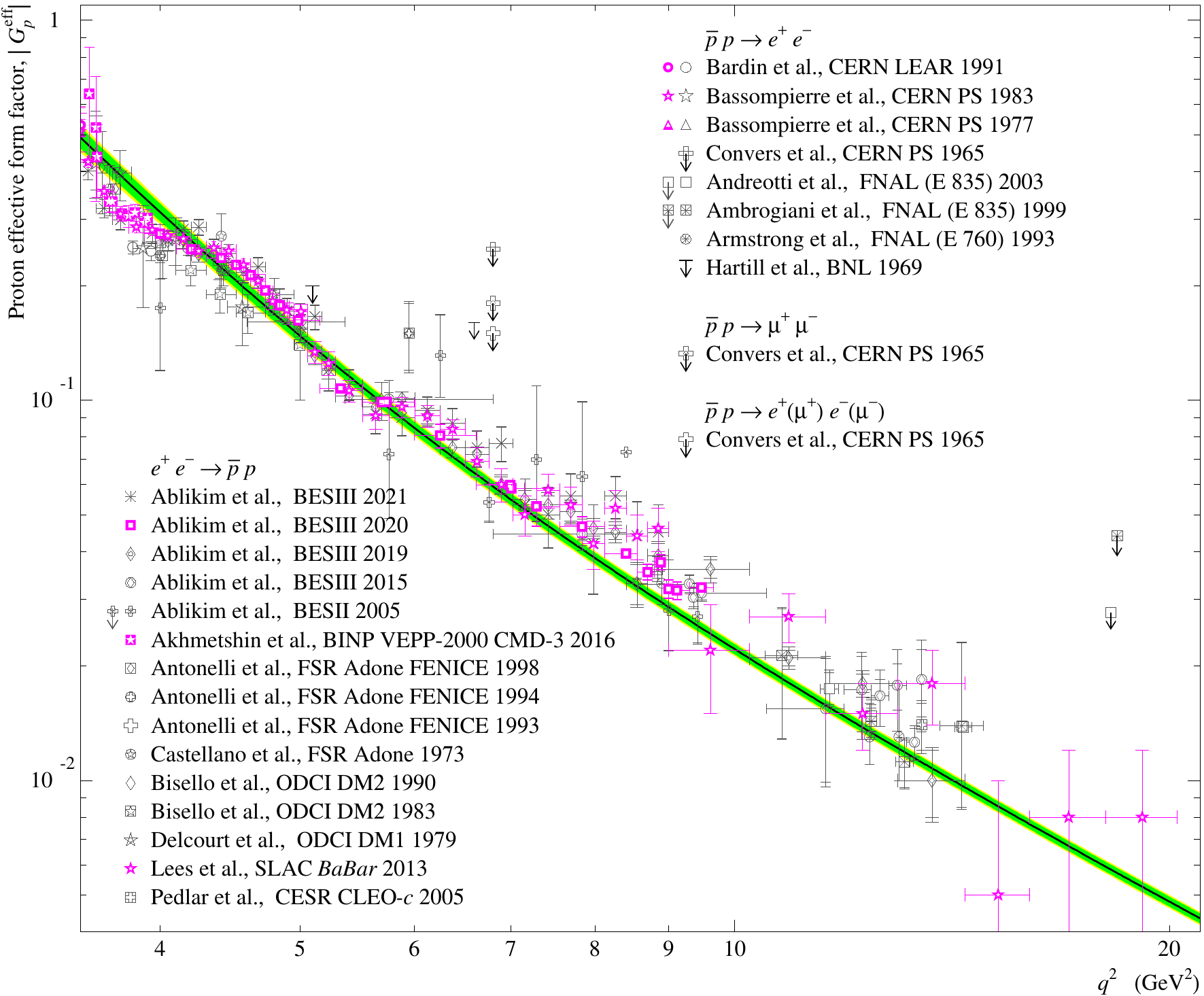}
  \caption{Proton effective form factor~\eqref{Eq:Geff} measured
           for the timelike region with
           BESIII 2021~\cite{BESIII:2021rqk},
                  2020~\cite{BESIII:2019hdp},
                  2019~\cite{BESIII:2019tgo},
                  2015~\cite{BESIII:2015axk},
           BESII 2005~\cite{BES:2005lpy},
           BINP VEPP-2000 CMD-3~\cite{CMD-3:2015fvi},
           FSR Adone FENICE 1998~\cite{Antonelli:1998fv},
                            1994~\cite{Antonelli:1994kq},
                            1993~\cite{Antonelli:1993vz},
               Adone 1973~\cite{Castellano:1973wh},
           OSCI ISR DM2 1990~\cite{DM2:1990tut,DM2:1988rej},
                        1983~\cite{Bisello:1983at},
                    DM1 1979~\cite{Delcourt:1979ed},                         
           SLAC {\it BaBar}~\cite{BaBar:2013ves}, and
           CESR CLEO-$c$ 2005~\cite{CLEO:2005tiu}
           from the analysis of $e^+e^-{\to}\,\overline{p}p$ reaction,
           CERN LEAR 1991~\cite{Bardin:1991rz},
                  PS 1983~\cite{Bassompierre:1983kt},
                     1977~\cite{Bassompierre:1977ks},
                     1965~\cite{Conversi:1965nn},
           FNAL (E 835) 2003~\cite{Andreotti:2003bt},
                        1999~\cite{E835:1999ml},
                (E 760) 1993~\cite{E760:1992rvj}, and
           BNL 1969~\cite{Hartill:1969ahu}
           from the analysis of $\overline{p}p\,{\to}\,e^+e^-$ reaction
           in comparison with the eVMD-VI model.
           The data of BNL 1969~\cite{Hartill:1969ahu},
           BES 2005~\cite{BES:2005lpy} measured at $\sqrt{s}= 2.9\,\text{GeV}$,
           and the last data points of
           FNAL (E 835)~\cite{Andreotti:2003bt,E835:1999ml}
           are constraints for the form factor.
           The constraints obtained with the pioneer experiment
           CERN PS 1965~\cite{Conversi:1965nn}
           from the analysis of $\overline{p}p\,{\to}\,e^+e^-$ and/or 
           $\overline{p}p\,{\to}\,\mu^+\mu^-$ reactions are also shown.
           The data published in~\cite{Delcourt:1979ed,%
                                            DM2:1988rej,%
                                            DM2:1990tut}
           are recalculated from $\left|G^2_p\right|$ data.}
  \label{Fig:Gp_011_2_1_PRD}
  \end{figure*}
  \begin{figure*}[htb!]
  \includegraphics[width=0.97\linewidth]{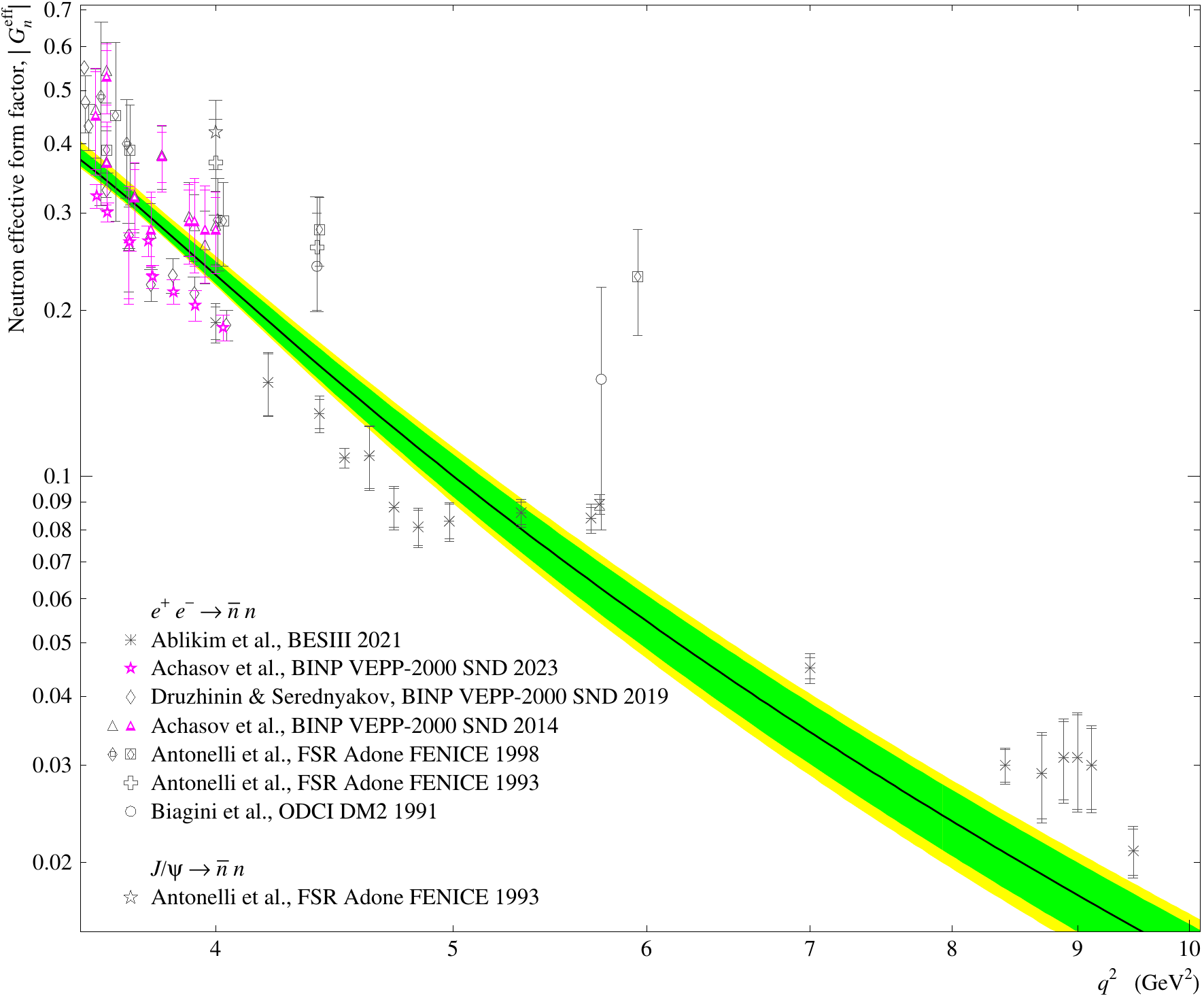}
  \caption{Neutron effective form factor~\eqref{Eq:Geff} measured 
           for the timelike region with
           BESIII 2021~\cite{BESIII:2021tbq},
           BINP VEPP-2000 SND 2023~\cite{SND:2023fos,Achasov:2023hic},
                              2019~\cite{Druzhinin:2019gpo},
                              2014~\cite{Achasov:2014ncd},
           FSR Adone FENICE 1998~\cite{Antonelli:1998fv}
                            (data point measured at $q^2= 3.61\,\text{GeV}^2$
                             as the 68\% C.L. limit of the form factor
                             is borrowed from~\cite{Achasov:2014ncd} and
                             shown here for completeness),
                            1993~\cite{Antonelli:1993vz}, and
           OSCI ISR DM2~\cite{Biagini:1990nb}
           from the analysis of $e^+e^-{\to}\,\overline{n}n$ reaction,
           FENICE 1993~\cite{Antonelli:1992ha}
           from the analysis of $J/\psi \to \overline{n} n$ reaction
           in comparison with the eVMD-VI model.
           Data published in~\cite{Druzhinin:2019gpo} and~\cite{Achasov:2014ncd}
           (marked as open triangle) are borrowed from~\cite{BESIII:2021tbq}.}
  \label{Fig:Gn_011_2_1_PRD}
  \end{figure*}
  \begin{figure*}[htb!]
  \includegraphics[width=0.97\linewidth]{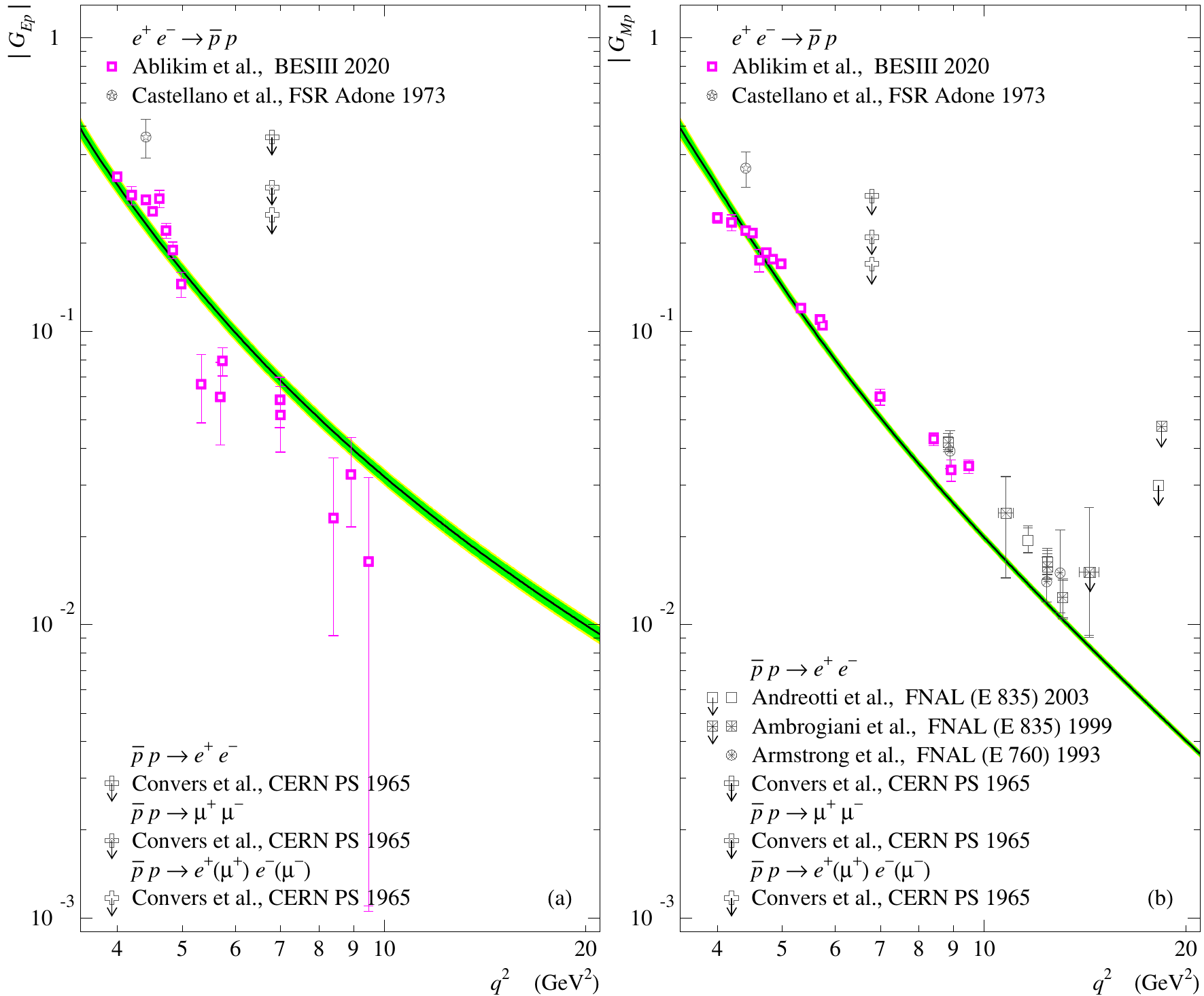}
  \caption{Modules of electric (a) and magnetic (b) form factors of the proton
           measured for the timelike region with
           BESIII 2020~\cite{BESIII:2019hdp} and
           FSR Adone 1973~\cite{Castellano:1973wh}
           from the analysis of $e^+e^-{\to}\,\overline{p}p$ reaction,
           FNAL (E 835) 2003~\cite{Andreotti:2003bt},
                        1999~\cite{E835:1999ml}, and
                (E 760) 1993~\cite{E760:1992rvj}
           from the analysis of $\overline{p}p\,{\to}\,e^+e^-$ reaction
           in comparison with the eVMD-VI model.
           The limits obtained with
           CERN PS 1965~\cite{Conversi:1965nn}
           from the analysis of $\overline{p}p\,{\to}\,e^+e^-$ or/and
           $\overline{p}p\,{\to}\,\mu^+\mu^-$ reactions also shown.}
  \label{Fig:GEp_GMp_011_2_1_PRD}
  \end{figure*}
  \begin{figure*}[htb!]
  \includegraphics[width=0.97\linewidth]{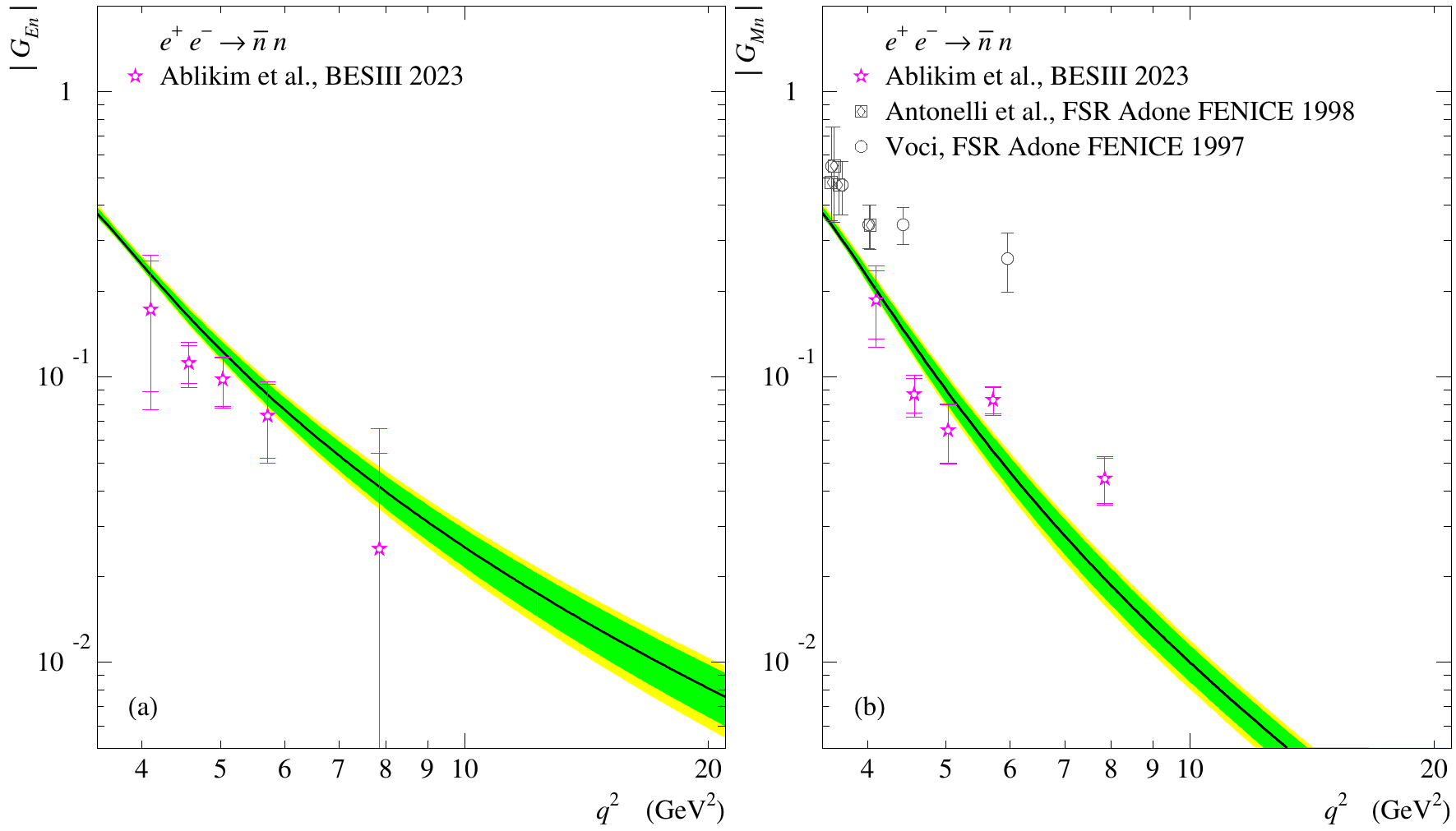}
  \caption{Modules of electric (a) and magnetic (b) form factor of the neutron
           measured for the timelike region with
           BESIII 2023~\cite{BESIII:2022rrg},
           Frascati SR Adone FENICE 1998~\cite{Antonelli:1998fv}
           (preliminary data of 1997~\cite{Voci:1997ku} are shown for completeness)
           from the analysis of $e^+e^-{\to}\,\overline{n}n$ reaction
           in comparison with the eVMD-VI model.}
  \label{Fig:GEn_GMn_011_2_1_PRD}
  \end{figure*}
  \begin{figure*}[htb!]
  \includegraphics[width=0.97\linewidth]{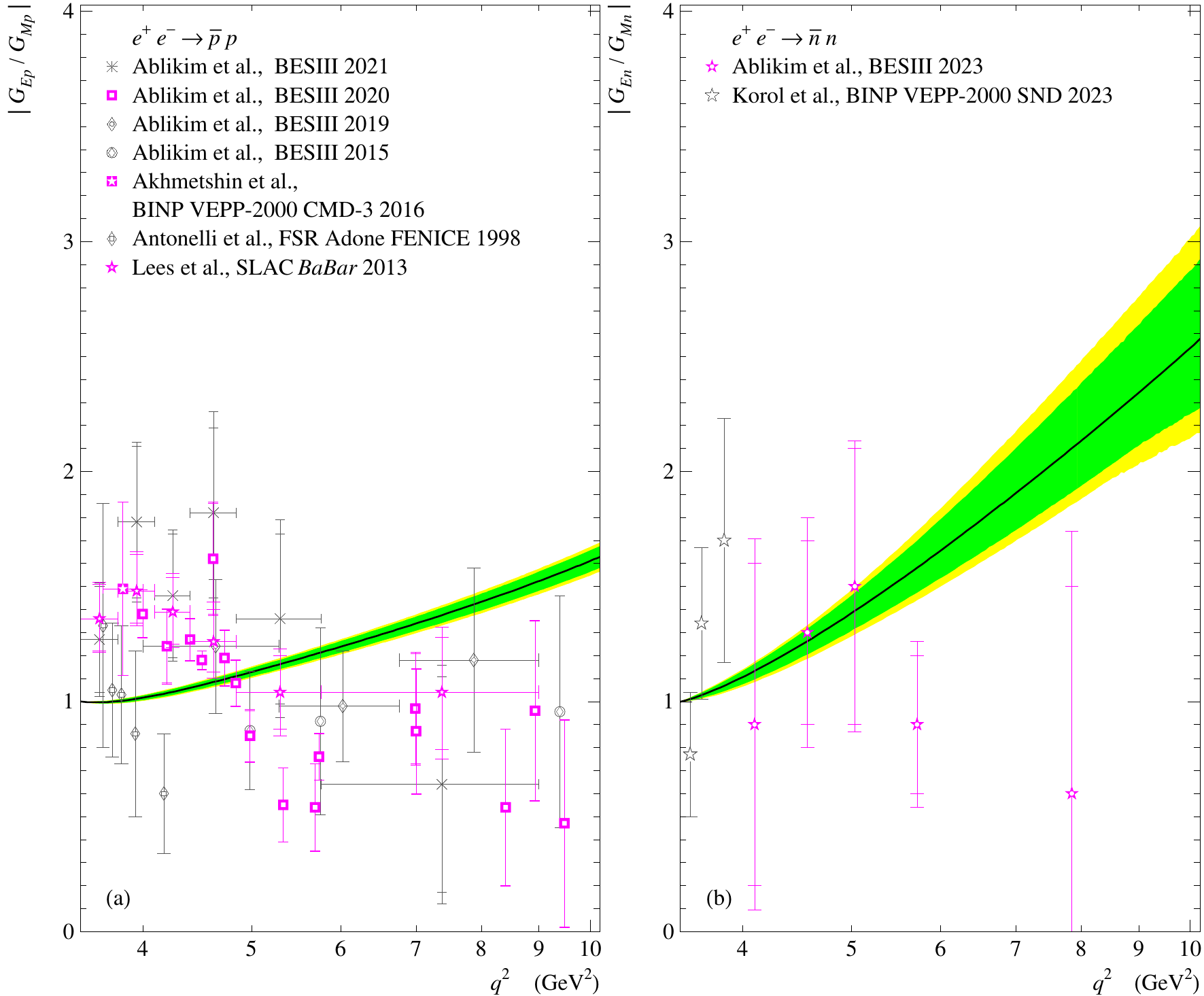}
  \caption{Modules of the ratios of electric to magnetic form factors
           of the proton (a) and neutron (b) measured for the timelike region with
           BESIII 2021~\cite{BESIII:2021rqk},
                  2020~\cite{BESIII:2019hdp},
                  2019~\cite{BESIII:2019tgo},
                  2015~\cite{BESIII:2015axk},
           BINP VEPP-2000 CMD-3 2016~\cite{CMD-3:2015fvi},
           Frascati SR FENICE 1998~\cite{Antonelli:1998fv,Lin:2021xrc},
           SLAC {\it BaBar} 2013~\cite{BaBar:2013ves}
           from the analysis of $e^+e^-{\to}\,\overline{p}p$ reaction and
           BESIII 2023~\cite{BESIII:2022rrg},
           BINP VEPP-2000 SND 2023~\cite{Korol:2023bop}
           from the analysis of $e^+e^-{\to}\,\overline{n}n$ reaction, respectively,
           in comparison with the eVMD-VI model.}
  \label{Fig:GEp_GMp_GEn_GMn_011_2_1_PRD}
  \end{figure*}

  We use the simplest form of the chi-squared function
  \begin{align}
  \chi^2\left(\boldsymbol{\theta}\right)
  &= \sum_{t_s < 0}\left(\dfrac{\mathcal{F}\left(t_s,\boldsymbol{\theta}\right)-
                                \mathcal{F}_s^{_{\exp}}}
                               {\Delta\mathcal{F}_s^{_{\exp}}}\right)^2               \nonumber \\
  &+ \sum_{t_s > 0}\left(\dfrac{\left|\mathcal{F}\left(t_s,\boldsymbol{\theta}\right)\right|-
                                \left|\mathcal{F}_s^{_{\exp}}\right|}
                               {\Delta\left|\mathcal{F}_s^{_{\exp}}\right|}\right)^2, \label{Eq:chi2}
  \end{align}
  where $\mathcal{F}\left(t,\boldsymbol{\theta}\right)$ is one of the form factors
  $F_{iN}(t)$, $G_{TN}(t)$, or $G^{\text{eff}}_N(t)$, and
  $\mathcal{F}_s^{_{\exp}}$ is its experimental value for $t= t_s$
  with experimental error $\Delta\mathcal{F}_s^{_{\exp}}$.
  The data points are enumerated by the index $s$, and the parameter set is 
  $\boldsymbol{\theta}= \left(c_p,\ d_p,\ h_p,\ c_n,\ d_n,\ h_n\right)$.
  The theoretical values of $\mathcal{F}\left(t_s,\boldsymbol{\theta}\right)$ are averaged
  over the experimental errors of the four-momentum transfer squared,
  as provided in the original papers. 
  When uncertainties are not supplied, the error is assumed to be the difference between
  neighboring $t_s$ values.
  In Eq.~\eqref{Eq:chi2}, the first term includes data from the spacelike region, while the second term
  includes data for the timelike region based on the measured absolute form factor values.

  When the upper and lower errors in form factors differ,
  the chi-squared estimation uses their root mean square values.
  The errors used in Eq.~\eqref{Eq:chi2} are calculated from statistical and systematic errors.
  The chi-squared function incorporates all data points with equal weights and no further constraints
  about possible normalization to specific form factor values are applied.
  The chi-squared minimization is performed using the CERN function minimization and
  error analysis package MINUIT (version 94.1)~\cite{James:1994vla,James:1975dr}.
  The errors of the parameters quoted below correspond
  to the one- and two-standard deviation ($1\sigma$ and $2\sigma$) errors.

  To assess the uncertainties in the model parameters, we first simulate a number of virtual experiments,
  the measurement results of which are determined by a normal distribution of the form factor values
  relative to the average and error of the form factors of the real experiment.
  The chi-square function is minimized to determine the model parameters for each simulation.
  At the final stage of the ``bootstrap'', based on the sample of
  $10^4$ virtual experiments, the average value and $1\sigma$ ($2\sigma$) variance of the parameters
  $\boldsymbol{\theta}$ are found to be~(given~in~GeV$^{-2}$)
  \begin{align*}
  c_p=         - 0.812{\pm}0.009(0.012), && c_n=\phantom{-}0.256{\pm}0.018(0.025), \\
  d_p=\phantom{-}0.221{\pm}0.003(0.004), && d_n=         - 0.110{\pm}0.011(0.015), \\
  h_p=         - 0.552{\pm}0.010(0.013), && h_n=\phantom{-}0.650{\pm}0.044(0.058).
  \end{align*}

  The uncertainties in the parameters $\boldsymbol{\theta}$ can alternatively be determined
  by directly generating an error ellipsoid from the error covariance matrix and, separately,
  using the standard package integrated into MINUIT.
  We use a flat prior probability density function of the parameters
  for evaluation of the posterior probability density function within the Bayesian method.
  The ellipsoids of errors in the parameter space are determined from the equation
  $\chi^2\left(\boldsymbol{\theta}\right)= 
   \chi^2\left(\boldsymbol{\theta}^\ast\right) +\text{constant}$,
  where $\boldsymbol{\theta}^*$ denotes the parameter set values found
  by minimizing the chi-squared function~\cite{Cowan:1998}.
  Our calculations based on the error covariance matrix data are entirely consistent with the MINUIT algorithm.
  The parameter values together with their $1\sigma$ and $2\sigma$ errors are shown in Tab.~\ref{Tab:Table}.
  Our estimates also agree with the ``bootstrap'' method.

  The minimal value of $\chi^2/\text{ndf}$~$= 4443.3/(395-6) \approx 11.4$ for
  the current model corresponds to the parameter values supplied with $1\sigma$ ($2\sigma$) errors.
  The eVMD-VI model reproduces the observational data for the space- and timelike regions with about equal precision,
  although the number of points for the timelike region is two times less.
  The $\chi^2/\text{ndf}$ values calculated for the space- and timelike data sets
  with the values of the eVMD-VI model obtained from the global fit are  
  $1948.5/227~{\approx}~8.6$ and $1543.3/156~{\approx}~9.9$, respectively.
  The $\chi^2/\text{ndf}$ values from separate fits of the space- and timelike data
  are $2794.4/227~{\approx}~12.3$ and $809.8/156~{\approx}~5.2$, respectively.
  Analysis shows that it is not possible to describe the data of the spacelike region
  using the model tuned to timelike data and vice versa.
  The parameters of the model and pairwise two-dimensional projections of error ellipsoids
  obtained from separate analyses of proton or neutron form factor data in the space-time domain only,
  or in both regions, are inconsistent with the corresponding values obtained from the global analysis.
  These findings point to the settings of the model parameters in a global analysis
  of all data for both $q^2$ regions.

  \subsection{Numerical results for moments of nucleon radii}

  The numerical values of the second moments obtained in the model are displayed in Tab.~\ref{Tab:Table}
  together with the experimental values.
  The coupling constants $f_i^{VNN}$ are calculated based on Eq.~\eqref{Eq:couplings}
  with the vector meson--photon coupling constants $g_{\omega \gamma }= 17.1$ and
  $g_{\rho\gamma}= 5.03$~\cite{Faessler:1999de} determined from
  the $\omega,\rho\,{\rightarrow}\,e^+e^-$ decays.
  The coupling constant errors of~\cite{Hoehler:1983} are evaluated
  from the claimed accuracy $\pm 15\%$ of the ${\pi}N$ scattering amplitudes.

  The Zemach moments are required for calculating hyperfine splitting and
  the Lamb shift in hydrogen and muonic atoms.
  The numerical values of the moments are
  $\langle{r\rangle}_{EE}= 1.034$ fm and
  $\langle{r\rangle}_{EM}= 1.015$ fm
  are found to be close to the empirical values
  $\langle{r\rangle}_{EE}^{\text{exp}}= 1.085 \pm 0.003$ fm and
  $\langle{r\rangle}_{EM}^{\text{exp}}= 1.045 \pm 0.004$ fm~\cite{Distler:2010zq}
  obtained from the analysis of the electron--proton scattering data~\cite{A1:2010nsl,Arrington:2007ux}.
  The~value of the third Zemach moment equals $\langle{r^3\rangle}_{EM}= 2.036$ fm$^3$.
  The experimental values derived from the electron--proton scattering data are 
  $\langle{r^3\rangle}_{EM}^{\text{exp}}= 2.71 \pm 0.13$ fm$^3$~\cite{Friar:2005jz} and 
  $2.85 \pm 0.08$ fm$^3$ (see \cite{Distler:2010zq} and references therein).
  The recent global analysis of the nucleon form factors gives
  $\langle{r\rangle}_{EE}^{\text{exp}}= 1.054_{-0.002-0.001}^{+0.003+0.000}$ fm and
  $\langle{r^3\rangle}_{EM}^{\text{exp}}= 2.310_{-0.014-0.015}^{+0.022+0.018}$ fm$^3$~\cite{Lin:2021xrc},
  where the first and second errors are statistical and systematic, respectively.

  Except for the neutron charge radius, which is small, the moments of nucleon radii
  in the model under consideration are several points of percentage underestimated.
  The logarithmic singularity in the spectral function ${\Im}F_{1,I=1}(t)$ below
  the two-pion threshold contributes positively to the nucleon isovector charge
  radius~\cite{Hohler:1974ht,Hoehler:1983}.
  This effect is beyond the scope of the eVMD models,
  although taking it into account could reduce some of the deviations from the experimental data.

  \subsection{Numerical results for nucleon form factors}

  The nucleon form factors as functions of the four-momentum transfer squared
  are presented in Figs.~\ref{Fig:GpE_GnE_GpM_GnM_011_2_3_PRD}~to~%
  \ref{Fig:GEp_GMp_GEn_GMn_011_2_1_PRD}.
  The experimental data points included in the fit
  are represented by shaded or partially shaded symbols,
  while the omitted data or data from the historical background are represented by open dark symbols.
  The horizontal error bars represent the experimental errors for the mean $Q^2= -q^2$ values.
  The inner and outer vertical error bars indicate statistical and total errors, respectively.
  The~colored uncertainty bands surrounding the curves
  represent the $1\sigma$ and $2\sigma$ confidence intervals of the form factors
  determined by the correlated statistical errors of the fitted parameters of the eVMD-VI model.
  Different form factors and ratios have different uncertainty band widths.
  The reason for this is that the fitted parameter contours for the correlated errors
  of the electric and magnetic proton and neutron form factors
  are not symmetrical and exhibit different ellipsoid features.

  Figure~\ref{Fig:GpE_GnE_GpM_GnM_011_2_3_PRD} shows modules
  of the electric and magnetic form factors of the proton and neutron
  predicted within the eVMD-VI model for positive and negative values of $t= q^2$
  in comparison with the world experimental data.
  The peaks generated by vector mesons
  determine the nontrivial behavior of the form factors in the region $0~<~t~<~4m^2_N$.
  The magnitudes and layout of the peaks are regulated by the masses and widths of the mesons.
  The eVMD-VI model predicts a smooth decline of the form factors at large positive values of $Q^2$
  to which the cross sections of lepton--nucleon interactions are sensitive.
  The non-physical region $0 < t < 4m_N^2$ is inaccessible for direct measurement of form factors.
  In this region, meson resonances generate narrow peaks that are represented by the Breit-Wigner formula
  with an energy-dependent width.
  Residues control the phases of the resonance contributions at the poles,
  which in turn influence the interference of vector mesons.

  Figures~\ref{Fig:GpE_GD_011_2_1_PRD} to \ref{Fig:GEp_GMp_GEn_GMn_011_2_1_PRD}
  show the eVMD-VI model predictions for the proton and neutron form factors
  in comparison with the experimental data on an enlarged scale
  to illustrate the absolute values and asymptotic behavior.

  The electric form factor of the proton normalized by the standard dipole function~\eqref{Eq:dipole_f}
  is shown in Fig.~\ref{Fig:GpE_GD_011_2_1_PRD} in comparison with the
  experimental data measured for record low values of $Q^2$.
  The experimental data and model predictions are well consistent with
  the dipole dependence of the form factors on $Q^2$ in this region.
  Additional data on measuring this ratio at high $Q^2$ values
  could allow to reliably fine-tune parameters of the eVMD models.

  In the space-like region, the eVMD-VI model qualitatively reproduces the $Q^2$ dependence of 
  the proton magnetic form factor, as shown in Fig.~\ref{Fig:GpM_mGD_mGpE_GpM_011_2_1_PRD}(a),
  for $Q^2 \lesssim 10$ GeV$^2$.
  The model curve has a local maximum at 1 GeV$^2$, while the experimental data corresponds
  to a maximum at around 2 GeV$^2$, with the latter being more pronounced. 
  The deviations from the model curve are not more than $5\%$. 
  Fig.~\ref{Fig:GnE_GnM_mGD_mGnE_GnM_011_2_3_PRD}(b) shows a similar structure in the neutron magnetic form factor,
  which could be associated with a more sophisticated version of the spectral function
  of nucleon form factors.
  The ratio of the proton electric and magnetic form factors shown
  in Fig.~\ref{Fig:GpM_mGD_mGpE_GpM_011_2_1_PRD}(b) is reproduced reasonably well
  for $Q^2 \lesssim 2$ GeV$^2$, but differs from that implied by the JLab data at higher momentum transfers.
  The deviation is partly due to the relatively large experimental errors and
  the resulting lower statistical weighting compared to other data included in the fit.
  New accurate measurements of the ratio at high values of $Q^2$ are essential to refine the analysis.
  Fig.~\ref{Fig:GpM_mGD_mGpE_GpM_011_2_1_PRD}(b) shows that the range of applicability
  of the SR's is limited by the values of $Q^2\lesssim 2$ GeV$^2$.

  As demonstrated in Fig.~\ref{Fig:GnE_GnM_mGD_mGnE_GnM_011_2_3_PRD}(c),
  the eVMD-VI model yields the neutron electric form factor
  consistent with the JLab data but not with the MAMI data for the spacelike region.
  The figure depicts the experimental findings on the neutron magnetic form factor.
  In the spacelike region of small $Q^2$ values,
  the eVMD-VI model agrees with the data obtained from the SLAC experiment and
  predicts a smoother dependence of $G_{Mn}$ than the MAMI experimental data.
  Panels (b) and (d) of the figure show that the eVMD-VI model accurately describes
  the ratio of the neutron electric and magnetic form factors.
  
  The magnetic and electric form factors coincide only at $t = 4m_N^2$,
  so the results of experiments, obtained assuming
  $\left|G_{ET}(t)\right|= \left|G_{MT}(t)\right|$ in the time-like region,
  are not included in the statistical analysis. 

  Recent experimental data from BESIII and BIND on the effective nucleon form factors
  indicate the presence of sinusoidal modulations in their behavior.
  In the case of a proton, sinusoidal modulations can be seen quite clearly in Fig.~\ref{Fig:Gp_011_2_1_PRD}. 
  Bianconi and Tomasi-Gustafsson~\cite{Bianconi:2015owa} discovered their existence
  by analyzing data from the {\it BaBar} Collaboration~\cite{BaBar:2013ves,BaBar:2013ukx}.
  The effective neutron form factor data shown in Fig.~\ref{Fig:Gn_011_2_1_PRD}
  has significantly higher experimental errors compared to the effective proton form factor data.
  The presence of an irregular structure at $q^2 \sim 5$~GeV$^2$ is clearly visible.
  The sinusoidal modulations observed by the BESIII Collaboration
  in the effective neutron form factor do not contradict
  the data from the SND Collaboration~\cite{SND:2022wdb,SND:2023fos}.
  The magnitude of the fluctuations is $10\%$ of the regular background dependence.
  The physical origin of the non-monotonic behavior of the proton and neutron form factors is not entirely clear.

  Bianconi and Tomasi-Gustafsson interpret the sinusoidal modulations as the result
  of a strong $N\overline{N}$ interaction at distances $0.7-1.5$ fm.
%
%
  The scattering of the pair is dominated by inelastic channels,
  which results in a large imaginary part in the amplitude and a cross section that
  is close to the unitary limit~\cite{Zhou:2012ui}.
  The oscillations in form factors can be caused by interference of the amplitudes associated 
  with different mechanisms, e.g., the meson and photon exchange within a $N\overline{N}$ pair
  on the one hand and the annihilation of a $N\overline{N}$ pair on the other.

  The $N\overline{N}$ potentials are evaluated in Ref.~\cite{Yang:2024iuc}
  using chiral perturbation theory for both the elastic and annihilation channels.


  The multiple formation of $\pi$ mesons is responsible for the annihilation component
  of the $N\overline{N}$ scattering amplitude.
  The ground state and the excited states of the $\rho$- and $\omega$-mesons contribute
  to saturating the annihilation channels.
  In Refs.~\cite{Lin:2021xrc,Yan:2023nlb}, the eVMD models include vector mesons
  with masses greater than 2 GeV. 
  The periodic modulation of the nucleon form factors is confirmed in Ref.~\cite{Lin:2021xrc},
  but not in Ref.~\cite{Yan:2023nlb}.
  The masses and widths of vector mesons above the $N\overline{N}$ threshold,
  as well as their contributions to the nucleon form factors, cannot yet be determined unambiguously.
  Quark models provide a valuable insight into the spectrum and decay widths of excited unflavored
  mesons~\cite{Wang:2021abg}, which can be useful for their experimental search.  
  The eVMD-VI model describes the regular background dependence of the nucleon form factors
  in the time-like region. 
  The non-monotonic structures (or so-called oscillating features) in the line shape
  of nucleon effective form factors cannot be reproduced without accounting
  for excited vector meson contributions, particularly for those with masses
  above the nucleon--anti-nucleon threshold. 

  Figures~\ref{Fig:GEp_GMp_011_2_1_PRD} and \ref{Fig:GEn_GMn_011_2_1_PRD}
  show a dozen data points for the modules of electric and magnetic form factors
  of the nucleons in the timelike region.
  The errors of the proton magnetic form factor data
  measured in the BESIII experiment are significantly lower than the errors of the electric form factor data.
  The data are not qualitatively at odds with the eVMD-VI model.
  The model agrees better with the data obtained at the near-threshold values and
  generally agrees with the data measured for high $q^2$-values
  within the experimental errors and uncertainties of the model.

  Figure~\ref{Fig:GEp_GMp_GEn_GMn_011_2_1_PRD} shows differences 
  between model predictions and observations.
  Experimental results show that the ratio of electric and magnetic form factors
  is independent of $q^2$ throughout a broad range, 
  which is one of the consequences of SR's.
  SR's are not followed accurately in the model under consideration.
  As a result, the eVMD-VI model does not follow the experimental trend well.

  The presented statistical analysis uses 60 experimental data sets
  obtained for the form factors and their ratios.
  The values of $\chi^2/\text{np}$ calculated within the framework of the eVMD-VI model,
  where $\text{np}$ is the number of data points in every data set,
  do not exceed $\sim 1$ for 12 experimental data sets
  measured with the completed MIT BLAC, MELA, CERN PS/LEAR and recent JLab experiments.
  This group of the data includes 44 ($\sim 10\%$ of the full set) data points for
  $G_{Ep}/G_D$ (18 data points),
  $G_{Mp}/\left(\mu_p G_D\right)$ (6),
  $\mu_p G_{Ep}/G_{Mp}$ (11),
  $G_{En}$ (6), and
  $\left|G^{\text{eff}}_p\right|$ (3).
  The values of $\chi^2/\text{np}$
  for every of 16 data sets with 62 data points ($\sim 16\%$) for
  $\mu_p G_{Ep}/G_{Mp}$ (22 data points),
  $G_{En}$ (9),
  $G_{Mn}/\left(\mu_n G_D\right)$ (4),
  $\mu_n G_{En}/G_{Mn}$ (5),
  $\left|G_{Ep}/G_{Mp}\right|$ (1),
  $\left|G_{En}\right|$ (5),
  $\left|G^{\text{eff}}_n\right|$ (11), and
  $\left|G_{En}/G_{Mn}\right|$ (5)
  measured with  
  BLAST,
  SLAC,
  JLab,
  MAMI,
  BINP, and
  BES experiments
  are less than $\sim 2$.
  The group of 20 data sets with 186 data points (47\%) for
  $G_{Ep}/G_D$ (22 data points),
  $G_{Mp}/\left(\mu_p G_D\right)$ (37),
  $\mu_p G_{Ep}/G_{Mp}$ (18),
  $G_{En}$ (7),
  $G_{Mn}/\left(\mu_n G_D\right)$ (10),
  $\mu_n G_{En}/G_{Mn}$ (3),
  $\left|G_{Ep}\right|$ (16),
  $\left|G^{\text{eff}}_p\right|$ (48),
  $\left|G_{Ep}/G_{Mp}\right|$ (22), and
  $\left|G_{Mn}\right|$ (5)
  measured with
  BLAST,
  SLAC,
  JLab,
  NIKHEF,
  MAMI,
  DESY,
  BINP, and
  BES experiments
  take values of $\chi^2/\text{np}$ exceed 10.
  For 12 experimental data sets with 101 data points (27\%)
  the values of $\chi^2/\text{np}$ are higher than 10.
  This group of data sets mostly includes the proton form factor measured in the spacelike region.


  \section{Conclusion}
  \label{Sec:Conclusion}

  The MFK model for electromagnetic nucleon form factors has only two fitting parameters.
  This aspect sets it apart from parameterizations and other phenomenological models,
  which typically have more than a dozen fitting parameters.
  The need to update the earlier MFK model is brought to light by the appearance
  in the last decade of new data from experiments on nucleon form factors.
  This paper offers an upgraded version of the MFK model.
  The upgraded version of the model includes now
  an additional pair of experimentally observed non-strange vector mesons.
  As a result, the number of parameters to be fitted grows from two to six.
  This number remains still low compared to the most popular parameterizations and models.
  Energy-dependent decay widths of vector mesons were also taken into account.
  The vector meson masses and on-shell widths are equal to their experimental values,
  with the sole exception of the $\omega(1250)$- and $\rho(1250)$-mesons, 
  the characteristics of which are not reliably determined yet.

  The presented model describes well the previous data
  used to calculate the parameters of the MFK model and the modern data.
  The revised version of the model agrees well with the experimental data
  on proton and neutron form factors, form factor ratios, and electric and magnetic nucleon radii.
  The Zemach radii are also consistent with the experimental data and alternative theoretical estimations. 
  The model's accuracy could be improved further by adding more parameters.
  However, as the number of parameters expands, their physical significance blurs.

  In the time-like region, the eVMD-VI model accurately reproduces the regular component of the nucleon form factors.
  The contribution of excited vector mesons with masses above the $N\overline{N}$ threshold
  appears to be essential for replicating the periodic modulations in the nucleon-form factors.

  The analysis of the nucleon form factors performed in this work
  is also aimed at identifying physical effects that influence the form factors' behavior.
  Such effects are the quark counting rules, the Okubo-Zweig-Iizuka rule,
  and the scaling laws of Sachs form factors for low and moderate momentum transfers.
  The coupling constants of $\rho$-mesons with nucleons found in the model are consistent
  with the coupling constants determined within the context of dispersion theory and Frazer-Fulco unitarity.
  The coupling constants of $\omega$- and $\rho$-mesons with nucleons also agree
  with phenomenological models of nucleon--nucleon interactions (Bonn model).
  The physical effects outlined above, as well as the model parameters associated with them, are
  considered to be rather well established.
  They can be recommended for physically justified parameterizations
  in numerical simulations of processes involving interactions with nucleons.

  \section*{Acknowledgements}
  The authors thank Mrs.~Galina Sandukovskaya for her help in preparing the manuscript.
  M.~I.~K. was supported by the Russian Science Foundation, project no. 23-22-00307.

  \section*{Appendix A. \\
            Derivation of Eqs.~(2.18) and (2.19)}
  \label{Sec:Appendix}
  \renewcommand{\theequation}{A.\arabic{equation}}
  \setcounter{equation}{0}

  The average powers of the nucleon radii are calculated using Eqs.~\eqref{Eq:IIIC1} and \eqref{Eq:IIIC2}.
  The density distribution being averaged over the angles of $\mathbf{Q}$ can be written in the form
  \begin{eqnarray*}
  \rho_{TN}\left(\mathbf{r}\right)=
  \int_0^{+\infty}\dfrac{4\pi Q^2dQ}{\left(2\pi\right)^3}
  \dfrac{\sin Qr}{Qr}\mathcal{G}_{TN}\left(-\mathbf{Q}^2\right).
  \end{eqnarray*}

\begin{figure}[htb!]
  \hspace*{-3mm}
  \includegraphics[width=0.382\textwidth]{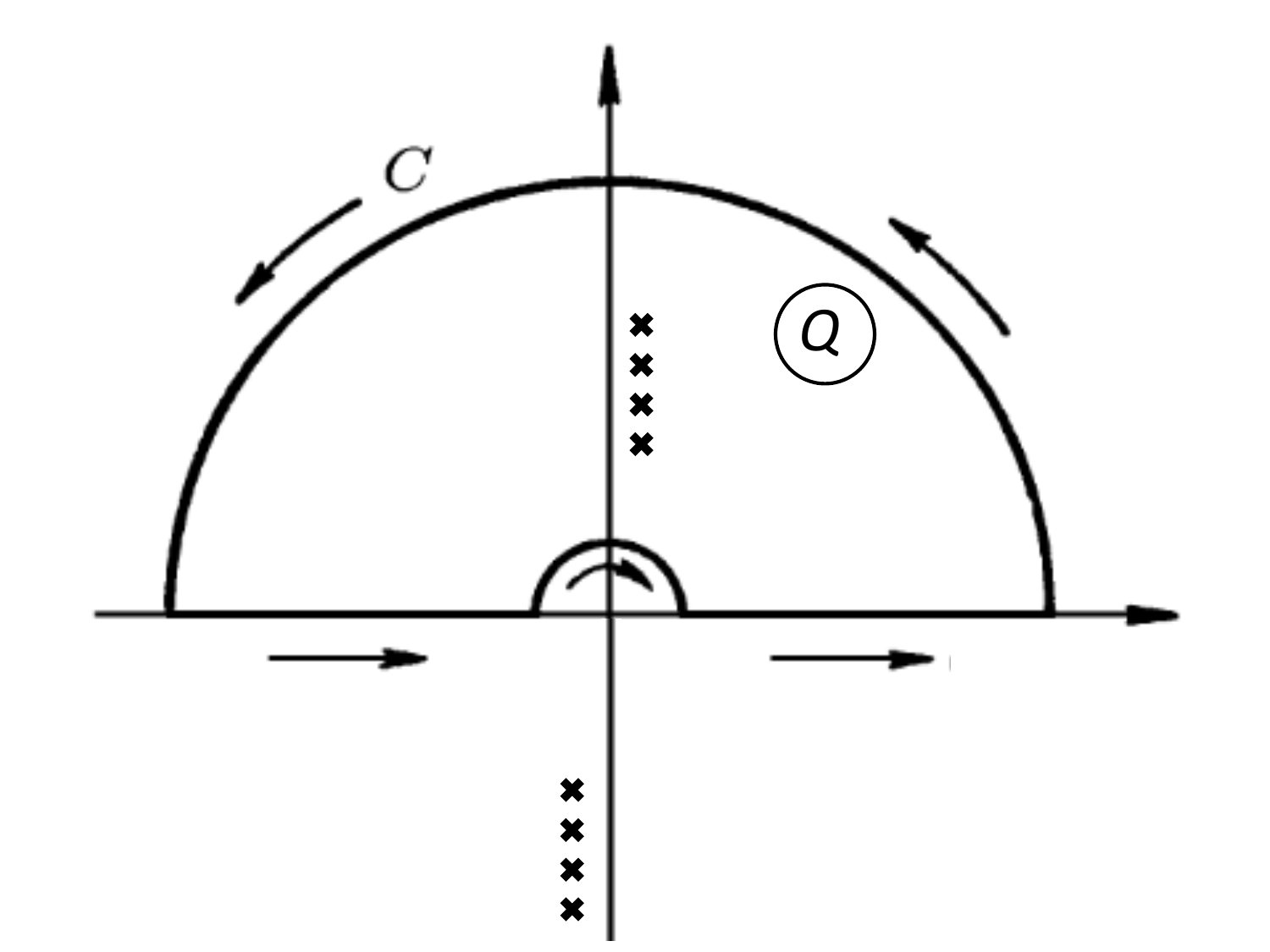}
  \caption{The directed closed curve, $C$, of the contour integral~\eqref{Eq:dirC} in the complex $Q$-plane.
           The crosses show simple poles of the form factor $\mathcal{G}_{TN}\left(-\mathbf{Q}^2\right)$
           corresponding to the vector meson masses and vector meson widths.}
  \label{fig:21}
  \end{figure}

  The integrand is the even function of $Q$.
  We thus extend the integration to the interval $Q \in (-\infty,+\infty)$ and
  place a factor $1/2$ in front of the integral.
  The integral is non-singular at $Q= 0$ and it converges as $Q \to \pm \infty$,
  where the form factor goes to vanish.
  We deform the contour as shown in Fig.~\ref{fig:21} and write $\sin Qr$ in terms of the exponents.
  The substitution $Q \rightarrow -Q$ in the second exponent gives, for the density distribution,
  \begin{equation}\label{Eq:dirC}
  \int_{C}\dfrac{4\pi Q^2 dQ}{\left(2\pi\right)^3}
  \dfrac{e^{iQr}}{2iQr}\mathcal{G}_{TN}\left(-\mathbf{Q}^2\right).
  \end{equation}

  The integrand is an analytical function in the complex $Q$-plane with $Q= \infty$
  being the unique essential singularity.
  The contour $C$ is closed in the upper half-plane, taking into account the fact
  that the exponential term makes the integral over the semicircle to vanish
  in the limit when radius of the semicircle goes to infinity.
  
  Now we change the order of integrals and obtain
  \begin{eqnarray}\label{Eq:FT}
  \int d\mathbf{r}\,r^{2s+1}\frac{e^{iQr}}{2iQr}=
  2\pi(-1)^{s+1}\frac{(2s+2)!}{Q^{2s+4}},
  \end{eqnarray}
  where $s= 0,1,\ldots$
  The convergence of the integral at $r=+\infty$ is provided by the condition of $\Im(Q) > 0$.

  The odd moments of the nucleon radii become
  \begin{equation*}
  \langle{r^{2s+1}\rangle}_{TN}= (-1)^{s+1}\dfrac{(2s+2)!}{\pi}\int_{C}
  \dfrac{dQ}{Q^{2s+2}}\mathcal{G}_{TN}\left(-\mathbf{Q}^2\right).
  \end{equation*}

  \begin{figure}[h!]
  \hspace*{-0.0cm}
  \includegraphics[width=0.435\textwidth]{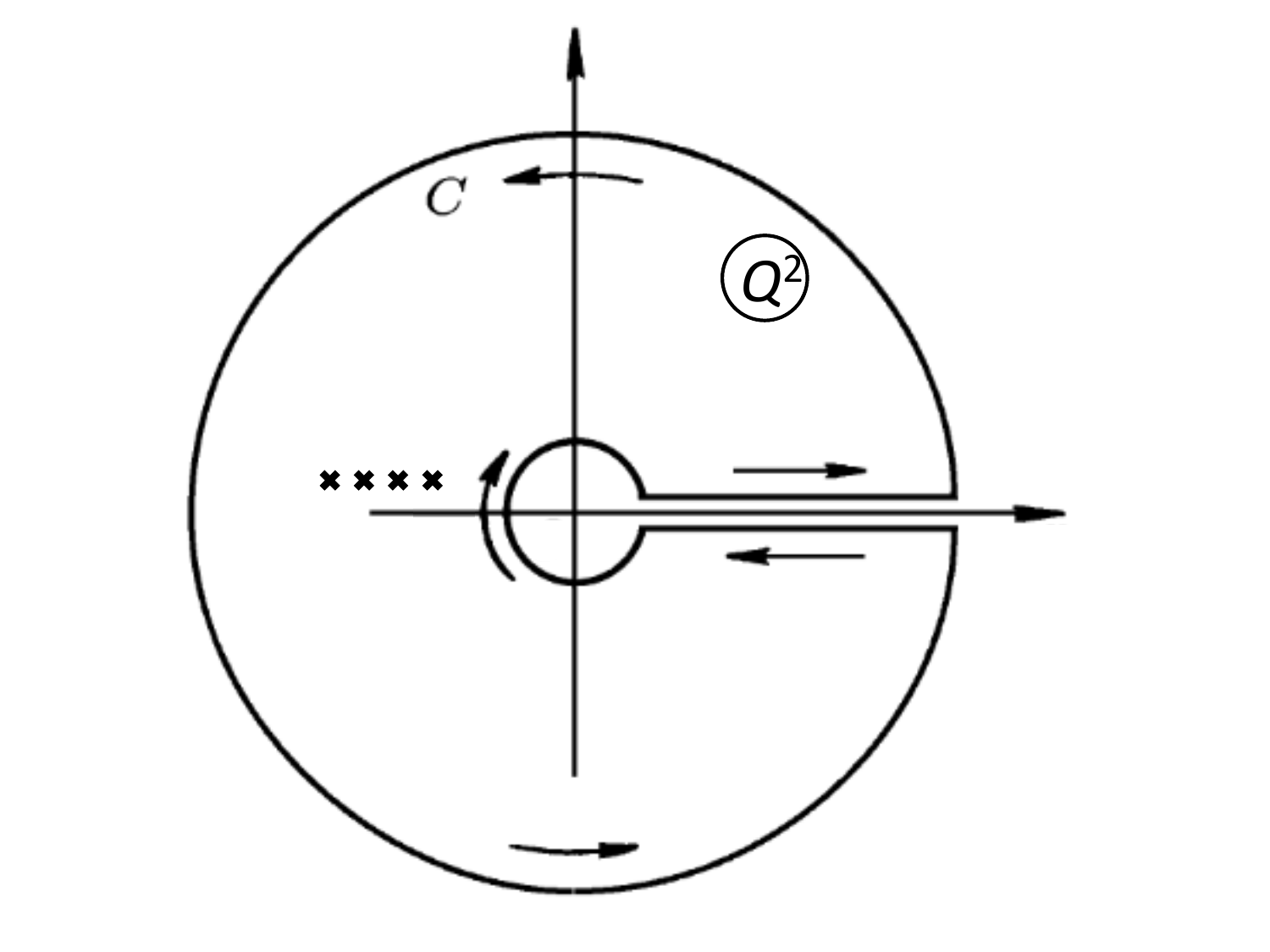}
  \caption{The contour for integration over the complex variable $Q^2$.
           The crosses show simple poles of the form factor $\mathcal{G}_{TN}\left(-\mathbf{Q}^2\right)$.}
  \label{fig:22}
  \end{figure}

  The contour integral with a degree-$s$ polynomial $\mathcal{P}_{s}\left(-\mathbf{Q}^2\right)$
  substituted in place of $\mathcal{G}_{TN}\left(-\mathbf{Q}^2\right)$, 
  vanishes identically, because there are no singularities inside of the contour $C$.
  The maximum admissible degree of $\mathcal{P}_s\left(-\mathbf{Q}^2\right)$ is determined
  by the condition of vanishing the integral over the infinitely distant semicircle.

  \begin{figure}[h!]
  \hspace*{-0.8cm}
  \includegraphics[width=0.458\textwidth]{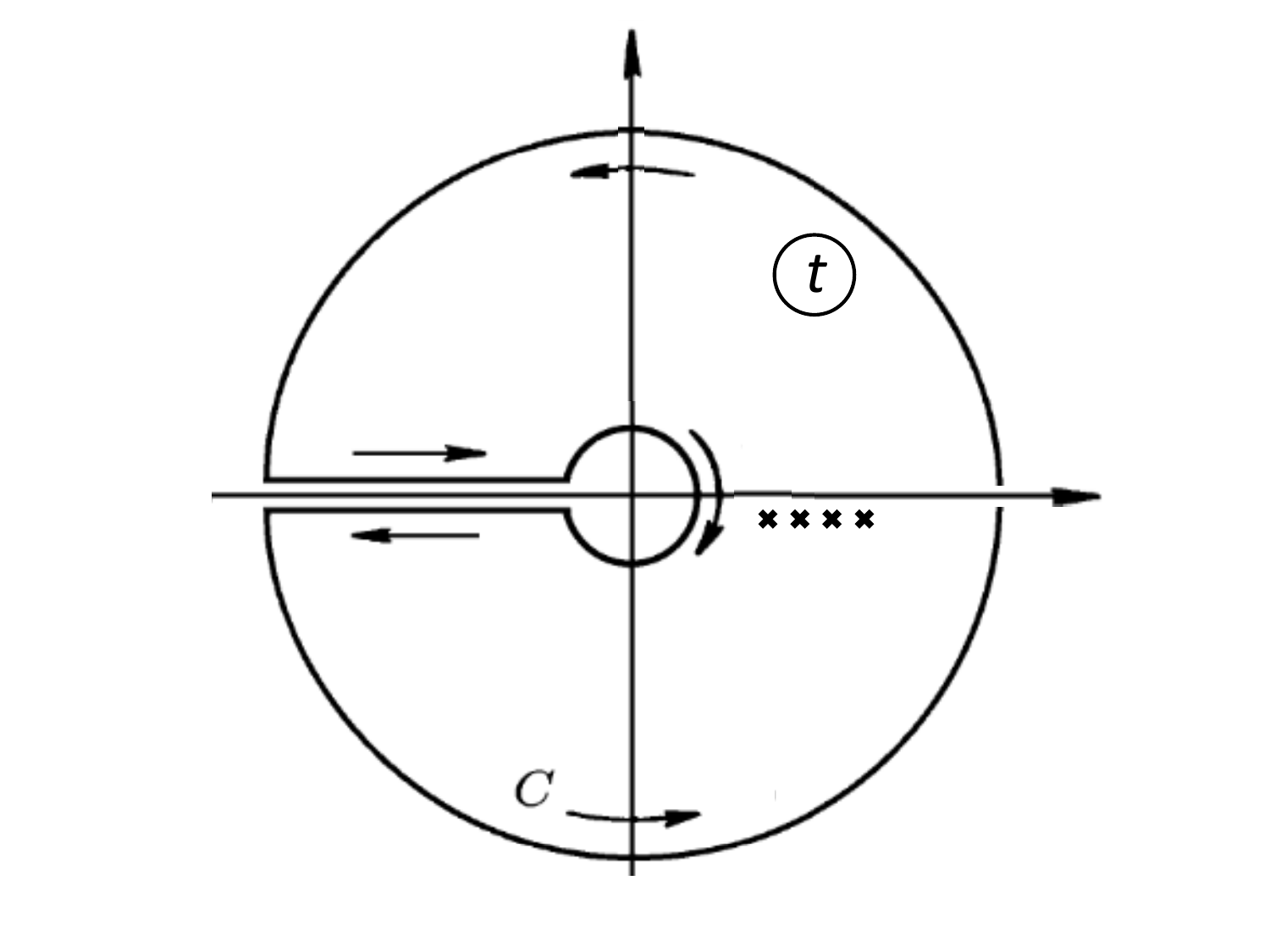}
  \caption{The contour of integration over the complex variable $t=-Q^2$.
           The crosses show simple poles of the form factor $\mathcal{G}_{TN}(t)$
           at $t= m^2_V -i m_V\Gamma_V$.}
  \label{fig:32}
  \end{figure}

  Subtracting the polynomial from the form factor,  
  $\mathcal{G}_{TN}\left(-\mathbf{Q}^2\right) \to
   \mathcal{G}_{TN}\left(-\mathbf{Q}^2\right) -
   \mathcal{P}_s\left(-\mathbf{Q}^2\right)$, does not change the integral.
  Next, we require the integrand to be non-singular at $t= 0$.
  The polynomial is thereby a Taylor series of $\mathcal{G}_{TN}\left(-\mathbf{Q}^2\right)$
  terminated at the highest permissible polynomial degree.

  By passing to the variable $Q^2$,
  the contour along which we integrate is altered, as illustrated in Fig.~\ref{fig:22}.
  Figure~\ref{fig:32} depicts the same contour in the complex plane of $t=-Q^2$.

  Equation~\eqref{Eq:odd_powers} represents the contour integral in the form of a definite integral of a real variable.

  Let us consider even powers of the nucleon radii. For a spherically symmetric density distribution,
  $\rho_{TN}(r)$, one has
  \begin{equation*}
  \mathcal{G}_{TN}(t)=
  \int d\mathbf{r}\dfrac{\sin Qr}{Qr}\rho _{TN}(r)=
  \sum_{s= 0}^{\infty}\dfrac{t^s}{(2s+1)!}\langle{r^{2s}\rangle}_{TN},
  \end{equation*}
  in agreement with Eq.~\eqref{Eq:even_powers}.

  The same arguments apply to the derivation of Eqs.~\eqref{Eq:firApp} and \eqref{Eq:r2s}.

  \bibliography{References}

\begin{thebibliography}{142}%
\makeatletter
\providecommand \@ifxundefined [1]{%
 \@ifx{#1\undefined}
}%
\providecommand \@ifnum [1]{%
 \ifnum #1\expandafter \@firstoftwo
 \else \expandafter \@secondoftwo
 \fi
}%
\providecommand \@ifx [1]{%
 \ifx #1\expandafter \@firstoftwo
 \else \expandafter \@secondoftwo
 \fi
}%
\providecommand \natexlab [1]{#1}%
\providecommand \enquote  [1]{``#1''}%
\providecommand \bibnamefont  [1]{#1}%
\providecommand \bibfnamefont [1]{#1}%
\providecommand \citenamefont [1]{#1}%
\providecommand \href@noop [0]{\@secondoftwo}%
\providecommand \href [0]{\begingroup \@sanitize@url \@href}%
\providecommand \@href[1]{\@@startlink{#1}\@@href}%
\providecommand \@@href[1]{\endgroup#1\@@endlink}%
\providecommand \@sanitize@url [0]{\catcode `\\12\catcode `\$12\catcode
  `\&12\catcode `\#12\catcode `\^12\catcode `\_12\catcode `\%12\relax}%
\providecommand \@@startlink[1]{}%
\providecommand \@@endlink[0]{}%
\providecommand \url  [0]{\begingroup\@sanitize@url \@url }%
\providecommand \@url [1]{\endgroup\@href {#1}{\urlprefix }}%
\providecommand \urlprefix  [0]{URL }%
\providecommand \Eprint [0]{\href }%
\providecommand \doibase [0]{https://doi.org/}%
\providecommand \selectlanguage [0]{\@gobble}%
\providecommand \bibinfo  [0]{\@secondoftwo}%
\providecommand \bibfield  [0]{\@secondoftwo}%
\providecommand \translation [1]{[#1]}%
\providecommand \BibitemOpen [0]{}%
\providecommand \bibitemStop [0]{}%
\providecommand \bibitemNoStop [0]{.\EOS\space}%
\providecommand \EOS [0]{\spacefactor3000\relax}%
\providecommand \BibitemShut  [1]{\csname bibitem#1\endcsname}%
\let\auto@bib@innerbib\@empty
\bibitem [{\citenamefont {Frazer}\ and\ \citenamefont
  {Fulco}(1959)}]{Frazer:1959gy}%
  \BibitemOpen
  \bibfield  {author} {\bibinfo {author} {\bibfnamefont {W.~R.}\ \bibnamefont
  {Frazer}}\ and\ \bibinfo {author} {\bibfnamefont {J.~R.}\ \bibnamefont
  {Fulco}},\ }\bibfield  {title} {\bibinfo {title} {{Effect of a pion-pion
  scattering resonance on nucleon structure}},\ }\href
  {https://doi.org/https://doi.org/10.1103/PhysRevLett.2.365} {\bibfield
  {journal} {\bibinfo  {journal} {Phys.\ Rev.\ Lett.}\ }\textbf {\bibinfo
  {volume} {2}},\ \bibinfo {pages} {365} (\bibinfo {year} {1959})}\BibitemShut
  {NoStop}%
\bibitem [{\citenamefont {Frazer}\ and\ \citenamefont
  {Fulco}(1960{\natexlab{a}})}]{Frazer:1960zzb}%
  \BibitemOpen
  \bibfield  {author} {\bibinfo {author} {\bibfnamefont {W.~R.}\ \bibnamefont
  {Frazer}}\ and\ \bibinfo {author} {\bibfnamefont {J.~R.}\ \bibnamefont
  {Fulco}},\ }\bibfield  {title} {\bibinfo {title} {{Effect of a pion-pion
  scattering resonance on nucleon structure. II}},\ }\href
  {https://doi.org/https://doi.org/10.1103/PhysRev.117.1609} {\bibfield
  {journal} {\bibinfo  {journal} {Phys.\ Rev.}\ }\textbf {\bibinfo {volume}
  {117}},\ \bibinfo {pages} {1609} (\bibinfo {year}
  {1960}{\natexlab{a}})}\BibitemShut {NoStop}%
\bibitem [{\citenamefont {Frazer}\ and\ \citenamefont
  {Fulco}(1960{\natexlab{b}})}]{Frazer:1960zza}%
  \BibitemOpen
  \bibfield  {author} {\bibinfo {author} {\bibfnamefont {W.~R.}\ \bibnamefont
  {Frazer}}\ and\ \bibinfo {author} {\bibfnamefont {J.~R.}\ \bibnamefont
  {Fulco}},\ }\bibfield  {title} {\bibinfo {title} {{Partial-wave dispersion
  relations for the process $\pi+\pi{\to}N +\overline{N}$}},\ }\href
  {https://doi.org/https://doi.org/10.1103/PhysRev.117.1603} {\bibfield
  {journal} {\bibinfo  {journal} {Phys.\ Rev.}\ }\textbf {\bibinfo {volume}
  {117}},\ \bibinfo {pages} {1603} (\bibinfo {year}
  {1960}{\natexlab{b}})}\BibitemShut {NoStop}%
\bibitem [{\citenamefont {Gounaris}\ and\ \citenamefont
  {Sakurai}(1968)}]{Gounaris:1968mw}%
  \BibitemOpen
  \bibfield  {author} {\bibinfo {author} {\bibfnamefont {G.}~\bibnamefont
  {Gounaris}}\ and\ \bibinfo {author} {\bibfnamefont {J.~J.}\ \bibnamefont
  {Sakurai}},\ }\bibfield  {title} {\bibinfo {title} {{Finite-width corrections
  to the vector-meson-dominance prediction for $\rho{\to}e^+e^-$}},\ }\href
  {https://doi.org/https://doi.org/10.1103/PhysRevLett.21.244} {\bibfield
  {journal} {\bibinfo  {journal} {Phys.\ Rev.\ Lett.}\ }\textbf {\bibinfo
  {volume} {21}},\ \bibinfo {pages} {244} (\bibinfo {year} {1968})}\BibitemShut
  {NoStop}%
\bibitem [{\citenamefont {Matveev}\ \emph {et~al.}(1973)\citenamefont
  {Matveev}, \citenamefont {Muradyan},\ and\ \citenamefont
  {Tavkhelidze}}]{Matveev:1973ra}%
  \BibitemOpen
  \bibfield  {author} {\bibinfo {author} {\bibfnamefont {V.~A.}\ \bibnamefont
  {Matveev}}, \bibinfo {author} {\bibfnamefont {R.~M.}\ \bibnamefont
  {Muradyan}},\ and\ \bibinfo {author} {\bibfnamefont {A.~V.}\ \bibnamefont
  {Tavkhelidze}},\ }\bibfield  {title} {\bibinfo {title} {{Automodelism in the
  large-angle elastic scattering and structure of hadrons}},\ }\href
  {https://doi.org/https://doi.org/10.1007/BF02728133} {\bibfield  {journal}
  {\bibinfo  {journal} {Lett.\ Nuovo Cim.}\ }\textbf {\bibinfo {volume} {7}},\
  \bibinfo {pages} {719} (\bibinfo {year} {1973})}\BibitemShut {NoStop}%
\bibitem [{\citenamefont {Brodsky}\ and\ \citenamefont
  {Farrar}(1973)}]{Brodsky:1973kr}%
  \BibitemOpen
  \bibfield  {author} {\bibinfo {author} {\bibfnamefont {S.~J.}\ \bibnamefont
  {Brodsky}}\ and\ \bibinfo {author} {\bibfnamefont {G.~R.}\ \bibnamefont
  {Farrar}},\ }\bibfield  {title} {\bibinfo {title} {{Scaling laws at large
  transverse momentum}},\ }\href
  {https://doi.org/https://doi.org/10.1103/PhysRevLett.31.1153} {\bibfield
  {journal} {\bibinfo  {journal} {Phys.\ Rev.\ Lett.}\ }\textbf {\bibinfo
  {volume} {31}},\ \bibinfo {pages} {1153} (\bibinfo {year}
  {1973})}\BibitemShut {NoStop}%
\bibitem [{\citenamefont {Brodsky}\ and\ \citenamefont
  {Farrar}(1975)}]{Brodsky:1974vy}%
  \BibitemOpen
  \bibfield  {author} {\bibinfo {author} {\bibfnamefont {S.~J.}\ \bibnamefont
  {Brodsky}}\ and\ \bibinfo {author} {\bibfnamefont {G.~R.}\ \bibnamefont
  {Farrar}},\ }\bibfield  {title} {\bibinfo {title} {{Scaling laws for
  large-momentum-transfer processes}},\ }\href
  {https://doi.org/https://doi.org/10.1103/PhysRevD.11.1309} {\bibfield
  {journal} {\bibinfo  {journal} {Phys.\ Rev.\ D}\ }\textbf {\bibinfo {volume}
  {11}},\ \bibinfo {pages} {1309} (\bibinfo {year} {1975})}\BibitemShut
  {NoStop}%
\bibitem [{\citenamefont {Vainshtein}\ and\ \citenamefont
  {Zakharov}(1978)}]{Vainshtein:1977db}%
  \BibitemOpen
  \bibfield  {author} {\bibinfo {author} {\bibfnamefont {A.~I.}\ \bibnamefont
  {Vainshtein}}\ and\ \bibinfo {author} {\bibfnamefont {V.~I.}\ \bibnamefont
  {Zakharov}},\ }\bibfield  {title} {\bibinfo {title} {{Remarks on the
  electromagnetic form factors of hadrons in the quark model}},\ }\href
  {https://doi.org/https://doi.org/10.1016/0370-2693(78)90140-5} {\bibfield
  {journal} {\bibinfo  {journal} {Phys.\ Lett.\ B}\ }\textbf {\bibinfo {volume}
  {72}},\ \bibinfo {pages} {368} (\bibinfo {year} {1978})}\BibitemShut
  {NoStop}%
\bibitem [{\citenamefont {Santini}\ \emph {et~al.}(2008)\citenamefont
  {Santini}, \citenamefont {Cozma}, \citenamefont {Faessler}, \citenamefont
  {Fuchs}, \citenamefont {Krivoruchenko},\ and\ \citenamefont
  {Martemyanov}}]{Santini:2008pk}%
  \BibitemOpen
  \bibfield  {author} {\bibinfo {author} {\bibfnamefont {E.}~\bibnamefont
  {Santini}}, \bibinfo {author} {\bibfnamefont {M.~D.}\ \bibnamefont {Cozma}},
  \bibinfo {author} {\bibfnamefont {A.}~\bibnamefont {Faessler}}, \bibinfo
  {author} {\bibfnamefont {C.}~\bibnamefont {Fuchs}}, \bibinfo {author}
  {\bibfnamefont {M.~I.}\ \bibnamefont {Krivoruchenko}},\ and\ \bibinfo
  {author} {\bibfnamefont {B.}~\bibnamefont {Martemyanov}},\ }\bibfield
  {title} {\bibinfo {title} {Dilepton production in heavy-ion collisions with
  in-medium spectral functions of vector mesons},\ }\href
  {https://doi.org/https://doi.org/10.1103/PhysRevC.78.034910} {\bibfield
  {journal} {\bibinfo  {journal} {Phys.\ Rev.\ C}\ }\textbf {\bibinfo {volume}
  {78}},\ \bibinfo {pages} {034910} (\bibinfo {year} {2008})}\BibitemShut
  {NoStop}%
\bibitem [{\citenamefont {Feynman}(1972)}]{Feynman:1972}%
  \BibitemOpen
  \bibfield  {author} {\bibinfo {author} {\bibfnamefont {R.~P.}\ \bibnamefont
  {Feynman}},\ }\href@noop {} {\emph {\bibinfo {title} {{Photon-hadron
  interactions}}}}\ (\bibinfo  {publisher} {{\relax W.~A.~Benjamin, Inc.}},\
  \bibinfo {address} {Reading, Massachusetts, USA},\ \bibinfo {year}
  {1972})\BibitemShut {NoStop}%
\bibitem [{\citenamefont {K$\ddot{\text{o}}$rner}\ and\ \citenamefont
  {Kuroda}(1977)}]{Korner:1976hv}%
  \BibitemOpen
  \bibfield  {author} {\bibinfo {author} {\bibfnamefont {J.~G.}\ \bibnamefont
  {K$\ddot{\text{o}}$rner}}\ and\ \bibinfo {author} {\bibfnamefont
  {M.}~\bibnamefont {Kuroda}},\ }\bibfield  {title} {\bibinfo {title}
  {{$e^+e^-$ annihilation into baryon-antibaryon pairs}},\ }\href
  {https://doi.org/https://doi.org/10.1103/PhysRevD.16.2165} {\bibfield
  {journal} {\bibinfo  {journal} {Phys.\ Rev.\ D}\ }\textbf {\bibinfo {volume}
  {16}},\ \bibinfo {pages} {2165} (\bibinfo {year} {1977})}\BibitemShut
  {NoStop}%
\bibitem [{\citenamefont {Faessler}\ \emph {et~al.}(2000)\citenamefont
  {Faessler}, \citenamefont {Fuchs},\ and\ \citenamefont
  {Krivoruchenko}}]{Faessler:1999de}%
  \BibitemOpen
  \bibfield  {author} {\bibinfo {author} {\bibfnamefont {A.}~\bibnamefont
  {Faessler}}, \bibinfo {author} {\bibfnamefont {C.}~\bibnamefont {Fuchs}},\
  and\ \bibinfo {author} {\bibfnamefont {M.~I.}\ \bibnamefont
  {Krivoruchenko}},\ }\bibfield  {title} {\bibinfo {title} {{Dilepton spectra
  from decays of light unflavored mesons}},\ }\href
  {https://doi.org/https://doi.org/10.1103/PhysRevC.61.035206} {\bibfield
  {journal} {\bibinfo  {journal} {Phys.\ Rev.\ C}\ }\textbf {\bibinfo {volume}
  {61}},\ \bibinfo {pages} {035206} (\bibinfo {year} {2000})},\ \Eprint
  {https://arxiv.org/abs/nucl-th/9904024} {arXiv:nucl-th/9904024} \BibitemShut
  {NoStop}%
\bibitem [{\citenamefont {Faessler}\ \emph {et~al.}(2003)\citenamefont
  {Faessler}, \citenamefont {Fuchs}, \citenamefont {Krivoruchenko},\ and\
  \citenamefont {Martemyanov}}]{Faessler:2000md}%
  \BibitemOpen
  \bibfield  {author} {\bibinfo {author} {\bibfnamefont {A.}~\bibnamefont
  {Faessler}}, \bibinfo {author} {\bibfnamefont {C.}~\bibnamefont {Fuchs}},
  \bibinfo {author} {\bibfnamefont {M.~I.}\ \bibnamefont {Krivoruchenko}},\
  and\ \bibinfo {author} {\bibfnamefont {B.~V.}\ \bibnamefont {Martemyanov}},\
  }\bibfield  {title} {\bibinfo {title} {{Dilepton production in proton--proton
  collisions at BEVALAC energies}},\ }\href
  {https://doi.org/https://doi.org/10.1088/0954-3899/29/4/302} {\bibfield
  {journal} {\bibinfo  {journal} {J.\ Phys.\ G}\ }\textbf {\bibinfo {volume}
  {29}},\ \bibinfo {pages} {603} (\bibinfo {year} {2003})},\ \Eprint
  {https://arxiv.org/abs/nucl-th/0010056} {arXiv:nucl-th/0010056} \BibitemShut
  {NoStop}%
\bibitem [{\citenamefont {Martemyanov}\ \emph {et~al.}(2010)\citenamefont
  {Martemyanov}, \citenamefont {Faessler},\ and\ \citenamefont
  {Krivoruchenko}}]{Faessler:2009tn}%
  \BibitemOpen
  \bibfield  {author} {\bibinfo {author} {\bibfnamefont {B.~V.}\ \bibnamefont
  {Martemyanov}}, \bibinfo {author} {\bibfnamefont {A.}~\bibnamefont
  {Faessler}},\ and\ \bibinfo {author} {\bibfnamefont {M.~I.}\ \bibnamefont
  {Krivoruchenko}},\ }\bibfield  {title} {\bibinfo {title} {{Electromagnetic
  form factors of nucleons in the extended vector meson dominance model}},\
  }\href {https://doi.org/https://doi.org/10.1103/PhysRevC.82.038201}
  {\bibfield  {journal} {\bibinfo  {journal} {Phys.\ Rev.\ C}\ }\textbf
  {\bibinfo {volume} {82}},\ \bibinfo {pages} {038201} (\bibinfo {year}
  {2010})},\ \Eprint {https://arxiv.org/abs/0910.5589} {0910.5589 [hep-ph]}
  \BibitemShut {NoStop}%
\bibitem [{\citenamefont {Krivoruchenko}\ \emph {et~al.}(2002)\citenamefont
  {Krivoruchenko}, \citenamefont {Martemyanov}, \citenamefont {Faessler},\ and\
  \citenamefont {Fuchs}}]{Krivoruchenko:2001jk}%
  \BibitemOpen
  \bibfield  {author} {\bibinfo {author} {\bibfnamefont {M.~I.}\ \bibnamefont
  {Krivoruchenko}}, \bibinfo {author} {\bibfnamefont {B.~V.}\ \bibnamefont
  {Martemyanov}}, \bibinfo {author} {\bibfnamefont {A.}~\bibnamefont
  {Faessler}},\ and\ \bibinfo {author} {\bibfnamefont {C.}~\bibnamefont
  {Fuchs}},\ }\bibfield  {title} {\bibinfo {title} {{Electromagnetic transition
  form factors and dilepton decay rates of nucleon resonances}},\ }\href
  {https://doi.org/https://doi.org/10.1006/aphy.2002.6223} {\bibfield
  {journal} {\bibinfo  {journal} {Annals Phys.}\ }\textbf {\bibinfo {volume}
  {296}},\ \bibinfo {pages} {299} (\bibinfo {year} {2002})},\ \Eprint
  {https://arxiv.org/abs/nucl-th/0110066} {arXiv:nucl-th/0110066} \BibitemShut
  {NoStop}%
\bibitem [{\citenamefont {Lin}\ \emph {et~al.}(2022)\citenamefont {Lin},
  \citenamefont {Hammer},\ and\ \citenamefont {Mei\ss{}ner}}]{Lin:2021xrc}%
  \BibitemOpen
  \bibfield  {author} {\bibinfo {author} {\bibfnamefont {Y.-H.}\ \bibnamefont
  {Lin}}, \bibinfo {author} {\bibfnamefont {H.-W.}\ \bibnamefont {Hammer}},\
  and\ \bibinfo {author} {\bibfnamefont {U.-G.}\ \bibnamefont {Mei\ss{}ner}},\
  }\bibfield  {title} {\bibinfo {title} {{New insights into the nucleon's
  electromagnetic structure}},\ }\href
  {https://doi.org/https://doi.org/10.1103/PhysRevLett.128.052002} {\bibfield
  {journal} {\bibinfo  {journal} {Phys.\ Rev.\ Lett.}\ }\textbf {\bibinfo
  {volume} {128}},\ \bibinfo {pages} {052002} (\bibinfo {year} {2022})},\
  \Eprint {https://arxiv.org/abs/2109.12961} {2109.12961 [hep-ph]} \BibitemShut
  {NoStop}%
\bibitem [{\citenamefont {Kelly}(2004)}]{Kelly:2004hm}%
  \BibitemOpen
  \bibfield  {author} {\bibinfo {author} {\bibfnamefont {J.~J.}\ \bibnamefont
  {Kelly}},\ }\bibfield  {title} {\bibinfo {title} {{Simple parametrization of
  nucleon form factors}},\ }\href
  {https://doi.org/https://doi.org/10.1103/PhysRevC.70.068202} {\bibfield
  {journal} {\bibinfo  {journal} {Phys.\ Rev.\ C}\ }\textbf {\bibinfo {volume}
  {70}},\ \bibinfo {pages} {068202} (\bibinfo {year} {2004})}\BibitemShut
  {NoStop}%
\bibitem [{\citenamefont {Galster}\ \emph {et~al.}(1971)\citenamefont {Galster}
  \emph {et~al.}}]{Galster:1971kv}%
  \BibitemOpen
  \bibfield  {author} {\bibinfo {author} {\bibfnamefont {S.}~\bibnamefont
  {Galster}} \emph {et~al.},\ }\bibfield  {title} {\bibinfo {title} {{Elastic
  electron--deuteron scattering and the electric neutron form factor at
  four-momentum transfers $5\,\text{fm}^{-2} < q^2 < 14\,\text{fm}^{-2}$}},\
  }\href {https://doi.org/https://doi.org/10.1016/0550-3213(71)90068-X}
  {\bibfield  {journal} {\bibinfo  {journal} {Nucl.\ Phys.\ B}\ }\textbf
  {\bibinfo {volume} {32}},\ \bibinfo {pages} {221} (\bibinfo {year}
  {1971})}\BibitemShut {NoStop}%
\bibitem [{\citenamefont {Bradford}\ \emph {et~al.}(2006)\citenamefont
  {Bradford}, \citenamefont {Bodek}, \citenamefont {Budd},\ and\ \citenamefont
  {Arrington}}]{Bradford:2006yz}%
  \BibitemOpen
  \bibfield  {author} {\bibinfo {author} {\bibfnamefont {R.~K.~{\relax Jr}.}\
  \bibnamefont {Bradford}}, \bibinfo {author} {\bibfnamefont {A.}~\bibnamefont
  {Bodek}}, \bibinfo {author} {\bibfnamefont {H.~S.}\ \bibnamefont {Budd}},\
  and\ \bibinfo {author} {\bibfnamefont {J.~R.}\ \bibnamefont {Arrington}},\
  }\bibfield  {title} {\bibinfo {title} {{A new parameterization of the nucleon
  elastic form factors}},\ }\bibfield  {booktitle} {\emph {\bibinfo {booktitle}
  {{Proceedings of the 4th International Workshop on Neutrino--Nucleus
  Interactions in the Few GeV Region (NuInt\,2005), Okayama, Japan, September
  26--29, 2005}}},\ }\href
  {https://doi.org/https://doi.org/10.1016/j.nuclphysbps.2006.08.028}
  {\bibfield  {journal} {\bibinfo  {journal} {Nucl.\ Phys.\ (Proc.\ Suppl.)}\
  }\textbf {\bibinfo {volume} {159}},\ \bibinfo {pages} {127} (\bibinfo {year}
  {2006})},\ \Eprint {https://arxiv.org/abs/hep-ex/0602017}
  {arXiv:hep-ex/0602017} \BibitemShut {NoStop}%
\bibitem [{\citenamefont {Bodek}\ \emph {et~al.}(2008)\citenamefont {Bodek},
  \citenamefont {Avvakumov}, \citenamefont {Bradford},\ and\ \citenamefont
  {Budd}}]{Bodek:2007ym}%
  \BibitemOpen
  \bibfield  {author} {\bibinfo {author} {\bibfnamefont {A.}~\bibnamefont
  {Bodek}}, \bibinfo {author} {\bibfnamefont {S.}~\bibnamefont {Avvakumov}},
  \bibinfo {author} {\bibfnamefont {R.}~\bibnamefont {Bradford}},\ and\
  \bibinfo {author} {\bibfnamefont {H.~S.}\ \bibnamefont {Budd}},\ }\bibfield
  {title} {\bibinfo {title} {{Vector and axial nucleon form factors: A duality
  constrained parameterization}},\ }\href
  {https://doi.org/https://doi.org/10.1140/epjc/s10052-007-0491-4} {\bibfield
  {journal} {\bibinfo  {journal} {Eur.\ Phys.\ J.\ C}\ }\textbf {\bibinfo
  {volume} {53}},\ \bibinfo {pages} {349} (\bibinfo {year} {2008})},\ \Eprint
  {https://arxiv.org/abs/0708.1946} {0708.1946 [hep-ex]} \BibitemShut {NoStop}%
\bibitem [{\citenamefont {Castellano}\ \emph {et~al.}(1973)\citenamefont
  {Castellano} \emph {et~al.}}]{Castellano:1973wh}%
  \BibitemOpen
  \bibfield  {author} {\bibinfo {author} {\bibfnamefont {M.}~\bibnamefont
  {Castellano}} \emph {et~al.},\ }\bibfield  {title} {\bibinfo {title} {{The
  reaction $e^+e^-{\to}\,\overline{p}p$ at a total energy of 2.1 GeV}},\ }\href
  {https://doi.org/https://doi.org/10.1007/BF02734600} {\bibfield  {journal}
  {\bibinfo  {journal} {Il Nuovo Cim.}\ }\textbf {\bibinfo {volume} {14 A}},\
  \bibinfo {pages} {1} (\bibinfo {year} {1973})}\BibitemShut {NoStop}%
\bibitem [{\citenamefont {Voci}(1997)}]{Voci:1997ku}%
  \BibitemOpen
  \bibfield  {author} {\bibinfo {author} {\bibfnamefont {C.}~\bibnamefont
  {Voci}} (\bibinfo {collaboration} {on behalf of FENICE Collaboration}),\
  }\bibfield  {title} {\bibinfo {title} {{The first measurement of the neutron
  electromagnetic form factors in the time-like region}},\ }\href
  {https://doi.org/https://doi.org/10.1016/S0375-9474(97)00452-1} {\bibfield
  {journal} {\bibinfo  {journal} {Nucl.\ Phys.\ A}\ }\textbf {\bibinfo {volume}
  {623}},\ \bibinfo {pages} {333c} (\bibinfo {year} {1997})}\BibitemShut
  {NoStop}%
\bibitem [{\citenamefont {Bassompierre}(1977)}]{Bassompierre:1977ks}%
  \BibitemOpen
  \bibfield  {author} {\bibinfo {author} {\bibfnamefont {G.}~\bibnamefont
  {Bassompierre}} (\bibinfo {collaboration} {Mulhouse--Strasbourg--Turin
  Collaboration}),\ }\bibfield  {title} {\bibinfo {title} {{First determination
  of the proton electromagnetic form factors at the threshold of the time-like
  region}},\ }\href
  {https://doi.org/https://doi.org/10.1016/0370-2693(77)90475-0} {\bibfield
  {journal} {\bibinfo  {journal} {Phys.\ Lett.}\ }\textbf {\bibinfo {volume}
  {68B}},\ \bibinfo {pages} {477} (\bibinfo {year} {1977})}\BibitemShut
  {NoStop}%
\bibitem [{\citenamefont {Conversi}\ \emph {et~al.}(1965)\citenamefont
  {Conversi}, \citenamefont {Massam}, \citenamefont {Zichichi},\ and\
  \citenamefont {Muller}}]{Conversi:1965nn}%
  \BibitemOpen
  \bibfield  {author} {\bibinfo {author} {\bibfnamefont {M.}~\bibnamefont
  {Conversi}}, \bibinfo {author} {\bibfnamefont {T.}~\bibnamefont {Massam}},
  \bibinfo {author} {\bibfnamefont {A.}~\bibnamefont {Zichichi}},\ and\
  \bibinfo {author} {\bibfnamefont {T.}~\bibnamefont {Muller}},\ }\bibfield
  {title} {\bibinfo {title} {{The leptonic annihilation modes of the
  proton-antiproton system, at $6.8\left(\text{GeV}/c\right)^2$ timelike
  four-momentum transfer}},\ }\href
  {https://doi.org/https://doi.org/10.1007/BF02721062} {\bibfield  {journal}
  {\bibinfo  {journal} {Il Nuovo Cim.}\ }\textbf {\bibinfo {volume} {XL A}},\
  \bibinfo {pages} {690} (\bibinfo {year} {1965})}\BibitemShut {NoStop}%
\bibitem [{\citenamefont {Hartill}\ \emph {et~al.}(1969)\citenamefont {Hartill}
  \emph {et~al.}}]{Hartill:1969ahu}%
  \BibitemOpen
  \bibfield  {author} {\bibinfo {author} {\bibfnamefont {D.~L.}\ \bibnamefont
  {Hartill}} \emph {et~al.},\ }\bibfield  {title} {\bibinfo {title}
  {{Antiproton-proton annihilation into electron-positron pairs and gamma-ray
  pairs}},\ }\href {https://doi.org/https://doi.org/10.1103/PhysRev.184.1415}
  {\bibfield  {journal} {\bibinfo  {journal} {Phys.\ Rev.}\ }\textbf {\bibinfo
  {volume} {184}},\ \bibinfo {pages} {1415} (\bibinfo {year}
  {1969})}\BibitemShut {NoStop}%
\bibitem [{\citenamefont {Bisello}\ \emph {et~al.}(1983)\citenamefont {Bisello}
  \emph {et~al.}}]{Bisello:1983at}%
  \BibitemOpen
  \bibfield  {author} {\bibinfo {author} {\bibfnamefont {D.}~\bibnamefont
  {Bisello}} \emph {et~al.},\ }\bibfield  {title} {\bibinfo {title} {{A
  measurement of $e^+e^-{\to}\,\overline{p}p$ for $\left(1975 \leqslant
  \sqrt{s} \leqslant 2250\right)$ MeV}},\ }\href
  {https://doi.org/https://doi.org/10.1016/0550-3213(83)90381-4} {\bibfield
  {journal} {\bibinfo  {journal} {Nucl.\ Phys. B}\ }\textbf {\bibinfo {volume}
  {224}},\ \bibinfo {pages} {379} (\bibinfo {year} {1983})}\BibitemShut
  {NoStop}%
\bibitem [{\citenamefont {Delcourt}\ \emph {et~al.}(1979)\citenamefont
  {Delcourt} \emph {et~al.}}]{Delcourt:1979ed}%
  \BibitemOpen
  \bibfield  {author} {\bibinfo {author} {\bibfnamefont {B.}~\bibnamefont
  {Delcourt}} \emph {et~al.},\ }\bibfield  {title} {\bibinfo {title} {{Study of
  the reaction $e^+e^-\,{\to}\,p\overline{p}$ in the total energy range
  $1925-2180$ MeV}},\ }\href
  {https://doi.org/https://doi.org/10.1016/0370-2693(79)90864-5} {\bibfield
  {journal} {\bibinfo  {journal} {Phys.\ Lett.}\ }\textbf {\bibinfo {volume}
  {86 B}},\ \bibinfo {pages} {395} (\bibinfo {year} {1979})}\BibitemShut
  {NoStop}%
\bibitem [{\citenamefont {Antonelli}\ \emph
  {et~al.}(1993{\natexlab{a}})\citenamefont {Antonelli} \emph
  {et~al.}}]{Antonelli:1993vz}%
  \BibitemOpen
  \bibfield  {author} {\bibinfo {author} {\bibfnamefont {A.}~\bibnamefont
  {Antonelli}} \emph {et~al.},\ }\bibfield  {title} {\bibinfo {title} {{First
  measurement of the neutron electromagnetic form factor in the time-like
  region}},\ }\href
  {https://doi.org/https://doi.org/10.1016/0370-2693(93)91225-C} {\bibfield
  {journal} {\bibinfo  {journal} {Phys.\ Lett.\ B}\ }\textbf {\bibinfo {volume}
  {313}},\ \bibinfo {pages} {283} (\bibinfo {year}
  {1993}{\natexlab{a}})}\BibitemShut {NoStop}%
\bibitem [{\citenamefont {Antonelli}\ \emph {et~al.}(1994)\citenamefont
  {Antonelli} \emph {et~al.}}]{Antonelli:1994kq}%
  \BibitemOpen
  \bibfield  {author} {\bibinfo {author} {\bibfnamefont {A.}~\bibnamefont
  {Antonelli}} \emph {et~al.},\ }\bibfield  {title} {\bibinfo {title}
  {{Measurement of the electromagnetic form factor of the proton in the
  time-like region}},\ }\href
  {https://doi.org/https://doi.org/10.1016/0370-2693(94)90710-2} {\bibfield
  {journal} {\bibinfo  {journal} {Phys.\ Lett.\ B}\ }\textbf {\bibinfo {volume}
  {334}},\ \bibinfo {pages} {431} (\bibinfo {year} {1994})}\BibitemShut
  {NoStop}%
\bibitem [{\citenamefont {Punjabi}\ \emph {et~al.}(2015)\citenamefont {Punjabi}
  \emph {et~al.}}]{Punjabi:2015bba}%
  \BibitemOpen
  \bibfield  {author} {\bibinfo {author} {\bibfnamefont {V.}~\bibnamefont
  {Punjabi}} \emph {et~al.},\ }\bibfield  {title} {\bibinfo {title} {{The
  structure of the nucleon: Elastic electromagnetic form factors}},\ }\href
  {https://doi.org/https://doi.org/10.1140/epja/i2015-15079-x} {\bibfield
  {journal} {\bibinfo  {journal} {Eur.\ Phys.\ J.\ A}\ }\textbf {\bibinfo
  {volume} {51}},\ \bibinfo {pages} {79} (\bibinfo {year} {2015})},\ \Eprint
  {https://arxiv.org/abs/1503.01452} {1503.01452 [nucl-ex]} \BibitemShut
  {NoStop}%
\bibitem [{\citenamefont {Denig}\ and\ \citenamefont
  {Salme}(2013)}]{Denig:2012by}%
  \BibitemOpen
  \bibfield  {author} {\bibinfo {author} {\bibfnamefont {A.}~\bibnamefont
  {Denig}}\ and\ \bibinfo {author} {\bibfnamefont {G.}~\bibnamefont {Salme}},\
  }\bibfield  {title} {\bibinfo {title} {{Nucleon electromagnetic form factors
  in the timelike region}},\ }\href
  {https://doi.org/https://doi.org/10.1016/j.ppnp.2012.09.005} {\bibfield
  {journal} {\bibinfo  {journal} {Prog.\ Part.\ Nucl.\ Phys.}\ }\textbf
  {\bibinfo {volume} {68}},\ \bibinfo {pages} {113} (\bibinfo {year} {2013})},\
  \Eprint {https://arxiv.org/abs/1210.4689} {1210.4689 [hep-ex]} \BibitemShut
  {NoStop}%
\bibitem [{\citenamefont {Pacetti}\ \emph {et~al.}(2015)\citenamefont
  {Pacetti}, \citenamefont {Baldini~Ferroli},\ and\ \citenamefont
  {Tomasi-Gustafsson}}]{Pacetti:2014jai}%
  \BibitemOpen
  \bibfield  {author} {\bibinfo {author} {\bibfnamefont {S.}~\bibnamefont
  {Pacetti}}, \bibinfo {author} {\bibfnamefont {R.}~\bibnamefont
  {Baldini~Ferroli}},\ and\ \bibinfo {author} {\bibfnamefont {E.}~\bibnamefont
  {Tomasi-Gustafsson}},\ }\bibfield  {title} {\bibinfo {title} {{Proton
  electromagnetic form factors: Basic notions, present achievements and future
  perspectives}},\ }\href
  {https://doi.org/https://doi.org/10.1016/j.physrep.2014.09.005} {\bibfield
  {journal} {\bibinfo  {journal} {Phys.\ Rep.}\ }\textbf {\bibinfo {volume}
  {550--551}},\ \bibinfo {pages} {1} (\bibinfo {year} {2015})}\BibitemShut
  {NoStop}%
\bibitem [{\citenamefont {Bjorken}\ and\ \citenamefont
  {Drell}(1964)}]{Bjorken:1964}%
  \BibitemOpen
  \bibfield  {author} {\bibinfo {author} {\bibfnamefont {J.~D.}\ \bibnamefont
  {Bjorken}}\ and\ \bibinfo {author} {\bibfnamefont {S.~D.}\ \bibnamefont
  {Drell}},\ }\href@noop {} {\emph {\bibinfo {title} {{Relativistic quantum
  mechanics}}}}\ (\bibinfo  {publisher} {McGraw-Hill Inc.},\ \bibinfo {address}
  {New York City, New York, USA},\ \bibinfo {year} {1964})\BibitemShut
  {NoStop}%
\bibitem [{\citenamefont {Geshkenbein}(1989)}]{Geshkenbein:1988ef}%
  \BibitemOpen
  \bibfield  {author} {\bibinfo {author} {\bibfnamefont {B.~V.}\ \bibnamefont
  {Geshkenbein}},\ }\bibfield  {title} {\bibinfo {title} {{Analysis of
  experiments on the measurement of the pion electromagnetic formfactor}},\
  }\href {https://doi.org/https://doi.org/10.1007/BF01674468} {\bibfield
  {journal} {\bibinfo  {journal} {Z.\ Phys.\ C}\ }\textbf {\bibinfo {volume}
  {45}},\ \bibinfo {pages} {351} (\bibinfo {year} {1989})}\BibitemShut
  {NoStop}%
\bibitem [{\citenamefont {Ablikim}\ \emph
  {et~al.}(2021{\natexlab{a}})\citenamefont {Ablikim} \emph
  {et~al.}}]{BESIII:2021rqk}%
  \BibitemOpen
  \bibfield  {author} {\bibinfo {author} {\bibfnamefont {M.}~\bibnamefont
  {Ablikim}} \emph {et~al.} (\bibinfo {collaboration} {BESIII Collaboration}),\
  }\bibfield  {title} {\bibinfo {title} {{Measurement of proton electromagnetic
  form factors in the time-like region using initial state radiation at
  BESIII}},\ }\href
  {https://doi.org/https://doi.org/10.1016/j.physletb.2021.136328} {\bibfield
  {journal} {\bibinfo  {journal} {Phys.\ Lett.\ B}\ }\textbf {\bibinfo {volume}
  {817}},\ \bibinfo {pages} {136328} (\bibinfo {year} {2021}{\natexlab{a}})},\
  \Eprint {https://arxiv.org/abs/2102.10337} {2102.10337 [hep-ex]} \BibitemShut
  {NoStop}%
\bibitem [{\citenamefont {Achasov}\ \emph
  {et~al.}(2023{\natexlab{a}})\citenamefont {Achasov} \emph
  {et~al.}}]{SND:2023fos}%
  \BibitemOpen
  \bibfield  {author} {\bibinfo {author} {\bibfnamefont {M.~N.}\ \bibnamefont
  {Achasov}} \emph {et~al.},\ }\bibfield  {title} {\bibinfo {title}
  {{Measurements of the neutron timelike electromagnetic formfactor with the
  SND detector}},\ }\href
  {https://doi.org/https://doi.org/10.1134/S1063778823060054} {\bibfield
  {journal} {\bibinfo  {journal} {Phys.\ Atom.\ Nucl.}\ }\textbf {\bibinfo
  {volume} {86}},\ \bibinfo {pages} {1165} (\bibinfo {year}
  {2023}{\natexlab{a}})},\ \Eprint {https://arxiv.org/abs/2309.05241}
  {2309.05241 [hep-ex]} \BibitemShut {NoStop}%
\bibitem [{\citenamefont {Achasov}\ \emph {et~al.}(2022)\citenamefont {Achasov}
  \emph {et~al.}}]{SND:2022wdb}%
  \BibitemOpen
  \bibfield  {author} {\bibinfo {author} {\bibfnamefont {M.~N.}\ \bibnamefont
  {Achasov}} \emph {et~al.} (\bibinfo {collaboration} {SND Collaboration}),\
  }\bibfield  {title} {\bibinfo {title} {{Experimental study of the
  $e^+e^-\rightarrow n{\overline{n}}$ process at the VEPP-2000 $e^+e^-$
  Collider with the SND detector}},\ }\href
  {https://doi.org/10.1140/epjc/s10052-022-10696-0} {\bibfield  {journal}
  {\bibinfo  {journal} {Eur.\ Phys.\ J.\ C}\ }\textbf {\bibinfo {volume}
  {82}},\ \bibinfo {pages} {761} (\bibinfo {year} {2022})},\ \Eprint
  {https://arxiv.org/abs/2206.13047} {2206.13047 [hep-ex]} \BibitemShut
  {NoStop}%
\bibitem [{\citenamefont {Griffiths}(d ed)}]{Griffiths:2008}%
  \BibitemOpen
  \bibfield  {author} {\bibinfo {author} {\bibfnamefont {D.~J.}\ \bibnamefont
  {Griffiths}},\ }\href@noop {} {\emph {\bibinfo {title} {{Introduction to
  elementary particles}}}}\ (\bibinfo  {publisher} {WILEY-VCH Verlag GmbH \&
  Co. KGaA},\ \bibinfo {address} {Weinheim, Gernamy},\ \bibinfo {year} {2008
  (second, revised ed.)})\BibitemShut {NoStop}%
\bibitem [{\citenamefont {Shirkov}\ \emph {et~al.}(1969)\citenamefont
  {Shirkov}, \citenamefont {Serebryakov},\ and\ \citenamefont
  {Mescheryakov}}]{Shirkov:1969}%
  \BibitemOpen
  \bibfield  {author} {\bibinfo {author} {\bibfnamefont {D.~V.}\ \bibnamefont
  {Shirkov}}, \bibinfo {author} {\bibfnamefont {V.~V.}\ \bibnamefont
  {Serebryakov}},\ and\ \bibinfo {author} {\bibfnamefont {V.~A.}\ \bibnamefont
  {Mescheryakov}},\ }\href@noop {} {\emph {\bibinfo {title} {{Dispersion
  theories of strong interactions at low energy}}}}\ (\bibinfo  {publisher}
  {North-Holland Publishing Co.},\ \bibinfo {address} {Amsterdam, The
  Netherlands},\ \bibinfo {year} {1969})\BibitemShut {NoStop}%
\bibitem [{\citenamefont {H$\ddot{\text{o}}$hler}(1983)}]{Hoehler:1983}%
  \BibitemOpen
  \bibfield  {author} {\bibinfo {author} {\bibfnamefont {G.}~\bibnamefont
  {H$\ddot{\text{o}}$hler}},\ }\href@noop {} {\emph {\bibinfo {title}
  {{Pion--nucleon scattering}}}}\ (\bibinfo  {publisher} {Springer,
  Landolt-B$\ddot{\text{o}}$rnstein},\ \bibinfo {address} {Switzerland},\
  \bibinfo {year} {1983})\BibitemShut {NoStop}%
\bibitem [{\citenamefont {Lin}\ \emph {et~al.}(2021)\citenamefont {Lin},
  \citenamefont {Hammer},\ and\ \citenamefont {Mei\ss{}ner}}]{Lin:2021umz}%
  \BibitemOpen
  \bibfield  {author} {\bibinfo {author} {\bibfnamefont {Y.-H.}\ \bibnamefont
  {Lin}}, \bibinfo {author} {\bibfnamefont {H.-W.}\ \bibnamefont {Hammer}},\
  and\ \bibinfo {author} {\bibfnamefont {U.-G.}\ \bibnamefont {Mei\ss{}ner}},\
  }\bibfield  {title} {\bibinfo {title} {{Dispersion-theoretical analysis of
  the electromagnetic form factors of the nucleon: Past, present and future}},\
  }\href {https://doi.org/https://doi.org/10.1140/epja/s10050-021-00562-0}
  {\bibfield  {journal} {\bibinfo  {journal} {Eur.\ Phys.\ J.\ A}\ }\textbf
  {\bibinfo {volume} {57}},\ \bibinfo {pages} {255} (\bibinfo {year} {2021})},\
  \Eprint {https://arxiv.org/abs/2106.06357} {2106.06357 [hep-ph]} \BibitemShut
  {NoStop}%
\bibitem [{\citenamefont {Collins}(2023)}]{Collins:1977jy}%
  \BibitemOpen
  \bibfield  {author} {\bibinfo {author} {\bibfnamefont {P.~D.~B.}\
  \bibnamefont {Collins}},\ }\href
  {https://doi.org/https://doi.org/10.1017/9781009403269} {\emph {\bibinfo
  {title} {{An introduction to Regge theory and high energy physics}}}}\
  (\bibinfo  {publisher} {Cambridge University Press},\ \bibinfo {address}
  {Cambridge, England},\ \bibinfo {year} {2023})\BibitemShut {NoStop}%
\bibitem [{\citenamefont {Sakurai}\ and\ \citenamefont
  {Schildknecht}(1972)}]{Sakurai:1972wk}%
  \BibitemOpen
  \bibfield  {author} {\bibinfo {author} {\bibfnamefont {J.~J.}\ \bibnamefont
  {Sakurai}}\ and\ \bibinfo {author} {\bibfnamefont {D.}~\bibnamefont
  {Schildknecht}},\ }\bibfield  {title} {\bibinfo {title} {{Generalized vector
  dominance and inelastic electron--proton scattering}},\ }\href
  {https://doi.org/https://doi.org/10.1016/0370-2693(72)90300-0} {\bibfield
  {journal} {\bibinfo  {journal} {Phys.\ Lett.}\ }\textbf {\bibinfo {volume}
  {40B}},\ \bibinfo {pages} {121} (\bibinfo {year} {1972})}\BibitemShut
  {NoStop}%
\bibitem [{\citenamefont {H$\ddot{\text{o}}$hler}\ \emph
  {et~al.}(1976)\citenamefont {H$\ddot{\text{o}}$hler} \emph
  {et~al.}}]{Hohler:1976ax}%
  \BibitemOpen
  \bibfield  {author} {\bibinfo {author} {\bibfnamefont {G.}~\bibnamefont
  {H$\ddot{\text{o}}$hler}} \emph {et~al.},\ }\bibfield  {title} {\bibinfo
  {title} {{Analysis of electromagnetic nucleon form factors}},\ }\href
  {https://doi.org/https://doi.org/10.1016/0550-3213(76)90449-1} {\bibfield
  {journal} {\bibinfo  {journal} {Nucl.\ Phys.\ B}\ }\textbf {\bibinfo {volume}
  {114}},\ \bibinfo {pages} {505} (\bibinfo {year} {1976})}\BibitemShut
  {NoStop}%
\bibitem [{\citenamefont {Belushkin}\ \emph {et~al.}(2007)\citenamefont
  {Belushkin}, \citenamefont {Hammer},\ and\ \citenamefont
  {Mei\ss{}ner}}]{Belushkin:2006qa}%
  \BibitemOpen
  \bibfield  {author} {\bibinfo {author} {\bibfnamefont {M.~A.}\ \bibnamefont
  {Belushkin}}, \bibinfo {author} {\bibfnamefont {H.-W.}\ \bibnamefont
  {Hammer}},\ and\ \bibinfo {author} {\bibfnamefont {U.-G.}\ \bibnamefont
  {Mei\ss{}ner}},\ }\bibfield  {title} {\bibinfo {title} {{Dispersion analysis
  of the nucleon form factors including meson continua}},\ }\href
  {https://doi.org/https://doi.org/10.1103/PhysRevC.75.035202} {\bibfield
  {journal} {\bibinfo  {journal} {Phys.\ Rev.\ C}\ }\textbf {\bibinfo {volume}
  {75}},\ \bibinfo {pages} {035202} (\bibinfo {year} {2007})},\ \Eprint
  {https://arxiv.org/abs/hep-ph/0608337} {arXiv:hep-ph/0608337} \BibitemShut
  {NoStop}%
\bibitem [{\citenamefont {Yan}\ \emph {et~al.}(2023)\citenamefont {Yan},
  \citenamefont {Chen},\ and\ \citenamefont {Xie}}]{Yan:2023yff}%
  \BibitemOpen
  \bibfield  {author} {\bibinfo {author} {\bibfnamefont {B.}~\bibnamefont
  {Yan}}, \bibinfo {author} {\bibfnamefont {C.}~\bibnamefont {Chen}},\ and\
  \bibinfo {author} {\bibfnamefont {J.-J.}\ \bibnamefont {Xie}},\ }\bibfield
  {title} {\bibinfo {title} {{$\Sigma$ and $\Xi$ electromagnetic form factors
  in the extended vector meson dominance model}},\ }\href
  {https://doi.org/https://doi.org/10.1103/PhysRevD.107.076008} {\bibfield
  {journal} {\bibinfo  {journal} {Phys.\ Rev.\ D}\ }\textbf {\bibinfo {volume}
  {107}},\ \bibinfo {pages} {076008} (\bibinfo {year} {2023})},\ \Eprint
  {https://arxiv.org/abs/2301.00976} {2301.00976 [hep-ph]} \BibitemShut
  {NoStop}%
\bibitem [{\citenamefont {Ye}\ \emph {et~al.}(2018)\citenamefont {Ye},
  \citenamefont {Arrington}, \citenamefont {Hill},\ and\ \citenamefont
  {Lee}}]{Ye:2017gyb}%
  \BibitemOpen
  \bibfield  {author} {\bibinfo {author} {\bibfnamefont {Z.}~\bibnamefont
  {Ye}}, \bibinfo {author} {\bibfnamefont {J.~R.}\ \bibnamefont {Arrington}},
  \bibinfo {author} {\bibfnamefont {R.~J.}\ \bibnamefont {Hill}},\ and\
  \bibinfo {author} {\bibfnamefont {G.}~\bibnamefont {Lee}},\ }\bibfield
  {title} {\bibinfo {title} {{Proton and neutron electromagnetic form factors
  and uncertainties}},\ }\href
  {https://doi.org/https://doi.org/10.1016/j.physletb.2017.11.023} {\bibfield
  {journal} {\bibinfo  {journal} {Phys.\ Lett.\ B}\ }\textbf {\bibinfo {volume}
  {777}},\ \bibinfo {pages} {8} (\bibinfo {year} {2018})},\ \Eprint
  {https://arxiv.org/abs/1707.09063} {1707.09063 [nucl-ex]} \BibitemShut
  {NoStop}%
\bibitem [{\citenamefont {Fuchs}\ \emph {et~al.}(2003)\citenamefont {Fuchs},
  \citenamefont {Krivoruchenko}, \citenamefont {Yadav}, \citenamefont
  {Faessler}, \citenamefont {Martemyanov},\ and\ \citenamefont
  {Shekhter}}]{Fuchs:2002vs}%
  \BibitemOpen
  \bibfield  {author} {\bibinfo {author} {\bibfnamefont {C.}~\bibnamefont
  {Fuchs}}, \bibinfo {author} {\bibfnamefont {M.~I.}\ \bibnamefont
  {Krivoruchenko}}, \bibinfo {author} {\bibfnamefont {H.~L.}\ \bibnamefont
  {Yadav}}, \bibinfo {author} {\bibfnamefont {A.}~\bibnamefont {Faessler}},
  \bibinfo {author} {\bibfnamefont {B.~V.}\ \bibnamefont {Martemyanov}},\ and\
  \bibinfo {author} {\bibfnamefont {K.}~\bibnamefont {Shekhter}},\ }\bibfield
  {title} {\bibinfo {title} {{Off-shell $\omega$ production in proton--proton
  collisions near threshold}},\ }\href
  {https://doi.org/https://doi.org/10.1103/PhysRevC.67.025202} {\bibfield
  {journal} {\bibinfo  {journal} {Phys.\ Rev.\ C}\ }\textbf {\bibinfo {volume}
  {67}},\ \bibinfo {pages} {025202} (\bibinfo {year} {2003})},\ \Eprint
  {https://arxiv.org/abs/nucl-th/0208022} {arXiv:nucl-th/0208022} \BibitemShut
  {NoStop}%
\bibitem [{\citenamefont {Gell-Mann}\ \emph {et~al.}(1962)\citenamefont
  {Gell-Mann}, \citenamefont {Sharp},\ and\ \citenamefont
  {Wagner}}]{Gell-Mann:1962hpq}%
  \BibitemOpen
  \bibfield  {author} {\bibinfo {author} {\bibfnamefont {M.}~\bibnamefont
  {Gell-Mann}}, \bibinfo {author} {\bibfnamefont {D.}~\bibnamefont {Sharp}},\
  and\ \bibinfo {author} {\bibfnamefont {W.~G.}\ \bibnamefont {Wagner}},\
  }\bibfield  {title} {\bibinfo {title} {{Decay rates of neutral mesons}},\
  }\href {https://doi.org/https://doi.org/10.1103/PhysRevLett.8.261} {\bibfield
   {journal} {\bibinfo  {journal} {Phys.\ Rev.\ Lett.}\ }\textbf {\bibinfo
  {volume} {8}},\ \bibinfo {pages} {261} (\bibinfo {year} {1962})}\BibitemShut
  {NoStop}%
\bibitem [{\citenamefont {Navas}\ \emph {et~al.}(2024)\citenamefont {Navas}
  \emph {et~al.}}]{ParticleDataGroup:2024}%
  \BibitemOpen
  \bibfield  {author} {\bibinfo {author} {\bibfnamefont {S.}~\bibnamefont
  {Navas}} \emph {et~al.} (\bibinfo {collaboration} {Particle Data Group}),\
  }\bibfield  {title} {\bibinfo {title} {{Review of particle physics}},\ }\href
  {https://doi.org/https://doi.org/10.1103/PhysRevD.110.030001} {\bibfield
  {journal} {\bibinfo  {journal} {Phys.\ Rev.\ D}\ }\textbf {\bibinfo {volume}
  {110}},\ \bibinfo {pages} {030001} (\bibinfo {year} {2024})}\BibitemShut
  {NoStop}%
\bibitem [{\citenamefont {Frenkiel}\ \emph {et~al.}(1972)\citenamefont
  {Frenkiel} \emph {et~al.}}]{Frenkiel:1972ngp}%
  \BibitemOpen
  \bibfield  {author} {\bibinfo {author} {\bibfnamefont {P.}~\bibnamefont
  {Frenkiel}} \emph {et~al.},\ }\bibfield  {title} {\bibinfo {title}
  {{$\omega\pi$ resonances and $\pi\pi$ s-wave structures as observed in
  $\overline{p}p$ annihilations at rest}},\ }\href
  {https://doi.org/https://doi.org/10.1016/0550-3213(72)90101-0} {\bibfield
  {journal} {\bibinfo  {journal} {Nucl.\ Phys.\ B}\ }\textbf {\bibinfo {volume}
  {47}},\ \bibinfo {pages} {61} (\bibinfo {year} {1972})}\BibitemShut {NoStop}%
\bibitem [{\citenamefont {Bartalucci}\ \emph {et~al.}(1979)\citenamefont
  {Bartalucci} \emph {et~al.}}]{Bartalucci:1978gy}%
  \BibitemOpen
  \bibfield  {author} {\bibinfo {author} {\bibfnamefont {S.}~\bibnamefont
  {Bartalucci}} \emph {et~al.},\ }\bibfield  {title} {\bibinfo {title}
  {{Experimental confirmation of the 1100 structure and first observation of
  the leptonic decay of the $\rho^\prime(1250)$}},\ }\href
  {https://doi.org/https://doi.org/10.1007/BF02896723} {\bibfield  {journal}
  {\bibinfo  {journal} {Nuovo Cim.}\ }\textbf {\bibinfo {volume} {49\ A}},\
  \bibinfo {pages} {207} (\bibinfo {year} {1979})}\BibitemShut {NoStop}%
\bibitem [{\citenamefont {Barber}\ \emph {et~al.}(1980)\citenamefont {Barber}
  \emph {et~al.}}]{LAMP2Group:1979ibr}%
  \BibitemOpen
  \bibfield  {author} {\bibinfo {author} {\bibfnamefont {D.~P.}\ \bibnamefont
  {Barber}} \emph {et~al.},\ }\bibfield  {title} {\bibinfo {title}
  {{Photoproduction of $\rho^\prime(1.2)$ and $\rho^\prime(1.6)$ in the final
  states $\pi^+\pi^-\pi^+\pi^-$ and $\pi^+\pi^-\pi^0\pi^0$}},\ }\href
  {https://doi.org/https://doi.org/10.1007/BF01421795} {\bibfield  {journal}
  {\bibinfo  {journal} {Z.\ Phys.\ C}\ }\textbf {\bibinfo {volume} {4}},\
  \bibinfo {pages} {169} (\bibinfo {year} {1980})}\BibitemShut {NoStop}%
\bibitem [{\citenamefont {Hammoud}\ \emph {et~al.}(2020)\citenamefont
  {Hammoud}, \citenamefont {Kami\'nski}, \citenamefont {Nazari},\ and\
  \citenamefont {Rupp}}]{Hammoud:2020aqi}%
  \BibitemOpen
  \bibfield  {author} {\bibinfo {author} {\bibfnamefont {N.}~\bibnamefont
  {Hammoud}}, \bibinfo {author} {\bibfnamefont {R.}~\bibnamefont {Kami\'nski}},
  \bibinfo {author} {\bibfnamefont {V.}~\bibnamefont {Nazari}},\ and\ \bibinfo
  {author} {\bibfnamefont {G.}~\bibnamefont {Rupp}},\ }\bibfield  {title}
  {\bibinfo {title} {{Strong evidence of the $\rho(1250)$ from a unitary
  multichannel reanalysis of elastic scattering data with crossing-symmetry
  constraints}},\ }\href {https://doi.org/10.1103/PhysRevD.102.054029}
  {\bibfield  {journal} {\bibinfo  {journal} {Phys.\ Rev.\ D}\ }\textbf
  {\bibinfo {volume} {102}},\ \bibinfo {pages} {054029} (\bibinfo {year}
  {2020})},\ \Eprint {https://arxiv.org/abs/2009.06317} {2009.06317 [hep-ph]}
  \BibitemShut {NoStop}%
\bibitem [{\citenamefont {Christy}\ \emph {et~al.}(2004)\citenamefont {Christy}
  \emph {et~al.}}]{E94110:2004lsx}%
  \BibitemOpen
  \bibfield  {author} {\bibinfo {author} {\bibfnamefont {M.~E.}\ \bibnamefont
  {Christy}} \emph {et~al.} (\bibinfo {collaboration} {Jefferson Lab E94-110
  Collaboration}),\ }\bibfield  {title} {\bibinfo {title} {{Measurements of
  electron--proton elastic cross sections for $0.4 < Q^2 <
  5.5\,\left(\text{GeV}/c\right)^2$}},\ }\href
  {https://doi.org/https://doi.org/10.1103/PhysRevC.70.015206} {\bibfield
  {journal} {\bibinfo  {journal} {Phys.\ Rev.\ C}\ }\textbf {\bibinfo {volume}
  {70}},\ \bibinfo {pages} {015206} (\bibinfo {year} {2004})},\ \Eprint
  {https://arxiv.org/abs/nucl-ex/0401030} {arXiv:nucl-ex/0401030} \BibitemShut
  {NoStop}%
\bibitem [{\citenamefont {Qattan}\ \emph {et~al.}(2005)\citenamefont {Qattan}
  \emph {et~al.}}]{Qattan:2004ht}%
  \BibitemOpen
  \bibfield  {author} {\bibinfo {author} {\bibfnamefont {I.~A.}\ \bibnamefont
  {Qattan}} \emph {et~al.},\ }\bibfield  {title} {\bibinfo {title} {{Precision
  Rosenbluth measurement of the proton elastic form factors}},\ }\href
  {https://doi.org/https://doi.org/10.1103/PhysRevLett.94.142301} {\bibfield
  {journal} {\bibinfo  {journal} {Phys.\ Rev.\ Lett.}\ }\textbf {\bibinfo
  {volume} {94}},\ \bibinfo {pages} {142301} (\bibinfo {year} {2005})},\
  \Eprint {https://arxiv.org/abs/nucl-ex/0410010} {arXiv:nucl-ex/0410010}
  \BibitemShut {NoStop}%
\bibitem [{\citenamefont {Warren}\ \emph {et~al.}(2004)\citenamefont {Warren}
  \emph {et~al.}}]{JeffersonLabE93-026:2003tty}%
  \BibitemOpen
  \bibfield  {author} {\bibinfo {author} {\bibfnamefont {G.}~\bibnamefont
  {Warren}} \emph {et~al.} (\bibinfo {collaboration} {Jefferson Lab E93-026
  Collaboration}),\ }\bibfield  {title} {\bibinfo {title} {{Measurement of the
  electric form factor of the neutron at $Q^2 = 0.5$ and $1.0$ GeV$^2/c^2$}},\
  }\href {https://doi.org/https://doi.org/10.1103/PhysRevLett.92.042301}
  {\bibfield  {journal} {\bibinfo  {journal} {Phys.\ Rev.\ Lett.}\ }\textbf
  {\bibinfo {volume} {92}},\ \bibinfo {pages} {042301} (\bibinfo {year}
  {2004})},\ \Eprint {https://arxiv.org/abs/nucl-ex/0308021}
  {arXiv:nucl-ex/0308021} \BibitemShut {NoStop}%
\bibitem [{\citenamefont {Punjabi}\ \emph {et~al.}(2005)\citenamefont {Punjabi}
  \emph {et~al.}}]{Punjabi:2005wq}%
  \BibitemOpen
  \bibfield  {author} {\bibinfo {author} {\bibfnamefont {V.}~\bibnamefont
  {Punjabi}} \emph {et~al.},\ }\bibfield  {title} {\bibinfo {title} {{Proton
  elastic form factor ratios to $Q^2 = 3.5$ GeV$^2$ by polarization
  transfer}},\ }\href
  {https://doi.org/https://doi.org/10.1103/PhysRevC.71.055202,
  https://doi.org/10.1103/PhysRevC.71.069902} {\bibfield  {journal} {\bibinfo
  {journal} {Phys.\ Rev.\ C}\ }\textbf {\bibinfo {volume} {71}},\ \bibinfo
  {pages} {055202} (\bibinfo {year} {2005})},\ \bibinfo {note} {{\relax
  Erratum: {\emph{ibid.}} {\bf{71}}, 069902 (2005)}},\ \Eprint
  {https://arxiv.org/abs/nucl-ex/0501018} {arXiv:nucl-ex/0501018} \BibitemShut
  {NoStop}%
\bibitem [{\citenamefont {Hu}\ \emph {et~al.}(2006)\citenamefont {Hu} \emph
  {et~al.}}]{Hu:2006fy}%
  \BibitemOpen
  \bibfield  {author} {\bibinfo {author} {\bibfnamefont {B.}~\bibnamefont {Hu}}
  \emph {et~al.},\ }\bibfield  {title} {\bibinfo {title} {{Polarization
  transfer in the ${}^2\text{H}\left(\vec{e},e'\vec{p}\right)n$ reaction up to
  $Q^2 = 1.61$ $\left(\text{GeV}/c\right)^2$}},\ }\href
  {https://doi.org/https://doi.org/10.1103/PhysRevC.73.064004} {\bibfield
  {journal} {\bibinfo  {journal} {Phys.\ Rev.\ C}\ }\textbf {\bibinfo {volume}
  {73}},\ \bibinfo {pages} {064004} (\bibinfo {year} {2006})},\ \Eprint
  {https://arxiv.org/abs/nucl-ex/0601025} {arXiv:nucl-ex/0601025} \BibitemShut
  {NoStop}%
\bibitem [{\citenamefont {Jones}\ \emph {et~al.}(2006)\citenamefont {Jones}
  \emph {et~al.}}]{ResonanceSpinStructure:2006oim}%
  \BibitemOpen
  \bibfield  {author} {\bibinfo {author} {\bibfnamefont {M.~K.}\ \bibnamefont
  {Jones}} \emph {et~al.} (\bibinfo {collaboration} {Resonance Spin Structure
  Collaboration}),\ }\bibfield  {title} {\bibinfo {title} {{Proton $G_E/G_M$
  from beam-target asymmetry}},\ }\href
  {https://doi.org/https://doi.org/10.1103/PhysRevC.74.035201} {\bibfield
  {journal} {\bibinfo  {journal} {Phys.\ Rev.\ C}\ }\textbf {\bibinfo {volume}
  {74}},\ \bibinfo {pages} {035201} (\bibinfo {year} {2006})},\ \Eprint
  {https://arxiv.org/abs/nucl-ex/0606015} {arXiv:nucl-ex/0606015} \BibitemShut
  {NoStop}%
\bibitem [{\citenamefont {Madey}\ \emph {et~al.}(2003)\citenamefont {Madey}
  \emph {et~al.}}]{E93-038:2003ixb}%
  \BibitemOpen
  \bibfield  {author} {\bibinfo {author} {\bibfnamefont {R.}~\bibnamefont
  {Madey}} \emph {et~al.} (\bibinfo {collaboration} {Jefferson Laboratory
  E93-038 Collaboration}),\ }\bibfield  {title} {\bibinfo {title}
  {{Measurements of $G^n_E / G^n_M$ from the
  ${}^2\text{H}\left(\vec{e},e'\vec{n}\right)$ reaction to $Q^2= 1.45
  \left(\text{GeV}/c\right)^2$}},\ }\href
  {https://doi.org/https://doi.org/10.1103/PhysRevLett.91.122002} {\bibfield
  {journal} {\bibinfo  {journal} {Phys.\ Rev.\ Lett.}\ }\textbf {\bibinfo
  {volume} {91}},\ \bibinfo {pages} {122002} (\bibinfo {year} {2003})},\
  \Eprint {https://arxiv.org/abs/nucl-ex/0308007} {arXiv:nucl-ex/0308007}
  \BibitemShut {NoStop}%
\bibitem [{\citenamefont {Plaster}\ \emph {et~al.}(2006)\citenamefont {Plaster}
  \emph {et~al.}}]{JeffersonLaboratoryE93-038:2005ryd}%
  \BibitemOpen
  \bibfield  {author} {\bibinfo {author} {\bibfnamefont {B.}~\bibnamefont
  {Plaster}} \emph {et~al.} (\bibinfo {collaboration} {Jefferson Laboratory
  E93-038 Collaboration}),\ }\bibfield  {title} {\bibinfo {title}
  {{Measurements of the neutron electric to magnetic form factor ratio
  $G_{En}/G_{Mn}$ via the
  ${}^2\text{H}\left(\vec{e},e',\vec{n}\right){}^1\text{H}$ reaction to $Q^2 =
  1.45\,\left(\text{GeV}/c\right)^2$}},\ }\href
  {https://doi.org/https://doi.org/10.1103/PhysRevC.73.025205} {\bibfield
  {journal} {\bibinfo  {journal} {Phys.\ Rev.\ C}\ }\textbf {\bibinfo {volume}
  {73}},\ \bibinfo {pages} {025205} (\bibinfo {year} {2006})},\ \Eprint
  {https://arxiv.org/abs/nucl-ex/0511025} {arXiv:nucl-ex/0511025} \BibitemShut
  {NoStop}%
\bibitem [{\citenamefont {Anderson}\ \emph {et~al.}(2007)\citenamefont
  {Anderson} \emph {et~al.}}]{JeffersonLabE95-001:2006dax}%
  \BibitemOpen
  \bibfield  {author} {\bibinfo {author} {\bibfnamefont {B.}~\bibnamefont
  {Anderson}} \emph {et~al.} (\bibinfo {collaboration} {Jefferson Lab E95-001
  Collaboration}),\ }\bibfield  {title} {\bibinfo {title} {{Extraction of the
  neutron magnetic form factor from quasielastic
  ${}^3\vec{\text{He}}\left(\vec{e},e'\right)$ at $Q^2 =
  0.1-0.6\,\left(\text{GeV}/c\right)^2$}},\ }\href
  {https://doi.org/https://doi.org/10.1103/PhysRevC.75.034003} {\bibfield
  {journal} {\bibinfo  {journal} {Phys.\ Rev.\ C}\ }\textbf {\bibinfo {volume}
  {75}},\ \bibinfo {pages} {034003} (\bibinfo {year} {2007})},\ \Eprint
  {https://arxiv.org/abs/nucl-ex/0605006} {arXiv:nucl-ex/0605006} \BibitemShut
  {NoStop}%
\bibitem [{\citenamefont {MacLachlan}\ \emph {et~al.}(2006)\citenamefont
  {MacLachlan} \emph {et~al.}}]{MacLachlan:2006vw}%
  \BibitemOpen
  \bibfield  {author} {\bibinfo {author} {\bibfnamefont {G.}~\bibnamefont
  {MacLachlan}} \emph {et~al.},\ }\bibfield  {title} {\bibinfo {title} {{The
  ratio of proton electromagnetic form factors via recoil polarimetry at $Q^2 =
  1.13\,\left(\text{GeV}/c\right)^2$}},\ }\href
  {https://doi.org/https://doi.org/10.1016/j.nuclphysa.2005.09.012} {\bibfield
  {journal} {\bibinfo  {journal} {Nucl.\ Phys.\ A}\ }\textbf {\bibinfo {volume}
  {764}},\ \bibinfo {pages} {261} (\bibinfo {year} {2006})}\BibitemShut
  {NoStop}%
\bibitem [{\citenamefont {Riordan}\ \emph {et~al.}(2010)\citenamefont {Riordan}
  \emph {et~al.}}]{Riordan:2010id}%
  \BibitemOpen
  \bibfield  {author} {\bibinfo {author} {\bibfnamefont {S.}~\bibnamefont
  {Riordan}} \emph {et~al.},\ }\bibfield  {title} {\bibinfo {title}
  {{Measurements of the electric form factor of the neutron up to $Q^2 =
  3.4\,\text{GeV}^2$ using the reaction
  ${}^3\vec{\text{He}}\left(\vec{e},e'n\right)pp$}},\ }\href
  {https://doi.org/https://doi.org/10.1103/PhysRevLett.105.262302} {\bibfield
  {journal} {\bibinfo  {journal} {Phys.\ Rev.\ Lett.}\ }\textbf {\bibinfo
  {volume} {105}},\ \bibinfo {pages} {262302} (\bibinfo {year} {2010})},\
  \Eprint {https://arxiv.org/abs/1008.1738} {1008.1738 [nucl-ex]} \BibitemShut
  {NoStop}%
\bibitem [{\citenamefont {Meziane}\ \emph {et~al.}(2011)\citenamefont {Meziane}
  \emph {et~al.}}]{GEp2gamma:2010gvp}%
  \BibitemOpen
  \bibfield  {author} {\bibinfo {author} {\bibfnamefont {M.}~\bibnamefont
  {Meziane}} \emph {et~al.} (\bibinfo {collaboration} {GEp2$\gamma$
  Collaboration}),\ }\bibfield  {title} {\bibinfo {title} {{Search for effects
  beyond the Born approximation in polarization transfer observables in
  $\vec{e}p$ elastic scattering}},\ }\href
  {https://doi.org/https://doi.org/10.1103/PhysRevLett.106.132501} {\bibfield
  {journal} {\bibinfo  {journal} {Phys.\ Rev.\ Lett.}\ }\textbf {\bibinfo
  {volume} {106}},\ \bibinfo {pages} {132501} (\bibinfo {year} {2011})},\
  \Eprint {https://arxiv.org/abs/1012.0339} {1012.0339 [nucl-ex]} \BibitemShut
  {NoStop}%
\bibitem [{\citenamefont {Puckett}\ \emph {et~al.}(2010)\citenamefont {Puckett}
  \emph {et~al.}}]{Puckett:2010ac}%
  \BibitemOpen
  \bibfield  {author} {\bibinfo {author} {\bibfnamefont {A.~J.~R.}\
  \bibnamefont {Puckett}} \emph {et~al.},\ }\bibfield  {title} {\bibinfo
  {title} {{Recoil polarization measurements of the proton electromagnetic form
  factor ratio to $Q^2 = 8.5~\text{GeV}^2$}},\ }\href
  {https://doi.org/https://doi.org/10.1103/PhysRevLett.104.242301} {\bibfield
  {journal} {\bibinfo  {journal} {Phys.\ Rev. Lett.}\ }\textbf {\bibinfo
  {volume} {104}},\ \bibinfo {pages} {242301} (\bibinfo {year} {2010})},\
  \Eprint {https://arxiv.org/abs/1005.3419} {arXiv:1005.3419 [nucl-ex]}
  \BibitemShut {NoStop}%
\bibitem [{\citenamefont {Paolone}\ \emph {et~al.}(2010)\citenamefont {Paolone}
  \emph {et~al.}}]{Paolone:2010qc}%
  \BibitemOpen
  \bibfield  {author} {\bibinfo {author} {\bibfnamefont {M.}~\bibnamefont
  {Paolone}} \emph {et~al.} (\bibinfo {collaboration} {E03-104
  Collaboration}),\ }\bibfield  {title} {\bibinfo {title} {{Polarization
  transfer in the ${}^4\text{He}\left(\vec{e},e'\vec{p}\right){}^3\text{H}$
  reaction at $Q^2 = 0.8$ and $1.3$ $\left(\text{GeV}/c\right)^2$}},\ }\href
  {https://doi.org/https://doi.org/10.1103/PhysRevLett.105.072001} {\bibfield
  {journal} {\bibinfo  {journal} {Phys.\ Rev.\ Lett.}\ }\textbf {\bibinfo
  {volume} {105}},\ \bibinfo {pages} {072001} (\bibinfo {year} {2010})},\
  \Eprint {https://arxiv.org/abs/1002.2188} {1002.2188 [nucl-ex]} \BibitemShut
  {NoStop}%
\bibitem [{\citenamefont {Zhan}\ \emph {et~al.}(2011)\citenamefont {Zhan} \emph
  {et~al.}}]{Zhan:2011ji}%
  \BibitemOpen
  \bibfield  {author} {\bibinfo {author} {\bibfnamefont {X.}~\bibnamefont
  {Zhan}} \emph {et~al.},\ }\bibfield  {title} {\bibinfo {title} {{High
  precision measurement of the proton elastic form factor ratio $\mu_p G_E/G_M$
  at low $Q^2$}},\ }\href
  {https://doi.org/https://doi.org/10.1016/j.physletb.2011.10.002} {\bibfield
  {journal} {\bibinfo  {journal} {Phys.\ Lett.\ B}\ }\textbf {\bibinfo {volume}
  {705}},\ \bibinfo {pages} {59} (\bibinfo {year} {2011})},\ \Eprint
  {https://arxiv.org/abs/1102.0318} {1102.0318 [nucl-ex]} \BibitemShut
  {NoStop}%
\bibitem [{\citenamefont {Ron}\ \emph {et~al.}(2011)\citenamefont {Ron} \emph
  {et~al.}}]{JeffersonLabHallA:2011yyi}%
  \BibitemOpen
  \bibfield  {author} {\bibinfo {author} {\bibfnamefont {G.}~\bibnamefont
  {Ron}} \emph {et~al.} (\bibinfo {collaboration} {Jefferson Lab Hall A
  Collaboration}),\ }\bibfield  {title} {\bibinfo {title} {{Low-$Q^2$
  measurements of the proton form factor ratio $\mu_p G_E/G_M$}},\ }\href
  {https://doi.org/https://doi.org/10.1103/PhysRevC.84.055204} {\bibfield
  {journal} {\bibinfo  {journal} {Phys.\ Rev.\ C}\ }\textbf {\bibinfo {volume}
  {84}},\ \bibinfo {pages} {055204} (\bibinfo {year} {2011})},\ \Eprint
  {https://arxiv.org/abs/1103.5784} {1103.5784 [nucl-ex]} \BibitemShut
  {NoStop}%
\bibitem [{\citenamefont {Puckett}\ \emph {et~al.}(2017)\citenamefont {Puckett}
  \emph {et~al.}}]{Puckett:2017flj}%
  \BibitemOpen
  \bibfield  {author} {\bibinfo {author} {\bibfnamefont {A.~J.~R.}\
  \bibnamefont {Puckett}} \emph {et~al.},\ }\bibfield  {title} {\bibinfo
  {title} {{Polarization transfer observables in elastic electron--proton
  scattering at $Q^2 = 2.5$, $5.2$, $6.8$ and $8.5$ $\text{GeV}^2$}},\ }\href
  {https://doi.org/https://doi.org/10.1103/PhysRevC.96.055203,
  https://doi.org/10.1103/PhysRevC.98.019907} {\bibfield  {journal} {\bibinfo
  {journal} {Phys.\ Rev.\ C}\ }\textbf {\bibinfo {volume} {96}},\ \bibinfo
  {pages} {055203} (\bibinfo {year} {2017})},\ \bibinfo {note} {{\relax
  Erratum: {\emph{ibid.}} {\bf{98}}, 019907 (2018)}},\ \Eprint
  {https://arxiv.org/abs/1707.08587} {1707.08587 [nucl-ex]} \BibitemShut
  {NoStop}%
\bibitem [{\citenamefont {Sulkosky}\ \emph {et~al.}(2017)\citenamefont
  {Sulkosky} \emph {et~al.}}]{Sulkosky:2017prr}%
  \BibitemOpen
  \bibfield  {author} {\bibinfo {author} {\bibfnamefont {V.}~\bibnamefont
  {Sulkosky}} \emph {et~al.} (\bibinfo {collaboration} {Jefferson Lab Hall A
  Collaboration}),\ }\bibfield  {title} {\bibinfo {title} {{Extraction of the
  neutron electric form factor from measurements of inclusive double spin
  asymmetries}},\ }\href
  {https://doi.org/https://doi.org/10.1103/PhysRevC.96.065206} {\bibfield
  {journal} {\bibinfo  {journal} {Phys.\ Rev.\ C}\ }\textbf {\bibinfo {volume}
  {96}},\ \bibinfo {pages} {065206} (\bibinfo {year} {2017})},\ \Eprint
  {https://arxiv.org/abs/1704.06253} {1704.06253 [nucl-ex]} \BibitemShut
  {NoStop}%
\bibitem [{\citenamefont {Liyanage}\ \emph {et~al.}(2020)\citenamefont
  {Liyanage} \emph {et~al.}}]{SANE:2018cub}%
  \BibitemOpen
  \bibfield  {author} {\bibinfo {author} {\bibfnamefont {A.}~\bibnamefont
  {Liyanage}} \emph {et~al.} (\bibinfo {collaboration} {SANE Collaboration}),\
  }\bibfield  {title} {\bibinfo {title} {{Proton form factor ratio $\mu_p
  G_E^p/G_M^p$ from double spin asymmetry}},\ }\href
  {https://doi.org/https://doi.org/10.1103/PhysRevC.101.035206} {\bibfield
  {journal} {\bibinfo  {journal} {Phys.\ Rev.\ C}\ }\textbf {\bibinfo {volume}
  {101}},\ \bibinfo {pages} {035206} (\bibinfo {year} {2020})},\ \Eprint
  {https://arxiv.org/abs/1806.11156} {1806.11156 [nucl-ex]} \BibitemShut
  {NoStop}%
\bibitem [{\citenamefont {Christy}\ \emph {et~al.}(2022)\citenamefont {Christy}
  \emph {et~al.}}]{Christy:2021snt}%
  \BibitemOpen
  \bibfield  {author} {\bibinfo {author} {\bibfnamefont {M.~E.}\ \bibnamefont
  {Christy}} \emph {et~al.},\ }\bibfield  {title} {\bibinfo {title} {{Form
  factors and two-photon exchange in high-energy elastic electron--proton
  scattering}},\ }\href
  {https://doi.org/https://doi.org/10.1103/PhysRevLett.128.102002} {\bibfield
  {journal} {\bibinfo  {journal} {Phys.\ Rev.\ Lett.}\ }\textbf {\bibinfo
  {volume} {128}},\ \bibinfo {pages} {102002} (\bibinfo {year} {2022})},\
  \Eprint {https://arxiv.org/abs/2103.01842} {2103.01842 [nucl-ex]}
  \BibitemShut {NoStop}%
\bibitem [{\citenamefont {Jones}\ \emph {et~al.}(2000)\citenamefont {Jones}
  \emph {et~al.}}]{JeffersonLabHallA:1999epl}%
  \BibitemOpen
  \bibfield  {author} {\bibinfo {author} {\bibfnamefont {M.~K.}\ \bibnamefont
  {Jones}} \emph {et~al.} (\bibinfo {collaboration} {Jefferson Lab Hall A
  Collaboration}),\ }\bibfield  {title} {\bibinfo {title} {{$G_{E_p}/G_{M_p}$
  ratio by polarization transfer in $\vec{e}p\to e\vec{p}$}},\ }\href
  {https://doi.org/https://doi.org/10.1103/PhysRevLett.84.1398} {\bibfield
  {journal} {\bibinfo  {journal} {Phys.\ Rev.\ Lett.}\ }\textbf {\bibinfo
  {volume} {84}},\ \bibinfo {pages} {1398} (\bibinfo {year} {2000})},\ \Eprint
  {https://arxiv.org/abs/nucl-ex/9910005} {arXiv:nucl-ex/9910005} \BibitemShut
  {NoStop}%
\bibitem [{\citenamefont {Janssens}\ \emph {et~al.}(1966)\citenamefont
  {Janssens}, \citenamefont {Hofstadter}, \citenamefont {Hughes},\ and\
  \citenamefont {Yearian}}]{Janssens:1965kd}%
  \BibitemOpen
  \bibfield  {author} {\bibinfo {author} {\bibfnamefont {T.}~\bibnamefont
  {Janssens}}, \bibinfo {author} {\bibfnamefont {R.}~\bibnamefont
  {Hofstadter}}, \bibinfo {author} {\bibfnamefont {E.~B.}\ \bibnamefont
  {Hughes}},\ and\ \bibinfo {author} {\bibfnamefont {M.~R.}\ \bibnamefont
  {Yearian}},\ }\bibfield  {title} {\bibinfo {title} {{Proton form factors from
  elastic electron--proton scattering}},\ }\href
  {https://doi.org/https://doi.org/10.1103/PhysRev.142.922} {\bibfield
  {journal} {\bibinfo  {journal} {Phys.\ Rev.}\ }\textbf {\bibinfo {volume}
  {142}},\ \bibinfo {pages} {922} (\bibinfo {year} {1966})}\BibitemShut
  {NoStop}%
\bibitem [{\citenamefont {Litt}\ \emph {et~al.}(1970)\citenamefont {Litt} \emph
  {et~al.}}]{Litt:1969my}%
  \BibitemOpen
  \bibfield  {author} {\bibinfo {author} {\bibfnamefont {J.}~\bibnamefont
  {Litt}} \emph {et~al.},\ }\bibfield  {title} {\bibinfo {title} {{Measurements
  of the ratio of the proton form factors, $G_E/G_M$ at high momentum transfers
  and the question of scaling}},\ }\href
  {https://doi.org/https://doi.org/10.1016/0370-2693(70)90015-8} {\bibfield
  {journal} {\bibinfo  {journal} {Phys.\ Lett.}\ }\textbf {\bibinfo {volume}
  {31B}},\ \bibinfo {pages} {40} (\bibinfo {year} {1970})}\BibitemShut
  {NoStop}%
\bibitem [{\citenamefont {Walker}\ \emph {et~al.}(1989)\citenamefont {Walker}
  \emph {et~al.}}]{Walker:1989af}%
  \BibitemOpen
  \bibfield  {author} {\bibinfo {author} {\bibfnamefont {R.~C.}\ \bibnamefont
  {Walker}} \emph {et~al.},\ }\bibfield  {title} {\bibinfo {title}
  {{Measurement of the proton elastic form factors for $Q^2 = 1-3$
  $\left(\text{GeV}/c\right)^2$}},\ }\href
  {https://doi.org/https://doi.org/10.1016/0370-2693(89)91245-8,
  https://doi.org/10.1016/0370-2693(90)91140-7} {\bibfield  {journal} {\bibinfo
   {journal} {Phys.\ Lett.\ B}\ }\textbf {\bibinfo {volume} {224}},\ \bibinfo
  {pages} {353} (\bibinfo {year} {1989})},\ \bibinfo {note} {{\relax Erratum:
  {\emph{ibid.}} {\bf{240}}, 522 (1990)}}\BibitemShut {NoStop}%
\bibitem [{\citenamefont {Sill}\ \emph {et~al.}(1993)\citenamefont {Sill} \emph
  {et~al.}}]{Sill:1992qw}%
  \BibitemOpen
  \bibfield  {author} {\bibinfo {author} {\bibfnamefont {A.~F.}\ \bibnamefont
  {Sill}} \emph {et~al.},\ }\bibfield  {title} {\bibinfo {title} {{Measurements
  of elastic electron--proton scattering at large momentum transfer}},\ }\href
  {https://doi.org/https://doi.org/10.1103/PhysRevD.48.29} {\bibfield
  {journal} {\bibinfo  {journal} {Phys.\ Rev.\ D}\ }\textbf {\bibinfo {volume}
  {48}},\ \bibinfo {pages} {29} (\bibinfo {year} {1993})}\BibitemShut {NoStop}%
\bibitem [{\citenamefont {Lung}\ \emph {et~al.}(1993)\citenamefont {Lung} \emph
  {et~al.}}]{Lung:1992bu}%
  \BibitemOpen
  \bibfield  {author} {\bibinfo {author} {\bibfnamefont {A.}~\bibnamefont
  {Lung}} \emph {et~al.},\ }\bibfield  {title} {\bibinfo {title} {{Measurements
  of the electric and magnetic form factors of the neutron from $Q^2= 1.75$ to
  $4.00 \left(\text{GeV}/c\right)^2$}},\ }\href
  {https://doi.org/https://doi.org/10.1103/PhysRevLett.70.718} {\bibfield
  {journal} {\bibinfo  {journal} {Phys.\ Rev.\ Lett.}\ }\textbf {\bibinfo
  {volume} {70}},\ \bibinfo {pages} {718} (\bibinfo {year} {1993})}\BibitemShut
  {NoStop}%
\bibitem [{\citenamefont {Andivahis}\ \emph {et~al.}(1994)\citenamefont
  {Andivahis} \emph {et~al.}}]{Andivahis:1994rq}%
  \BibitemOpen
  \bibfield  {author} {\bibinfo {author} {\bibfnamefont {L.}~\bibnamefont
  {Andivahis}} \emph {et~al.},\ }\bibfield  {title} {\bibinfo {title}
  {{Measurements of the electric and magnetic form factors of the proton from
  $Q^2= 1.75$ to $8.83\,\left(\text{GeV}/c\right)^2$}},\ }\href
  {https://doi.org/https://doi.org/10.1103/PhysRevD.50.5491} {\bibfield
  {journal} {\bibinfo  {journal} {Phys.\ Rev.\ D}\ }\textbf {\bibinfo {volume}
  {50}},\ \bibinfo {pages} {5491} (\bibinfo {year} {1994})}\BibitemShut
  {NoStop}%
\bibitem [{\citenamefont {Andreotti}\ \emph {et~al.}(2003)\citenamefont
  {Andreotti} \emph {et~al.}}]{Andreotti:2003bt}%
  \BibitemOpen
  \bibfield  {author} {\bibinfo {author} {\bibfnamefont {M.}~\bibnamefont
  {Andreotti}} \emph {et~al.},\ }\bibfield  {title} {\bibinfo {title}
  {{Measurements of the magnetic form factor of the proton for timelike
  momentum transfers}},\ }\href
  {https://doi.org/https://doi.org/10.1016/S0370-2693(03)00300-9} {\bibfield
  {journal} {\bibinfo  {journal} {Phys.\ Lett.\ B}\ }\textbf {\bibinfo {volume}
  {559}},\ \bibinfo {pages} {20} (\bibinfo {year} {2003})}\BibitemShut
  {NoStop}%
\bibitem [{\citenamefont {Armstrong}\ \emph {et~al.}(1993)\citenamefont
  {Armstrong} \emph {et~al.}}]{E760:1992rvj}%
  \BibitemOpen
  \bibfield  {author} {\bibinfo {author} {\bibfnamefont {T.~A.}\ \bibnamefont
  {Armstrong}} \emph {et~al.} (\bibinfo {collaboration} {E-760
  Collaboration}),\ }\bibfield  {title} {\bibinfo {title} {{Proton
  electromagnetic form factors in the timelike region from $8.9$ to
  $13.0\,\text{GeV}^2$}},\ }\href
  {https://doi.org/https://doi.org/10.1103/PhysRevLett.70.1212} {\bibfield
  {journal} {\bibinfo  {journal} {Phys.\ Rev.\ Lett.}\ }\textbf {\bibinfo
  {volume} {70}},\ \bibinfo {pages} {1212} (\bibinfo {year}
  {1993})}\BibitemShut {NoStop}%
\bibitem [{\citenamefont {Ambrogiani}\ \emph {et~al.}(1999)\citenamefont
  {Ambrogiani} \emph {et~al.}}]{E835:1999ml}%
  \BibitemOpen
  \bibfield  {author} {\bibinfo {author} {\bibfnamefont {M.}~\bibnamefont
  {Ambrogiani}} \emph {et~al.},\ }\bibfield  {title} {\bibinfo {title}
  {{Measurements of the magnetic form factor of the proton in the timelike
  region at large momentum transfer}},\ }\href
  {https://doi.org/https://doi.org/10.1103/PhysRevD.60.032002} {\bibfield
  {journal} {\bibinfo  {journal} {Phys.\ Rev.\ D}\ }\textbf {\bibinfo {volume}
  {60}},\ \bibinfo {pages} {032002} (\bibinfo {year} {1999})}\BibitemShut
  {NoStop}%
\bibitem [{\citenamefont {Lees}\ \emph
  {et~al.}(2013{\natexlab{a}})\citenamefont {Lees} \emph
  {et~al.}}]{BaBar:2013ves}%
  \BibitemOpen
  \bibfield  {author} {\bibinfo {author} {\bibfnamefont {J.~P.}\ \bibnamefont
  {Lees}} \emph {et~al.} (\bibinfo {collaboration} {{\emph{BaBar}
  Collaboration}}),\ }\bibfield  {title} {\bibinfo {title} {{Study of
  $e^+e^-\,{\to}\,p\overline{p}$ via initial-state radiation at
  {\emph{BABAR}}}},\ }\href
  {https://doi.org/https://doi.org/10.1103/PhysRevD.87.092005} {\bibfield
  {journal} {\bibinfo  {journal} {Phys.\ Rev.\ D}\ }\textbf {\bibinfo {volume}
  {87}},\ \bibinfo {pages} {092005} (\bibinfo {year} {2013}{\natexlab{a}})},\
  \Eprint {https://arxiv.org/abs/1302.0055} {1302.0055 [hep-ex]} \BibitemShut
  {NoStop}%
\bibitem [{\citenamefont {Eden}\ \emph {et~al.}(1994)\citenamefont {Eden} \emph
  {et~al.}}]{Eden:1994ji}%
  \BibitemOpen
  \bibfield  {author} {\bibinfo {author} {\bibfnamefont {T.}~\bibnamefont
  {Eden}} \emph {et~al.},\ }\bibfield  {title} {\bibinfo {title} {{Electric
  form factor of the neutron from the
  ${}^2\text{H}\left(\vec{e},e'\vec{n}\right){}^1\text{H}$ reaction at $Q^2 =
  0.255\,\left(\text{GeV}/c\right)^2$}},\ }\href
  {https://doi.org/https://doi.org/10.1103/PhysRevC.50.R1749} {\bibfield
  {journal} {\bibinfo  {journal} {Phys.\ Rev.\ C}\ }\textbf {\bibinfo {volume}
  {50}},\ \bibinfo {pages} {R1749} (\bibinfo {year} {1994})}\BibitemShut
  {NoStop}%
\bibitem [{\citenamefont {Milbrath}\ \emph {et~al.}(1998)\citenamefont
  {Milbrath} \emph {et~al.}}]{BatesFPP:1997rpw}%
  \BibitemOpen
  \bibfield  {author} {\bibinfo {author} {\bibfnamefont {B.~D.}\ \bibnamefont
  {Milbrath}} \emph {et~al.} (\bibinfo {collaboration} {Bates FPP
  Collaboration}),\ }\bibfield  {title} {\bibinfo {title} {{Comparison of
  polarization observables in electron scattering from the proton and
  deuteron}},\ }\href
  {https://doi.org/https://doi.org/10.1103/PhysRevLett.80.452,
  https://doi.org/10.1103/PhysRevLett.82.2221} {\bibfield  {journal} {\bibinfo
  {journal} {Phys.\ Rev.\ Lett.}\ }\textbf {\bibinfo {volume} {80}},\ \bibinfo
  {pages} {452} (\bibinfo {year} {1998})},\ \bibinfo {note} {{\relax Erratum:
  {\emph{ibid.}} {\bf{82}}, 2221 (1999)}},\ \Eprint
  {https://arxiv.org/abs/nucl-ex/9712006} {arXiv:nucl-ex/9712006} \BibitemShut
  {NoStop}%
\bibitem [{\citenamefont {Crawford}\ \emph {et~al.}(2007)\citenamefont
  {Crawford} \emph {et~al.}}]{Crawford:2006rz}%
  \BibitemOpen
  \bibfield  {author} {\bibinfo {author} {\bibfnamefont {C.~B.}\ \bibnamefont
  {Crawford}} \emph {et~al.},\ }\bibfield  {title} {\bibinfo {title}
  {{Measurement of the proton's electric to magnetic form factor ratio from
  ${}^1\vec{\text{H}}\left(\vec{e},e'p\right)$}},\ }\href
  {https://doi.org/https://doi.org/10.1103/PhysRevLett.98.052301} {\bibfield
  {journal} {\bibinfo  {journal} {Phys.\ Rev.\ Lett.}\ }\textbf {\bibinfo
  {volume} {98}},\ \bibinfo {pages} {052301} (\bibinfo {year} {2007})},\
  \Eprint {https://arxiv.org/abs/nucl-ex/0609007} {arXiv:nucl-ex/0609007}
  \BibitemShut {NoStop}%
\bibitem [{\citenamefont {Geis}\ \emph {et~al.}(2008)\citenamefont {Geis} \emph
  {et~al.}}]{BLAST:2008bub}%
  \BibitemOpen
  \bibfield  {author} {\bibinfo {author} {\bibfnamefont {E.}~\bibnamefont
  {Geis}} \emph {et~al.} (\bibinfo {collaboration} {BLAST Collaboration}),\
  }\bibfield  {title} {\bibinfo {title} {{Charge form factor of the neutron at
  low momentum transfer from the
  ${}^2\vec{\text{H}}\left(\vec{e},e'n\right){}^1\text{H}$ reaction}},\ }\href
  {https://doi.org/https://doi.org/10.1103/PhysRevLett.101.042501} {\bibfield
  {journal} {\bibinfo  {journal} {Phys.\ Rev.\ Lett.}\ }\textbf {\bibinfo
  {volume} {101}},\ \bibinfo {pages} {042501} (\bibinfo {year} {2008})},\
  \Eprint {https://arxiv.org/abs/0803.3827} {0803.3827 [nucl-ex]} \BibitemShut
  {NoStop}%
\bibitem [{\citenamefont {Pedlar}\ \emph {et~al.}(2005)\citenamefont {Pedlar}
  \emph {et~al.}}]{CLEO:2005tiu}%
  \BibitemOpen
  \bibfield  {author} {\bibinfo {author} {\bibfnamefont {T.~K.}\ \bibnamefont
  {Pedlar}} \emph {et~al.} (\bibinfo {collaboration} {CLEO Collaboration}),\
  }\bibfield  {title} {\bibinfo {title} {{Precision measurements of the
  timelike electromagnetic form factors of pion, kaon, and proton}},\ }\href
  {https://doi.org/https://doi.org/10.1103/PhysRevLett.95.261803} {\bibfield
  {journal} {\bibinfo  {journal} {Phys.\ Rev.\ Lett.}\ }\textbf {\bibinfo
  {volume} {95}},\ \bibinfo {pages} {261803} (\bibinfo {year} {2005})},\
  \Eprint {https://arxiv.org/abs/hep-ex/0510005} {arXiv:hep-ex/0510005}
  \BibitemShut {NoStop}%
\bibitem [{\citenamefont {Anklin}\ \emph {et~al.}(1994)\citenamefont {Anklin}
  \emph {et~al.}}]{Anklin:1994ae}%
  \BibitemOpen
  \bibfield  {author} {\bibinfo {author} {\bibfnamefont {H.}~\bibnamefont
  {Anklin}} \emph {et~al.},\ }\bibfield  {title} {\bibinfo {title} {{Precision
  measurement of the neutron magnetic form factor}},\ }\href
  {https://doi.org/https://doi.org/10.1016/0370-2693(94)90538-X} {\bibfield
  {journal} {\bibinfo  {journal} {Phys.\ Lett.\ B}\ }\textbf {\bibinfo {volume}
  {336}},\ \bibinfo {pages} {313} (\bibinfo {year} {1994})}\BibitemShut
  {NoStop}%
\bibitem [{\citenamefont {Passchier}\ \emph {et~al.}(1999)\citenamefont
  {Passchier} \emph {et~al.}}]{Passchier:1999cj}%
  \BibitemOpen
  \bibfield  {author} {\bibinfo {author} {\bibfnamefont {I.}~\bibnamefont
  {Passchier}} \emph {et~al.},\ }\bibfield  {title} {\bibinfo {title} {{Charge
  form factor of the neutron from the reaction
  ${}^2\vec{\text{H}}\left(\vec{e},e'n\right)p$}},\ }\href
  {https://doi.org/https://doi.org/10.1103/PhysRevLett.82.4988} {\bibfield
  {journal} {\bibinfo  {journal} {Phys.\ Rev.\ Lett.}\ }\textbf {\bibinfo
  {volume} {82}},\ \bibinfo {pages} {4988} (\bibinfo {year} {1999})},\ \Eprint
  {https://arxiv.org/abs/nucl-ex/9907012} {arXiv:nucl-ex/9907012} \BibitemShut
  {NoStop}%
\bibitem [{\citenamefont {Passchier}\ \emph {et~al.}(2000)\citenamefont
  {Passchier} \emph {et~al.}}]{Passchier:1999ju}%
  \BibitemOpen
  \bibfield  {author} {\bibinfo {author} {\bibfnamefont {I.}~\bibnamefont
  {Passchier}} \emph {et~al.},\ }\bibfield  {title} {\bibinfo {title} {{The
  charge form factor of the neutron from
  ${}^2\vec{\text{H}}\left(\vec{e},e'n\right)p$}},\ }\href
  {https://doi.org/https://doi.org/10.1016/S0375-9474(99)00673-9} {\bibfield
  {journal} {\bibinfo  {journal} {Nucl.\ Phys.\ A}\ }\textbf {\bibinfo {volume}
  {663\&664}},\ \bibinfo {pages} {421c} (\bibinfo {year} {2000})},\ \Eprint
  {https://arxiv.org/abs/nucl-ex/9908002} {arXiv:nucl-ex/9908002} \BibitemShut
  {NoStop}%
\bibitem [{\citenamefont {Bartel}\ \emph {et~al.}(1973)\citenamefont {Bartel}
  \emph {et~al.}}]{Bartel:1973rf}%
  \BibitemOpen
  \bibfield  {author} {\bibinfo {author} {\bibfnamefont {W.}~\bibnamefont
  {Bartel}} \emph {et~al.},\ }\bibfield  {title} {\bibinfo {title}
  {{Measurement of proton and neutron electromagnetic form factors at squared
  four-momentum transfers up to $3~\left(\text{GeV}/c\right)^2$}},\ }\href
  {https://doi.org/https://doi.org/10.1016/0550-3213(73)90594-4} {\bibfield
  {journal} {\bibinfo  {journal} {Nucl.\ Phys.\ B}\ }\textbf {\bibinfo {volume}
  {58}},\ \bibinfo {pages} {429} (\bibinfo {year} {1973})}\BibitemShut
  {NoStop}%
\bibitem [{\citenamefont {Berger}\ \emph {et~al.}(1971)\citenamefont {Berger}
  \emph {et~al.}}]{Berger:1971kr}%
  \BibitemOpen
  \bibfield  {author} {\bibinfo {author} {\bibfnamefont {C.}~\bibnamefont
  {Berger}} \emph {et~al.},\ }\bibfield  {title} {\bibinfo {title}
  {{Electromagnetic form factors of the proton at squared four-momentum
  transfers between 10 and 50~fm$^{-2}$}},\ }\href
  {https://doi.org/https://doi.org/10.1016/0370-2693(71)90448-5} {\bibfield
  {journal} {\bibinfo  {journal} {Phys.\ Lett.}\ }\textbf {\bibinfo {volume}
  {35B}},\ \bibinfo {pages} {87} (\bibinfo {year} {1971})}\BibitemShut
  {NoStop}%
\bibitem [{\citenamefont {Borkowski}\ \emph {et~al.}(1975)\citenamefont
  {Borkowski} \emph {et~al.}}]{Borkowski:1974mb}%
  \BibitemOpen
  \bibfield  {author} {\bibinfo {author} {\bibfnamefont {F.}~\bibnamefont
  {Borkowski}} \emph {et~al.},\ }\bibfield  {title} {\bibinfo {title}
  {{Electromagnetic form factors of the proton at low four-momentum transfer
  (II)}},\ }\href
  {https://doi.org/https://doi.org/10.1016/0550-3213(75)90514-3} {\bibfield
  {journal} {\bibinfo  {journal} {Nucl.\ Phys.\ B}\ }\textbf {\bibinfo {volume}
  {93}},\ \bibinfo {pages} {461} (\bibinfo {year} {1975})}\BibitemShut
  {NoStop}%
\bibitem [{\citenamefont {Anklin}\ \emph {et~al.}(1998)\citenamefont {Anklin}
  \emph {et~al.}}]{Anklin:1998ae}%
  \BibitemOpen
  \bibfield  {author} {\bibinfo {author} {\bibfnamefont {H.}~\bibnamefont
  {Anklin}} \emph {et~al.},\ }\bibfield  {title} {\bibinfo {title} {{Precise
  measurements of the neutron magnetic form factor}},\ }\href
  {https://doi.org/https://doi.org/10.1016/S0370-2693(98)00442-0} {\bibfield
  {journal} {\bibinfo  {journal} {Phys.\ Lett.\ B}\ }\textbf {\bibinfo {volume}
  {428}},\ \bibinfo {pages} {248} (\bibinfo {year} {1998})}\BibitemShut
  {NoStop}%
\bibitem [{\citenamefont {Herberg}\ \emph {et~al.}(1999)\citenamefont {Herberg}
  \emph {et~al.}}]{Herberg:1999ud}%
  \BibitemOpen
  \bibfield  {author} {\bibinfo {author} {\bibfnamefont {C.}~\bibnamefont
  {Herberg}} \emph {et~al.},\ }\bibfield  {title} {\bibinfo {title}
  {{Determination of the neutron electric form factor in the $D(e,e'n)p$
  reaction and the influence of nuclear binding}},\ }\href
  {https://doi.org/https://doi.org/10.1007/s100500050268} {\bibfield  {journal}
  {\bibinfo  {journal} {Eur.\ Phys.\ J.\ A}\ }\textbf {\bibinfo {volume} {5}},\
  \bibinfo {pages} {131} (\bibinfo {year} {1999})}\BibitemShut {NoStop}%
\bibitem [{\citenamefont {Rohe}\ \emph {et~al.}(1999)\citenamefont {Rohe} \emph
  {et~al.}}]{Rohe:1999sh}%
  \BibitemOpen
  \bibfield  {author} {\bibinfo {author} {\bibfnamefont {D.}~\bibnamefont
  {Rohe}} \emph {et~al.},\ }\bibfield  {title} {\bibinfo {title} {{Measurement
  of the neutron electric form factor $G_{en}$ at
  $0.67~\left(\text{GeV}/c\right)^2$ via
  ${}^3\vec{\text{He}}\left(\vec{e},e'n\right)$}},\ }\href
  {https://doi.org/https://doi.org/10.1103/PhysRevLett.83.4257} {\bibfield
  {journal} {\bibinfo  {journal} {Phys.\ Rev.\ Lett.}\ }\textbf {\bibinfo
  {volume} {83}},\ \bibinfo {pages} {4257} (\bibinfo {year}
  {1999})}\BibitemShut {NoStop}%
\bibitem [{\citenamefont {Pospischil}\ \emph {et~al.}(2001)\citenamefont
  {Pospischil} \emph {et~al.}}]{A1:2001xxy}%
  \BibitemOpen
  \bibfield  {author} {\bibinfo {author} {\bibfnamefont {T.}~\bibnamefont
  {Pospischil}} \emph {et~al.},\ }\bibfield  {title} {\bibinfo {title}
  {{Measurement of $G_{Ep}/G_{Mp}$ via polarization transfer at $Q^2 =
  0.4\,\text{GeV}/c^2$}},\ }\href
  {https://doi.org/https://doi.org/10.1007/s100500170046} {\bibfield  {journal}
  {\bibinfo  {journal} {Eur.\ Phys.\ J.\ A}\ }\textbf {\bibinfo {volume}
  {12}},\ \bibinfo {pages} {125} (\bibinfo {year} {2001})}\BibitemShut
  {NoStop}%
\bibitem [{\citenamefont {Kubon}\ \emph {et~al.}(2002)\citenamefont {Kubon}
  \emph {et~al.}}]{Kubon:2001rj}%
  \BibitemOpen
  \bibfield  {author} {\bibinfo {author} {\bibfnamefont {G.}~\bibnamefont
  {Kubon}} \emph {et~al.},\ }\bibfield  {title} {\bibinfo {title} {{Precise
  neutron magnetic form factors}},\ }\href
  {https://doi.org/https://doi.org/10.1016/S0370-2693(01)01386-7} {\bibfield
  {journal} {\bibinfo  {journal} {Phys.\ Lett.\ B}\ }\textbf {\bibinfo {volume}
  {524}},\ \bibinfo {pages} {26} (\bibinfo {year} {2002})},\ \Eprint
  {https://arxiv.org/abs/nucl-ex/0107016} {arXiv:nucl-ex/0107016} \BibitemShut
  {NoStop}%
\bibitem [{\citenamefont {Schlimme}\ \emph {et~al.}(2013)\citenamefont
  {Schlimme} \emph {et~al.}}]{Schlimme:2013eoz}%
  \BibitemOpen
  \bibfield  {author} {\bibinfo {author} {\bibfnamefont {B.~S.}\ \bibnamefont
  {Schlimme}} \emph {et~al.},\ }\bibfield  {title} {\bibinfo {title}
  {{Measurement of the neutron electric to magnetic form factor ratio at $Q^2 =
  1.58$ GeV$^2$ using the reaction
  ${}^3\vec{\text{He}}\left(\vec{e},e'n\right)pp$}},\ }\href
  {https://doi.org/https://doi.org/10.1103/PhysRevLett.111.132504} {\bibfield
  {journal} {\bibinfo  {journal} {Phys.\ Rev.\ Lett.}\ }\textbf {\bibinfo
  {volume} {111}},\ \bibinfo {pages} {132504} (\bibinfo {year} {2013})},\
  \Eprint {https://arxiv.org/abs/1307.7361} {1307.7361 [nucl-ex]} \BibitemShut
  {NoStop}%
\bibitem [{\citenamefont {Mihovilovi\v{c}}\ \emph {et~al.}(2017)\citenamefont
  {Mihovilovi\v{c}} \emph {et~al.}}]{Mihovilovic:2016rkr}%
  \BibitemOpen
  \bibfield  {author} {\bibinfo {author} {\bibfnamefont {M.}~\bibnamefont
  {Mihovilovi\v{c}}} \emph {et~al.},\ }\bibfield  {title} {\bibinfo {title}
  {{First measurement of proton's charge form factor at very low $Q^2$ with
  initial state radiation}},\ }\href
  {https://doi.org/https://doi.org/10.1016/j.physletb.2017.05.031} {\bibfield
  {journal} {\bibinfo  {journal} {Phys.\ Lett.\ B}\ }\textbf {\bibinfo {volume}
  {771}},\ \bibinfo {pages} {194} (\bibinfo {year} {2017})},\ \Eprint
  {https://arxiv.org/abs/1612.06707} {1612.06707 [nucl-ex]} \BibitemShut
  {NoStop}%
\bibitem [{\citenamefont {Borkowski}\ \emph {et~al.}(1974)\citenamefont
  {Borkowski} \emph {et~al.}}]{Borkowski:1974tm}%
  \BibitemOpen
  \bibfield  {author} {\bibinfo {author} {\bibfnamefont {F.}~\bibnamefont
  {Borkowski}} \emph {et~al.},\ }\bibfield  {title} {\bibinfo {title}
  {{Electromagnetic form factors of the proton at low four-momentum
  transfer}},\ }\href {https://doi.org/10.1016/0375-9474(74)90392-3} {\bibfield
   {journal} {\bibinfo  {journal} {Nucl.\ Phys.\ A}\ }\textbf {\bibinfo
  {volume} {222}},\ \bibinfo {pages} {269} (\bibinfo {year}
  {1974})}\BibitemShut {NoStop}%
\bibitem [{\citenamefont {Simon}\ \emph {et~al.}(1980)\citenamefont {Simon},
  \citenamefont {Schmitt}, \citenamefont {Borkowski},\ and\ \citenamefont
  {Walther}}]{Simon:1980hu}%
  \BibitemOpen
  \bibfield  {author} {\bibinfo {author} {\bibfnamefont {G.~G.}\ \bibnamefont
  {Simon}}, \bibinfo {author} {\bibfnamefont {C.}~\bibnamefont {Schmitt}},
  \bibinfo {author} {\bibfnamefont {F.}~\bibnamefont {Borkowski}},\ and\
  \bibinfo {author} {\bibfnamefont {V.~H.}\ \bibnamefont {Walther}},\
  }\bibfield  {title} {\bibinfo {title} {{Absolute electron--proton cross
  sections at low momentum transfer measured with a high pressure gas target
  system}},\ }\href
  {https://doi.org/https://doi.org/10.1016/0375-9474(80)90104-9} {\bibfield
  {journal} {\bibinfo  {journal} {Nucl.\ Phys.\ A}\ }\textbf {\bibinfo {volume}
  {333}},\ \bibinfo {pages} {381} (\bibinfo {year} {1980})}\BibitemShut
  {NoStop}%
\bibitem [{\citenamefont {Castro}\ \emph {et~al.}()\citenamefont {Castro} \emph
  {et~al.}}]{DM2:1988rej}%
  \BibitemOpen
  \bibfield  {author} {\bibinfo {author} {\bibfnamefont {A.}~\bibnamefont
  {Castro}} \emph {et~al.},\ }\bibfield  {title} {\bibinfo {title} {{The $\pi$,
  $K$, proton electromagnetic form factors and new related DM2 results}},\ }in\
  \href@noop {} {\emph {\bibinfo {booktitle} {Proceedings of the Nucleon
  Stucture Workshop: FENICE Experiment and Investigation of the Neutron Form
  Factor, 27--28 October 1988, Frascati, Italy, Printed and published by
  Servizio Documentazione dei Laboratori Nazionali di Frascati,
  Eq.~Mrs.~L.~Invidia, 1988}}},\ \bibinfo {note} {{\relax LAL
  88-58}}\BibitemShut {NoStop}%
\bibitem [{\citenamefont {Bisello}\ \emph {et~al.}(1990)\citenamefont {Bisello}
  \emph {et~al.}}]{DM2:1990tut}%
  \BibitemOpen
  \bibfield  {author} {\bibinfo {author} {\bibfnamefont {D.}~\bibnamefont
  {Bisello}} \emph {et~al.} (\bibinfo {collaboration} {DM2 Collaboration}),\
  }\bibfield  {title} {\bibinfo {title} {{Baryon pairs production in $e^+e^-$
  annihilation at $\sqrt{s}=2.4$ GeV}},\ }\href
  {https://doi.org/https://doi.org/10.1007/BF01565602} {\bibfield  {journal}
  {\bibinfo  {journal} {Z.\ Phys.\ C}\ }\textbf {\bibinfo {volume} {48}},\
  \bibinfo {pages} {23} (\bibinfo {year} {1990})}\BibitemShut {NoStop}%
\bibitem [{\citenamefont {Biagini}\ \emph {et~al.}(1991)\citenamefont
  {Biagini}, \citenamefont {Pasqualucci},\ and\ \citenamefont
  {Rotondo}}]{Biagini:1990nb}%
  \BibitemOpen
  \bibfield  {author} {\bibinfo {author} {\bibfnamefont {M.~E.}\ \bibnamefont
  {Biagini}}, \bibinfo {author} {\bibfnamefont {E.}~\bibnamefont
  {Pasqualucci}},\ and\ \bibinfo {author} {\bibfnamefont {A.}~\bibnamefont
  {Rotondo}},\ }\bibfield  {title} {\bibinfo {title} {{$U$-spin considerations
  to guess the unknown time-like neutron form factors}},\ }\href
  {https://doi.org/https://doi.org/10.1007/BF01562337} {\bibfield  {journal}
  {\bibinfo  {journal} {Z.\ Phys.\ C}\ }\textbf {\bibinfo {volume} {52}},\
  \bibinfo {pages} {631} (\bibinfo {year} {1991})}\BibitemShut {NoStop}%
\bibitem [{\citenamefont {Antonelli}\ \emph
  {et~al.}(1993{\natexlab{b}})\citenamefont {Antonelli} \emph
  {et~al.}}]{Antonelli:1992ha}%
  \BibitemOpen
  \bibfield  {author} {\bibinfo {author} {\bibfnamefont {A.}~\bibnamefont
  {Antonelli}} \emph {et~al.},\ }\bibfield  {title} {\bibinfo {title} {{A new
  measurement of $J/\psi\,{\to}\,n\overline{n}$}},\ }\href
  {https://doi.org/https://doi.org/10.1016/0370-2693(93)90708-P} {\bibfield
  {journal} {\bibinfo  {journal} {Phys.\ Lett.\ B}\ }\textbf {\bibinfo {volume}
  {301}},\ \bibinfo {pages} {317} (\bibinfo {year}
  {1993}{\natexlab{b}})}\BibitemShut {NoStop}%
\bibitem [{\citenamefont {Antonelli}\ \emph {et~al.}(1998)\citenamefont
  {Antonelli} \emph {et~al.}}]{Antonelli:1998fv}%
  \BibitemOpen
  \bibfield  {author} {\bibinfo {author} {\bibfnamefont {A.}~\bibnamefont
  {Antonelli}} \emph {et~al.},\ }\bibfield  {title} {\bibinfo {title} {{The
  first measurement of the neutron electromagnetic form factors in the
  time-like region}},\ }\href
  {https://doi.org/https://doi.org/10.1016/S0550-3213(98)00083-2} {\bibfield
  {journal} {\bibinfo  {journal} {Nucl.\ Phys.\ B}\ }\textbf {\bibinfo {volume}
  {517}},\ \bibinfo {pages} {3} (\bibinfo {year} {1998})}\BibitemShut {NoStop}%
\bibitem [{\citenamefont {Bassompierre}\ \emph {et~al.}(1983)\citenamefont
  {Bassompierre} \emph {et~al.}}]{Bassompierre:1983kt}%
  \BibitemOpen
  \bibfield  {author} {\bibinfo {author} {\bibfnamefont {G.}~\bibnamefont
  {Bassompierre}} \emph {et~al.},\ }\bibfield  {title} {\bibinfo {title}
  {{Electron--positron pair production in $\overline{p}p$ annihilation at rest
  and related determination of the electromagnetic form factor of the proton in
  the timelike region}},\ }\href
  {https://doi.org/https://doi.org/10.1007/BF02724235} {\bibfield  {journal}
  {\bibinfo  {journal} {Il Nuovo Cim.}\ }\textbf {\bibinfo {volume} {73A}},\
  \bibinfo {pages} {348} (\bibinfo {year} {1983})}\BibitemShut {NoStop}%
\bibitem [{\citenamefont {Bardin}\ \emph {et~al.}(1991)\citenamefont {Bardin}
  \emph {et~al.}}]{Bardin:1991rz}%
  \BibitemOpen
  \bibfield  {author} {\bibinfo {author} {\bibfnamefont {G.}~\bibnamefont
  {Bardin}} \emph {et~al.},\ }\bibfield  {title} {\bibinfo {title}
  {{Measurement of the proton electromagnetic form factor near threshold in the
  time-like region}},\ }\href
  {https://doi.org/https://doi.org/10.1016/0370-2693(91)91157-Q} {\bibfield
  {journal} {\bibinfo  {journal} {Phys.\ Lett.\ B}\ }\textbf {\bibinfo {volume}
  {255}},\ \bibinfo {pages} {149} (\bibinfo {year} {1991})}\BibitemShut
  {NoStop}%
\bibitem [{\citenamefont {Achasov}\ \emph {et~al.}(2014)\citenamefont {Achasov}
  \emph {et~al.}}]{Achasov:2014ncd}%
  \BibitemOpen
  \bibfield  {author} {\bibinfo {author} {\bibfnamefont {M.~N.}\ \bibnamefont
  {Achasov}} \emph {et~al.},\ }\bibfield  {title} {\bibinfo {title} {{Study of
  the process $e^+e^-\,{\to}\,n\overline{n}$ at the VEPP-2000 $e^+e^-$ Collider
  with the SND detector}},\ }\href
  {https://doi.org/https://doi.org/10.1103/PhysRevD.90.112007} {\bibfield
  {journal} {\bibinfo  {journal} {Phys.\ Rev.\ D}\ }\textbf {\bibinfo {volume}
  {90}},\ \bibinfo {pages} {112007} (\bibinfo {year} {2014})},\ \Eprint
  {https://arxiv.org/abs/1410.3188} {1410.3188 [hep-ex]} \BibitemShut {NoStop}%
\bibitem [{\citenamefont {Akhmetshin}\ \emph {et~al.}(2016)\citenamefont
  {Akhmetshin} \emph {et~al.}}]{CMD-3:2015fvi}%
  \BibitemOpen
  \bibfield  {author} {\bibinfo {author} {\bibfnamefont {R.~R.}\ \bibnamefont
  {Akhmetshin}} \emph {et~al.},\ }\bibfield  {title} {\bibinfo {title} {{Study
  of the process $e^+e^-\,{\to}\,p\overline{p}$ in the c.m. energy range from
  threshold to 2 GeV with the CMD-3 detector}},\ }\href
  {https://doi.org/https://doi.org/10.1016/j.physletb.2016.04.048} {\bibfield
  {journal} {\bibinfo  {journal} {Phys. Lett. B}\ }\textbf {\bibinfo {volume}
  {759}},\ \bibinfo {pages} {634} (\bibinfo {year} {2016})},\ \Eprint
  {https://arxiv.org/abs/1507.08013} {1507.08013 [hep-ex]} \BibitemShut
  {NoStop}%
\bibitem [{\citenamefont {Druzhinin}\ and\ \citenamefont
  {Serednyakov}(2019)}]{Druzhinin:2019gpo}%
  \BibitemOpen
  \bibfield  {author} {\bibinfo {author} {\bibfnamefont {V.~P.}\ \bibnamefont
  {Druzhinin}}\ and\ \bibinfo {author} {\bibfnamefont {S.~I.}\ \bibnamefont
  {Serednyakov}},\ }\bibfield  {title} {\bibinfo {title} {{Measurement of the
  $e^+e^-\,{\to}\,n\overline{n}$ cross section with the SND detector at the
  VEPP-2000 Collider}},\ }\href
  {https://doi.org/https://doi.org/10.1051/epjconf/201921207007} {\bibfield
  {journal} {\bibinfo  {journal} {Eur.\ Phys.\ J.\ Web of Conf.}\ }\textbf
  {\bibinfo {volume} {212}},\ \bibinfo {pages} {07007} (\bibinfo {year}
  {2019})}\BibitemShut {NoStop}%
\bibitem [{\citenamefont {Achasov}\ \emph
  {et~al.}(2023{\natexlab{b}})\citenamefont {Achasov} \emph
  {et~al.}}]{Achasov:2023hic}%
  \BibitemOpen
  \bibfield  {author} {\bibinfo {author} {\bibfnamefont {M.~N.}\ \bibnamefont
  {Achasov}} \emph {et~al.},\ }\bibfield  {title} {\bibinfo {title}
  {{Experimental study of the $e^+e^-\to n\overline{n}$ process at the
  VEPP-2000 Collider with the SND detector}},\ }\href
  {https://doi.org/https://doi.org/10.1134/S1063779623040020} {\bibfield
  {journal} {\bibinfo  {journal} {Phys.\ Part.\ Nucl.}\ }\textbf {\bibinfo
  {volume} {54}},\ \bibinfo {pages} {624} (\bibinfo {year}
  {2023}{\natexlab{b}})}\BibitemShut {NoStop}%
\bibitem [{\citenamefont {Korol}\ \emph {et~al.}(2023)\citenamefont {Korol}
  \emph {et~al.}}]{Korol:2023bop}%
  \BibitemOpen
  \bibfield  {author} {\bibinfo {author} {\bibfnamefont {A.~A.}\ \bibnamefont
  {Korol}} \emph {et~al.},\ }\bibfield  {title} {\bibinfo {title} {{Recent SND
  experiment results on $e^+e^-$ annihilation to hadrons}},\ }\href
  {https://doi.org/https://doi.org/10.1142/S2010194523610050} {\bibfield
  {journal} {\bibinfo  {journal} {Int.\ J.\ Mod.\ Phys.: Conf.\ Ser.}\ }\textbf
  {\bibinfo {volume} {51}},\ \bibinfo {pages} {2361005} (\bibinfo {year}
  {2023})}\BibitemShut {NoStop}%
\bibitem [{\citenamefont {Ablikim}\ \emph {et~al.}(2005)\citenamefont {Ablikim}
  \emph {et~al.}}]{BES:2005lpy}%
  \BibitemOpen
  \bibfield  {author} {\bibinfo {author} {\bibfnamefont {M.}~\bibnamefont
  {Ablikim}} \emph {et~al.} (\bibinfo {collaboration} {BES Collaboration}),\
  }\bibfield  {title} {\bibinfo {title} {{Measurement of the cross section for
  $e^+e^-\,{\to}\,p\overline{p}$ center-of-mass energies from $2.0$ to $3.07$
  GeV}},\ }\href
  {https://doi.org/https://doi.org/10.1016/j.physletb.2005.09.044} {\bibfield
  {journal} {\bibinfo  {journal} {Phys.\ Lett.\ B}\ }\textbf {\bibinfo {volume}
  {630}},\ \bibinfo {pages} {14} (\bibinfo {year} {2005})},\ \Eprint
  {https://arxiv.org/abs/hep-ex/0506059} {arXiv:hep-ex/0506059} \BibitemShut
  {NoStop}%
\bibitem [{\citenamefont {Ablikim}\ \emph {et~al.}(2015)\citenamefont {Ablikim}
  \emph {et~al.}}]{BESIII:2015axk}%
  \BibitemOpen
  \bibfield  {author} {\bibinfo {author} {\bibfnamefont {M.}~\bibnamefont
  {Ablikim}} \emph {et~al.} (\bibinfo {collaboration} {BESIII Collaboration}),\
  }\bibfield  {title} {\bibinfo {title} {{Measurement of the proton form factor
  by studying $e^+e^-\,{\to}\,p\overline{p}$}},\ }\href
  {https://doi.org/https://doi.org/10.1103/PhysRevD.91.112004} {\bibfield
  {journal} {\bibinfo  {journal} {Phys.\ Rev.\ D}\ }\textbf {\bibinfo {volume}
  {91}},\ \bibinfo {pages} {112004} (\bibinfo {year} {2015})},\ \Eprint
  {https://arxiv.org/abs/1504.02680} {1504.02680 [hep-ex]} \BibitemShut
  {NoStop}%
\bibitem [{\citenamefont {Ablikim}\ \emph {et~al.}(2019)\citenamefont {Ablikim}
  \emph {et~al.}}]{BESIII:2019tgo}%
  \BibitemOpen
  \bibfield  {author} {\bibinfo {author} {\bibfnamefont {M.}~\bibnamefont
  {Ablikim}} \emph {et~al.} (\bibinfo {collaboration} {BESIII Collaboration}),\
  }\bibfield  {title} {\bibinfo {title} {{Study of the process
  $e^+e^-\,{\to}\,p\overline{p}$ via initial state radiation at BESIII}},\
  }\href {https://doi.org/https://doi.org/10.1103/PhysRevD.99.092002}
  {\bibfield  {journal} {\bibinfo  {journal} {Phys.\ Rev.\ D}\ }\textbf
  {\bibinfo {volume} {99}},\ \bibinfo {pages} {092002} (\bibinfo {year}
  {2019})},\ \Eprint {https://arxiv.org/abs/1902.00665} {1902.00665 [hep-ex]}
  \BibitemShut {NoStop}%
\bibitem [{\citenamefont {Ablikim}\ \emph {et~al.}(2020)\citenamefont {Ablikim}
  \emph {et~al.}}]{BESIII:2019hdp}%
  \BibitemOpen
  \bibfield  {author} {\bibinfo {author} {\bibfnamefont {M.}~\bibnamefont
  {Ablikim}} \emph {et~al.} (\bibinfo {collaboration} {BESIII Collaboration}),\
  }\bibfield  {title} {\bibinfo {title} {{Measurement of proton electromagnetic
  form factors in $e^+e^-\,{\to}\,p\overline{p}$ in the energy region
  $2.00-3.08$ GeV}},\ }\href
  {https://doi.org/https://doi.org/10.1103/PhysRevLett.124.042001} {\bibfield
  {journal} {\bibinfo  {journal} {Phys.\ Rev.\ Lett.}\ }\textbf {\bibinfo
  {volume} {124}},\ \bibinfo {pages} {042001} (\bibinfo {year} {2020})},\
  \Eprint {https://arxiv.org/abs/1905.09001} {1905.09001 [hep-ex]} \BibitemShut
  {NoStop}%
\bibitem [{\citenamefont {Ablikim}\ \emph
  {et~al.}(2021{\natexlab{b}})\citenamefont {Ablikim} \emph
  {et~al.}}]{BESIII:2021tbq}%
  \BibitemOpen
  \bibfield  {author} {\bibinfo {author} {\bibfnamefont {M.}~\bibnamefont
  {Ablikim}} \emph {et~al.} (\bibinfo {collaboration} {BESIII Collaboration}),\
  }\bibfield  {title} {\bibinfo {title} {{Oscillating features in the
  electromagnetic structure of the neutron}},\ }\href
  {https://doi.org/https://doi.org/10.1038/s41567-021-01345-6} {\bibfield
  {journal} {\bibinfo  {journal} {Nature Phys.}\ }\textbf {\bibinfo {volume}
  {17}},\ \bibinfo {pages} {1200} (\bibinfo {year} {2021}{\natexlab{b}})},\
  \Eprint {https://arxiv.org/abs/2103.12486} {2103.12486 [hep-ex]} \BibitemShut
  {NoStop}%
\bibitem [{\citenamefont {Ablikim}\ \emph {et~al.}(2023)\citenamefont {Ablikim}
  \emph {et~al.}}]{BESIII:2022rrg}%
  \BibitemOpen
  \bibfield  {author} {\bibinfo {author} {\bibfnamefont {M.}~\bibnamefont
  {Ablikim}} \emph {et~al.} (\bibinfo {collaboration} {BESIII Collaboration}),\
  }\bibfield  {title} {\bibinfo {title} {{Measurements of the electric and
  magnetic form factors of the neutron for timelike momentum transfer}},\
  }\href {https://doi.org/https://doi.org/10.1103/PhysRevLett.130.151905}
  {\bibfield  {journal} {\bibinfo  {journal} {Phys.\ Rev.\ Lett.}\ }\textbf
  {\bibinfo {volume} {130}},\ \bibinfo {pages} {151905} (\bibinfo {year}
  {2023})},\ \Eprint {https://arxiv.org/abs/2212.07071} {2212.07071 [hep-ex]}
  \BibitemShut {NoStop}%
\bibitem [{\citenamefont {Brash}\ \emph {et~al.}(2002)\citenamefont {Brash},
  \citenamefont {Kozlov}, \citenamefont {Li},\ and\ \citenamefont
  {Huber}}]{Brash:2001qq}%
  \BibitemOpen
  \bibfield  {author} {\bibinfo {author} {\bibfnamefont {E.~J.}\ \bibnamefont
  {Brash}}, \bibinfo {author} {\bibfnamefont {A.}~\bibnamefont {Kozlov}},
  \bibinfo {author} {\bibfnamefont {S.}~\bibnamefont {Li}},\ and\ \bibinfo
  {author} {\bibfnamefont {G.~M.}\ \bibnamefont {Huber}},\ }\bibfield  {title}
  {\bibinfo {title} {{New empirical fits to the proton electromagnetic form
  factors}},\ }\href
  {https://doi.org/https://doi.org/10.1103/PhysRevC.65.051001} {\bibfield
  {journal} {\bibinfo  {journal} {Phys.\ Rev.\ C}\ }\textbf {\bibinfo {volume}
  {65}},\ \bibinfo {pages} {051001(R)} (\bibinfo {year} {2002})},\ \Eprint
  {https://arxiv.org/abs/hep-ex/0111038} {arXiv:hep-ex/0111038} \BibitemShut
  {NoStop}%
\bibitem [{\citenamefont {Arrington}(2003)}]{Arrington:2003df}%
  \BibitemOpen
  \bibfield  {author} {\bibinfo {author} {\bibfnamefont {J.~R.}\ \bibnamefont
  {Arrington}},\ }\bibfield  {title} {\bibinfo {title} {{How well do we know
  the electromagnetic form factors of the proton?}},\ }\href
  {https://doi.org/https://doi.org/10.1103/PhysRevC.68.034325} {\bibfield
  {journal} {\bibinfo  {journal} {Phys.\ Rev.\ C}\ }\textbf {\bibinfo {volume}
  {68}},\ \bibinfo {pages} {034325} (\bibinfo {year} {2003})},\ \Eprint
  {https://arxiv.org/abs/nucl-ex/0305009} {arXiv:nucl-ex/0305009} \BibitemShut
  {NoStop}%
\bibitem [{\citenamefont {Lendel}\ \emph {et~al.}(1966)\citenamefont {Lendel},
  \citenamefont {Lendyel}, \citenamefont {Meshcheryakov},\ and\ \citenamefont
  {Ernst}}]{Lendel:1966}%
  \BibitemOpen
  \bibfield  {author} {\bibinfo {author} {\bibfnamefont {A.~I.}\ \bibnamefont
  {Lendel}}, \bibinfo {author} {\bibfnamefont {V.~I.}\ \bibnamefont {Lendyel}},
  \bibinfo {author} {\bibfnamefont {V.~A.}\ \bibnamefont {Meshcheryakov}},\
  and\ \bibinfo {author} {\bibfnamefont {B.~M.}\ \bibnamefont {Ernst}},\
  }\bibfield  {title} {\bibinfo {title} {{Influence of $\pi\pi$-interactions on
  the electromagnetic structure of the nucleon}},\ }\href@noop {} {\bibfield
  {journal} {\bibinfo  {journal} {Yad.\ Fiz.}\ }\textbf {\bibinfo {volume}
  {3}},\ \bibinfo {pages} {1093} (\bibinfo {year} {1966})}\BibitemShut
  {NoStop}%
\bibitem [{\citenamefont {H$\ddot{\text{o}}$hler}\ and\ \citenamefont
  {Pietarinen}(1975)}]{Hohler:1974ht}%
  \BibitemOpen
  \bibfield  {author} {\bibinfo {author} {\bibfnamefont {G.}~\bibnamefont
  {H$\ddot{\text{o}}$hler}}\ and\ \bibinfo {author} {\bibfnamefont
  {E.}~\bibnamefont {Pietarinen}},\ }\bibfield  {title} {\bibinfo {title} {{The
  $\rho NN$ vertex in vector-dominance model}},\ }\href
  {https://doi.org/https://doi.org/10.1016/0550-3213(75)90042-5} {\bibfield
  {journal} {\bibinfo  {journal} {Nucl.\ Phys.\ B}\ }\textbf {\bibinfo {volume}
  {95}},\ \bibinfo {pages} {210} (\bibinfo {year} {1975})}\BibitemShut
  {NoStop}%
\bibitem [{\citenamefont {Krivoruchenko}()}]{Krivoruchenko:1997nb}%
  \BibitemOpen
  \bibfield  {author} {\bibinfo {author} {\bibfnamefont {M.~I.}\ \bibnamefont
  {Krivoruchenko}},\ }\href@noop {} {\bibinfo {title} {{Analytical extension of
  the Frazer-Fulco unitarity relations for isovector nucleon form factors to
  the complex $t$-plane}}},\ \Eprint {https://arxiv.org/abs/nucl-th/9710072}
  {arXiv:nucl-th/9710072} \BibitemShut {NoStop}%
\bibitem [{\citenamefont {Machleidt}\ \emph {et~al.}(1987)\citenamefont
  {Machleidt}, \citenamefont {Holinde},\ and\ \citenamefont
  {Elster}}]{Machleidt:1987hj}%
  \BibitemOpen
  \bibfield  {author} {\bibinfo {author} {\bibfnamefont {R.}~\bibnamefont
  {Machleidt}}, \bibinfo {author} {\bibfnamefont {K.}~\bibnamefont {Holinde}},\
  and\ \bibinfo {author} {\bibfnamefont {C.}~\bibnamefont {Elster}},\
  }\bibfield  {title} {\bibinfo {title} {{The Bonn meson exchange model for the
  nucleon--nucleon interaction}},\ }\href
  {https://doi.org/https://doi.org/10.1016/S0370-1573(87)80002-9} {\bibfield
  {journal} {\bibinfo  {journal} {Phys.\ Rept.}\ }\textbf {\bibinfo {volume}
  {149}},\ \bibinfo {pages} {1} (\bibinfo {year} {1987})}\BibitemShut {NoStop}%
\bibitem [{\citenamefont {James}(1994)}]{James:1994vla}%
  \BibitemOpen
  \bibfield  {author} {\bibinfo {author} {\bibfnamefont {F.}~\bibnamefont
  {James}},\ }\href@noop {} {\emph {\bibinfo {title} {{MINUIT -- function
  minimization and error analysis: Reference manual version 94.1}}}} (\bibinfo
  {year} {1994}),\ \bibinfo {note} {cERN-D-506}\BibitemShut {NoStop}%
\bibitem [{\citenamefont {James}\ and\ \citenamefont
  {Roos}(1975)}]{James:1975dr}%
  \BibitemOpen
  \bibfield  {author} {\bibinfo {author} {\bibfnamefont {F.}~\bibnamefont
  {James}}\ and\ \bibinfo {author} {\bibfnamefont {M.}~\bibnamefont {Roos}},\
  }\bibfield  {title} {\bibinfo {title} {{MINUIT -- a system for function
  minimization and analysis of the parameter errors and correlations}},\ }\href
  {https://doi.org/https://doi.org/10.1016/0010-4655(75)90039-9} {\bibfield
  {journal} {\bibinfo  {journal} {Comput.\ Phys.\ Commun.}\ }\textbf {\bibinfo
  {volume} {10}},\ \bibinfo {pages} {343} (\bibinfo {year} {1975})}\BibitemShut
  {NoStop}%
\bibitem [{\citenamefont {Cowan}(1998)}]{Cowan:1998}%
  \BibitemOpen
  \bibfield  {author} {\bibinfo {author} {\bibfnamefont {G.}~\bibnamefont
  {Cowan}},\ }\href@noop {} {\emph {\bibinfo {title} {{Statistical data
  analysis}}}}\ (\bibinfo  {publisher} {{\relax Clarendon Press Pub., Oxford
  University Press Inc.}},\ \bibinfo {address} {Oxford, England},\ \bibinfo
  {year} {1998})\BibitemShut {NoStop}%
\bibitem [{\citenamefont {Distler}\ \emph {et~al.}(2011)\citenamefont
  {Distler}, \citenamefont {Bernauer},\ and\ \citenamefont
  {Walcher}}]{Distler:2010zq}%
  \BibitemOpen
  \bibfield  {author} {\bibinfo {author} {\bibfnamefont {M.~O.}\ \bibnamefont
  {Distler}}, \bibinfo {author} {\bibfnamefont {J.~C.}\ \bibnamefont
  {Bernauer}},\ and\ \bibinfo {author} {\bibfnamefont {T.}~\bibnamefont
  {Walcher}},\ }\bibfield  {title} {\bibinfo {title} {{The RMS charge radius of
  the proton and Zemach moments}},\ }\href
  {https://doi.org/https://doi.org/10.1016/j.physletb.2010.12.067} {\bibfield
  {journal} {\bibinfo  {journal} {Phys.\ Lett.\ B}\ }\textbf {\bibinfo {volume}
  {696}},\ \bibinfo {pages} {343} (\bibinfo {year} {2011})},\ \Eprint
  {https://arxiv.org/abs/1011.1861} {1011.1861 [nucl-th]} \BibitemShut
  {NoStop}%
\bibitem [{\citenamefont {Bernauer}\ \emph {et~al.}(2010)\citenamefont
  {Bernauer} \emph {et~al.}}]{A1:2010nsl}%
  \BibitemOpen
  \bibfield  {author} {\bibinfo {author} {\bibfnamefont {J.~C.}\ \bibnamefont
  {Bernauer}} \emph {et~al.} (\bibinfo {collaboration} {A1 Collaboration}),\
  }\bibfield  {title} {\bibinfo {title} {{High-precision determination of the
  electric and magnetic form factors of the proton}},\ }\href
  {https://doi.org/https://doi.org/10.1103/PhysRevLett.105.242001} {\bibfield
  {journal} {\bibinfo  {journal} {Phys.\ Rev.\ Lett.}\ }\textbf {\bibinfo
  {volume} {105}},\ \bibinfo {pages} {242001} (\bibinfo {year} {2010})},\
  \Eprint {https://arxiv.org/abs/1007.5076} {24:1007.5076 [nucl-ex]}
  \BibitemShut {NoStop}%
\bibitem [{\citenamefont {Arrington}\ \emph {et~al.}(2007)\citenamefont
  {Arrington}, \citenamefont {Melnitchouk},\ and\ \citenamefont
  {Tjon}}]{Arrington:2007ux}%
  \BibitemOpen
  \bibfield  {author} {\bibinfo {author} {\bibfnamefont {J.}~\bibnamefont
  {Arrington}}, \bibinfo {author} {\bibfnamefont {W.}~\bibnamefont
  {Melnitchouk}},\ and\ \bibinfo {author} {\bibfnamefont {J.~A.}\ \bibnamefont
  {Tjon}},\ }\bibfield  {title} {\bibinfo {title} {{Global analysis of proton
  elastic form factor data with two-photon exchange corrections}},\ }\href
  {https://doi.org/https://doi.org/10.1103/PhysRevC.76.035205} {\bibfield
  {journal} {\bibinfo  {journal} {Phys.\ Rev.\ C}\ }\textbf {\bibinfo {volume}
  {76}},\ \bibinfo {pages} {035205} (\bibinfo {year} {2007})},\ \Eprint
  {https://arxiv.org/abs/0707.1861} {0707.1861 [nucl-ex]} \BibitemShut
  {NoStop}%
\bibitem [{\citenamefont {Friar}\ and\ \citenamefont
  {Sick}(2005)}]{Friar:2005jz}%
  \BibitemOpen
  \bibfield  {author} {\bibinfo {author} {\bibfnamefont {J.~L.}\ \bibnamefont
  {Friar}}\ and\ \bibinfo {author} {\bibfnamefont {I.}~\bibnamefont {Sick}},\
  }\bibfield  {title} {\bibinfo {title} {{Muonic hydrogen and the third Zemach
  moment}},\ }\href
  {https://doi.org/https://doi.org/10.1103/PhysRevA.72.040502} {\bibfield
  {journal} {\bibinfo  {journal} {Phys.\ Rev.\ A}\ }\textbf {\bibinfo {volume}
  {72}},\ \bibinfo {pages} {040502(R)} (\bibinfo {year} {2005})},\ \Eprint
  {https://arxiv.org/abs/nucl-th/0508025} {arXiv:nucl-th/0508025} \BibitemShut
  {NoStop}%
\bibitem [{\citenamefont {Bianconi}\ and\ \citenamefont
  {Tomasi-Gustafsson}(2015)}]{Bianconi:2015owa}%
  \BibitemOpen
  \bibfield  {author} {\bibinfo {author} {\bibfnamefont {A.}~\bibnamefont
  {Bianconi}}\ and\ \bibinfo {author} {\bibfnamefont {E.}~\bibnamefont
  {Tomasi-Gustafsson}},\ }\bibfield  {title} {\bibinfo {title} {{Periodic
  interference structures in the timelike proton form factor}},\ }\href
  {https://doi.org/https://doi.org/10.1103/PhysRevLett.114.232301} {\bibfield
  {journal} {\bibinfo  {journal} {Phys.\ Rev.\ Lett.}\ }\textbf {\bibinfo
  {volume} {114}},\ \bibinfo {pages} {232301} (\bibinfo {year} {2015})},\
  \Eprint {https://arxiv.org/abs/1503.02140} {1503.02140 [nucl-th]}
  \BibitemShut {NoStop}%
\bibitem [{\citenamefont {Lees}\ \emph
  {et~al.}(2013{\natexlab{b}})\citenamefont {Lees} \emph
  {et~al.}}]{BaBar:2013ukx}%
  \BibitemOpen
  \bibfield  {author} {\bibinfo {author} {\bibfnamefont {J.~P.}\ \bibnamefont
  {Lees}} \emph {et~al.} (\bibinfo {collaboration} {\emph{BaBar}
  Collaboration}),\ }\bibfield  {title} {\bibinfo {title} {{Measurement of the
  $e^+e^-\to p\overline{p}$ cross section in the energy range from 3.0 to 6.5
  GeV}},\ }\href {https://doi.org/https://doi.org/10.1103/PhysRevD.88.072009}
  {\bibfield  {journal} {\bibinfo  {journal} {Phys.\ Rev.\ D}\ }\textbf
  {\bibinfo {volume} {88}},\ \bibinfo {pages} {072009} (\bibinfo {year}
  {2013}{\natexlab{b}})},\ \Eprint {https://arxiv.org/abs/1308.1795} {1308.1795
  [hep-ex]} \BibitemShut {NoStop}%
\bibitem [{\citenamefont {Zhou}\ and\ \citenamefont
  {Timmermans}(2012)}]{Zhou:2012ui}%
  \BibitemOpen
  \bibfield  {author} {\bibinfo {author} {\bibfnamefont {D.}~\bibnamefont
  {Zhou}}\ and\ \bibinfo {author} {\bibfnamefont {R.~G.~E.}\ \bibnamefont
  {Timmermans}},\ }\bibfield  {title} {\bibinfo {title} {{Energy-dependent
  partial-wave analysis of all antiproton--proton scattering data below 925
  MeV$/c$}},\ }\href
  {https://doi.org/https://doi.org/10.1103/PhysRevC.86.044003} {\bibfield
  {journal} {\bibinfo  {journal} {Phys.\ Rev.\ C}\ }\textbf {\bibinfo {volume}
  {86}},\ \bibinfo {pages} {044003} (\bibinfo {year} {2012})},\ \Eprint
  {https://arxiv.org/abs/1210.7074} {1210.7074 [hep-ph]} \BibitemShut {NoStop}%
\bibitem [{\citenamefont {Yang}\ \emph {et~al.}(2024)\citenamefont {Yang},
  \citenamefont {Guo}, \citenamefont {Li}, \citenamefont {Dai}, \citenamefont
  {Haidenbauer},\ and\ \citenamefont {Mei\ss{}ner}}]{Yang:2024iuc}%
  \BibitemOpen
  \bibfield  {author} {\bibinfo {author} {\bibfnamefont {Q.-H.}\ \bibnamefont
  {Yang}}, \bibinfo {author} {\bibfnamefont {D.}~\bibnamefont {Guo}}, \bibinfo
  {author} {\bibfnamefont {M.-Y.}\ \bibnamefont {Li}}, \bibinfo {author}
  {\bibfnamefont {L.-Y.}\ \bibnamefont {Dai}}, \bibinfo {author} {\bibfnamefont
  {J.}~\bibnamefont {Haidenbauer}},\ and\ \bibinfo {author} {\bibfnamefont
  {U.-G.}\ \bibnamefont {Mei\ss{}ner}},\ }\bibfield  {title} {\bibinfo {title}
  {{Study of the electromagnetic form factors of the nucleons in the timelike
  region}},\ }\href {https://doi.org/https://doi.org/10.1007/JHEP08(2024)208}
  {\bibfield  {journal} {\bibinfo  {journal} {JHEP}\ }\textbf {\bibinfo
  {volume} {08}},\ \bibinfo {pages} {208}},\ \Eprint
  {https://arxiv.org/abs/2404.12448} {2404.12448 [nucl-th]} \BibitemShut
  {NoStop}%
\bibitem [{\citenamefont {Yan}\ \emph {et~al.}(2024)\citenamefont {Yan},
  \citenamefont {Chen}, \citenamefont {Li},\ and\ \citenamefont
  {Xie}}]{Yan:2023nlb}%
  \BibitemOpen
  \bibfield  {author} {\bibinfo {author} {\bibfnamefont {B.}~\bibnamefont
  {Yan}}, \bibinfo {author} {\bibfnamefont {C.}~\bibnamefont {Chen}}, \bibinfo
  {author} {\bibfnamefont {X.}~\bibnamefont {Li}},\ and\ \bibinfo {author}
  {\bibfnamefont {J.-J.}\ \bibnamefont {Xie}},\ }\bibfield  {title} {\bibinfo
  {title} {{Understanding oscillating features of the time-like nucleon
  electromagnetic form factors within the extending vector meson dominance
  model}},\ }\href
  {https://doi.org/https://doi.org/10.1103/PhysRevD.109.036033} {\bibfield
  {journal} {\bibinfo  {journal} {Phys.\ Rev.\ D}\ }\textbf {\bibinfo {volume}
  {109}},\ \bibinfo {pages} {036033} (\bibinfo {year} {2024})},\ \Eprint
  {https://arxiv.org/abs/2312.04866} {2312.04866 [nucl-th]} \BibitemShut
  {NoStop}%
\bibitem [{\citenamefont {Wang}\ \emph {et~al.}(2022)\citenamefont {Wang},
  \citenamefont {Luo},\ and\ \citenamefont {Liu}}]{Wang:2021abg}%
  \BibitemOpen
  \bibfield  {author} {\bibinfo {author} {\bibfnamefont {L.-M.}\ \bibnamefont
  {Wang}}, \bibinfo {author} {\bibfnamefont {S.-Q.}\ \bibnamefont {Luo}},\ and\
  \bibinfo {author} {\bibfnamefont {X.}~\bibnamefont {Liu}},\ }\bibfield
  {title} {\bibinfo {title} {{Light unflavored vector meson spectroscopy around
  the mass range of $2.4 \sim 3$~GeV and possible experimental evidence}},\
  }\href {https://doi.org/https://doi.org/10.1103/PhysRevD.105.034011}
  {\bibfield  {journal} {\bibinfo  {journal} {Phys.\ Rev.\ D}\ }\textbf
  {\bibinfo {volume} {105}},\ \bibinfo {pages} {034011} (\bibinfo {year}
  {2022})},\ \Eprint {https://arxiv.org/abs/2109.06617} {2109.06617 [hep-ph]}
  \BibitemShut {NoStop}%
\end{thebibliography}%

  \end{document}